\documentclass[aps,prd,twocolumn,floatfix,preprintnumbers,nofootinbib,superscriptaddress]{revtex4-1}
\usepackage{mathtools}
\usepackage{graphicx}
\usepackage{color}
\usepackage{natbib}
\usepackage{multirow}
\usepackage{amsmath}
\usepackage{amsfonts}
\usepackage[amssymb]{SIunits}
\usepackage{empheq}
\usepackage{color}
\usepackage[colorlinks=true,citecolor=blue,urlcolor=blue]{hyperref}

\newcommand{\barDe}{\bar{d}}
\newcommand{\bartHe}{\overline{{}^3{\rm He}}}

\newcommand{\barqHe}{\overline{{}^4{\rm He}}}
\begin{document}
\title{Where do the {\it AMS-02} anti-helium events come from?}
\author{Vivian Poulin}
\affiliation{Department of Physics and Astronomy, Johns Hopkins University, Baltimore, MD 21218, USA}
\author{Pierre Salati}
\affiliation{{LAPTh}, Universit\'e Savoie Mont Blanc \& CNRS, 74941 Annecy Cedex, France}
\author{Ilias Cholis}
\affiliation{Department of Physics, Oakland University, Rochester, MI 48309, USA}
\affiliation{Department of Physics and Astronomy, Johns Hopkins University, Baltimore, MD 21218, USA}
\author{Marc Kamionkowski}
\affiliation{Department of Physics and Astronomy, Johns Hopkins University, Baltimore, MD 21218, USA}
\author{Joseph Silk}
\affiliation{Department of Physics and Astronomy, Johns Hopkins University, Baltimore, MD 21218, USA}
\affiliation{Sorbonne Universit\'es, UPMC Univ. Paris 6 et CNRS, UMR 7095, Institut d’Astrophysique de Paris, 98 bis bd Arago, 75014 Paris, France}
\affiliation{Beecroft Institute of Particle Astrophysics and Cosmology, Department of Physics, University of Oxford, Denys Wilkinson Building, 1 Keble Road, Oxford OX1 3RH, UK}

\date{\today}

\preprint{LAPTH-034/18}
\begin{abstract}
We discuss the origin of the anti-helium-3 and -4 events possibly detected by {\it AMS-02}. Using up-to-date semi-analytical tools, we show that spallation from primary hydrogen and helium nuclei onto the ISM predicts a $\bartHe$ flux typically one to two orders of magnitude below the sensitivity of {\it AMS-02} after 5 years, and a $\barqHe$ flux roughly 5 orders of magnitude below the {\it AMS-02} sensitivity.  We argue that dark matter annihilations  face similar difficulties in explaining this event.  We then entertain the possibility that these events originate from anti-matter-dominated regions in the form of anti-clouds or anti-stars. In the case of anti-clouds, we show how the isotopic ratio of anti-helium nuclei might suggest that BBN has happened in an inhomogeneous manner, resulting in anti-regions with a anti-baryon-to-photon ratio  $\bar{\eta}\simeq10^{-3}\eta$. We discuss properties of these regions, as well as relevant constraints on the presence of anti-clouds in our Galaxy. We present constraints from the survival of anti-clouds in the Milky-Way and in the early Universe, as well as from CMB,  gamma-ray and cosmic-ray observations.  In particular, these require the anti-clouds to be almost free of normal matter. We also discuss an alternative where anti-domains are dominated by surviving anti-stars. We suggest that part of the unindentified sources in the 3FGL catalog can originate from  anti-clouds or anti-stars. {\it AMS-02} and {\it GAPS} data could further probe this scenario. 
\end{abstract}
\maketitle
%
\section{Introduction}
The origin of cosmic ray (CR) anti-matter is one of the many conundrums that {\it AMS-02} is trying to solve thanks to precise measurements of CR fluxes at the Earth. In over six years, {\it AMS-02} has accumulated several billion events, whose composition is mostly dominated by protons and helium nuclei. Moreover, positrons and antiprotons have been frequently observed and are the object of intense theoretical investigations in order to explain their spectral features. Indeed, anti-matter particles are believed to be mainly of secondary origin, i.e., they are created by primary CR nuclei (accelerated by supernova-driven shock waves) impinging onto the interstellar medium (ISM).  However, deviations from these standard predictions have been observed, hinting at a possible primary component. In the case of positrons, a very significant high-energy excess  has already been seen in {\it PAMELA} data \cite{Adriani:2010rc}. The main sources under investigation to explain this excess are DM and pulsars (see e.g. \cite{Bergstrom:2008gr, Cirelli:2008jk, Cirelli:2008pk, Nelson:2008hj, ArkaniHamed:2008qn, 
Harnik:2008uu, Fox:2008kb, Pospelov:2008jd, MarchRussell:2008tu, Dienes:2013xff, Kopp:2013eka, Yuksel:2008rf, Profumo:2008ms, Kawanaka:2009dk, Yuan:2013eja, Yin:2013vaa,Hooper:2008kg,Cholis:2008qq,Cholis:2008hb,Cholis:2013psa,Boudaud:2014dta,Boudaud:2016jvj,Hooper:2017gtd,Cholis:2017ccs,Cholis:2018izy}). In the case of antiprotons, a putative excess at the GeV-energy \cite{Cuoco:2016eej} is under discussion \cite{Reinert:2017aga}. Still, antiprotons represent one of the most promising probes to look for the presence of DM in our Galaxy through its annihilation.   

But the searches for anti-matter CR do not limit themselves to antiprotons and positrons. Hence, many theoretical and experimental efforts are devoted to detecting anti-deuterons, which are believed to be a very clean probe of DM annihilations especially at the lowest energies (below tens of GeV) \cite{Chardonnet:1997dv,Donato:1999gy,Duperray:2005si,vonDoetinchem:2015yva}. Similarly, measurement of the anti-helium nuclei CR flux is a very promising probe of new physics, that has been suggested to look for DM annihilations \cite{Duperray:2005si,Carlson:2014ssa,Cirelli:2014qia,Coogan:2017pwt} or other sources of primary CR, such as anti-matter stars or clouds \cite{Steigman:1976ev,Belotsky:1998kz,Bambi:2007cc}. Strikingly, {\it AMS-02} has recently reported the possible discovery of eight anti-helium events in the mass region from 0 to 10 GeV/c$^2$ with $Z= 2$ and rigidity $<50$ GV \cite{Choutko}. Six of the events are compatible with being anti-helium-3 and two events with anti-helium-4. The total event rate is roughly one anti-helium in a hundred million heliums. This preliminary sample includes one event with a momentum of $32.6\pm2.5$ GeV/c and a mass of $3.81\pm 0.29$ GeV/c$^2$ compatible with that of anti-helium-4. Earlier already, another event with a momentum of 40.3 $\pm$ 2.9 GeV and a mass compatible with anti-helium-3 had been reported \cite{TingTalk}.

In this paper, we discuss various possibilities for the origin of {\it AMS-02} anti-helium events. Should these events be confirmed, their detection would be a breakthrough discovery, with immediate and considerable implications onto our current understanding of cosmology. The discovery of a single anti-helium-4 nucleus is challenging to explain in terms of known physics.
In this article, we start stressing why such a discovery is unexpected. For this, we re-evaluate the secondary flux of anti-helium nuclei. In particular, we provide the first estimate of the $\barqHe$ flux at the Earth coming from the spallation of primary CR onto the ISM. We show that it is impossible to explain AMS results in terms of a pure secondary component, even though large uncertainties still affect the prediction. Moreover, we argue that the DM explanations of these events face similar difficulties, although given the virtually infinite freedom in the building of DM models, it is conceivable that a tuned scenario might succeed in explaining these events.

We then discuss the implications of the anti-helium observation. We essentially suggest that the putative detection of $\bartHe$ and $\barqHe$ by {\it AMS-02} indicates the existence of an anti-world, i.e., a world made of anti-matter, in the form of anti-stars or anti-clouds. We discuss properties of these regions, as well as relevant constraints on the presence of anti-clouds in our Galaxy. We present constraints from the survival of anti-clouds in the Milky-Way and in the early Universe, as well as from CMB, gamma-ray and cosmic-ray observations.  We show in particular that these require the anti-clouds to be almost free of normal matter. Moreover, we show how the isotopic ratio of anti-helium nuclei might suggest that BBN happened inhomogeneously, resulting in anti-regions with a anti-baryon-to-photon ratio  $\bar{\eta}\simeq10^{-3}\,\eta$. Given the very strong constraints applying to the existence and survival  of anti-clouds,  we also discuss an alternative scenario in which anti-domains are dominated by anti-stars.  We suggest that part of the unidentified sources in the 3FGL catalog can be anti-clouds or anti-stars.  Future {\it AMS-02} and {\it GAPS} data could further probe this scenario.

The paper is structured as follows. Section \ref{sec:secondary} is devoted to a thorough re-evaluation of the secondary astrophysical component from spallation within the coalescence scheme. A discussion on the possible limitations of our estimates and on the DM scenario is also provided. In section \ref{sec:antidomain}, we discuss the possibility of anti-domains in our Galaxy being responsible for {\it AMS-02} events. Properties of anti-clouds and their constraints are presented in sec~\ref{sec:anticloud}, while the alternative anti-star scenario is developed in section~\ref{sec:antistar}. Finally, we draw our conclusions in sec.~\ref{sec:discussion}. 
%
\section{Updated calculation of $\barDe$, $\bartHe$ and $\barqHe$ from spallation onto the ISM}\label{sec:secondary}
As for any secondaries,  the prediction of the $\bartHe$ flux at Earth is the result of two main processes affected by potentially large uncertainties: i) the {\em production} due to spallation of primary CR onto the ISM and ii) the {\em propagation} of cosmic rays in the magnetic field of our Galaxy, eventually modulated by the impact of the Sun.
In this section, we briefly review how to calculate the secondary flux of $\bartHe$ from spallation onto the ISM in a semi-analytical way.
%
\subsection{Source term for anti-nuclei in the coalescence scenario}

The spallation production cross-section of an anti-nucleus A from the collision of a primary CR species $i$ onto an ISM species $j$ can be computed within the {\em coalescence scenario} as follows:
\begin{equation}
\frac{E_A}{\sigma_{ij}} \frac{d^3\sigma^{ij}_A}{d^3k_A} = B_A \cdot
\bigg(\frac{E_p}{\sigma_{ij}} \frac{d^3\sigma^{ij}_p}{d^3k_p} \bigg)^{\! Z} \cdot
\bigg(\frac{E_n}{\sigma_{ij}} \frac{d^3\sigma^{ij}_n}{d^3k_n}\bigg)^{\! A-Z}\,,
\label{eq:coal_1}
\end{equation}
where $\sigma^{ij}$ is the total inelastic cross section for the $ij$ collision, and the constituent momenta are taken at $k_p=k_n=k_A/A$.
$B_A$ is the coalescence factor, whose role is to capture the probability for A anti-nucleons produced in a collision to merge into a composite anti-nucleus. It is often written as
\begin{equation}
B_A = \bigg(\frac{4\pi}{3} \frac{p_{\rm coal}^3}{8}\bigg)^{\! A-1}
\frac{m_A}{m_p^{Z} m_n^{A-Z}}\,,
\label{eq:BA}
\end{equation}
where $p_{\rm coal}$ is the diameter of a sphere in phase-space within which anti-nucleons have to lie in order to form an anti-nucleus.
The coalescence factor $B_A$ is a key quantity which can be estimated from $pp$-collision data, as  has been done recently by the ALICE collaboration \cite{Acharya:2017fvb} for anti-deuteron and anti-helium. We use the values measured at low transverse momentum as these are adequate for CR spallation, namely $B_2 \simeq (15\pm5)\times10^{-2}$ GeV$^2$ and $B_3\simeq(2\pm1)\times10^{-4}$ GeV$^4$. We extrapolate these values to $pA$ and $AA$ collisions. There is no measurement of $B_4$ yet available. Hence, we make use of eq.~\ref{eq:BA} in order to extract the coalescence momentum (common to each species in the coalescence model) from the $B_3$ measurement. This gives a coalesence momentum that varies between $0.218$ GeV and $0.262$ GeV. Using the measurement of $B_2$, the coalescence momentum varies between  $0.208$ GeV and $0.262$ GeV, which is in excellent agreement with the value extracted from $B_3$. We stress that the fact that the coalescence momenta extracted from both coalescence factors agree is far from trival. It indicates that the coalescence scenario is much more predictive and accurate than one might have naively expected from its apparent simplicity. To phrase this otherwise: from the $B_2$ measurement of ALICE, one can predict how many anti-helium-3 ALICE should measure; this turns out to be in very good agreement with the actual measurement, which is quite remarkable. The final step is thus to apply eq.~\ref{eq:BA} to the case of anti-helium-4. We find that $B_4$ varies between $7.7\times10^{-7}$ GeV$^{6}$ and $3.9\times10^{-6}$ GeV$^{6}$.

In the context of antiproton production, it has been found that \cite{diMauro:2014zea}
\begin{equation}
E_n\frac{d^3\sigma^{ij}_n}{d^3k_n} = \Delta_{np} \, E_p\frac{d^3\sigma^{ij}_p}{d^3k_p}\,,
\end{equation}
where $\Delta_{np}\simeq 1.3$ is introduced to model the isospin symmetry breaking. However, the ALICE experiment has extracted $B_2$ and $B_3$ assuming a perfect isospin symmetry between the antineutron and antiproton production. Hence, it would be wrong to make use of the factor $1.3$ in this context and we set $\Delta_{np} = 1$.
Additionally, we follow ref.~\cite{Chardonnet:1997dv} and compute the anti-deuterium production cross-section by evaluating the production cross-sections of the two anti-nucleons at respectively $\sqrt{s}$ and $\sqrt{s} -2E^{*}_p$ where $E^{*}_p$ denotes the anti-nucleon energy in the center of mass frame of the collision.
%
\begin{figure}[h!]
\hspace{-1.15cm}
\includegraphics[scale=0.42]{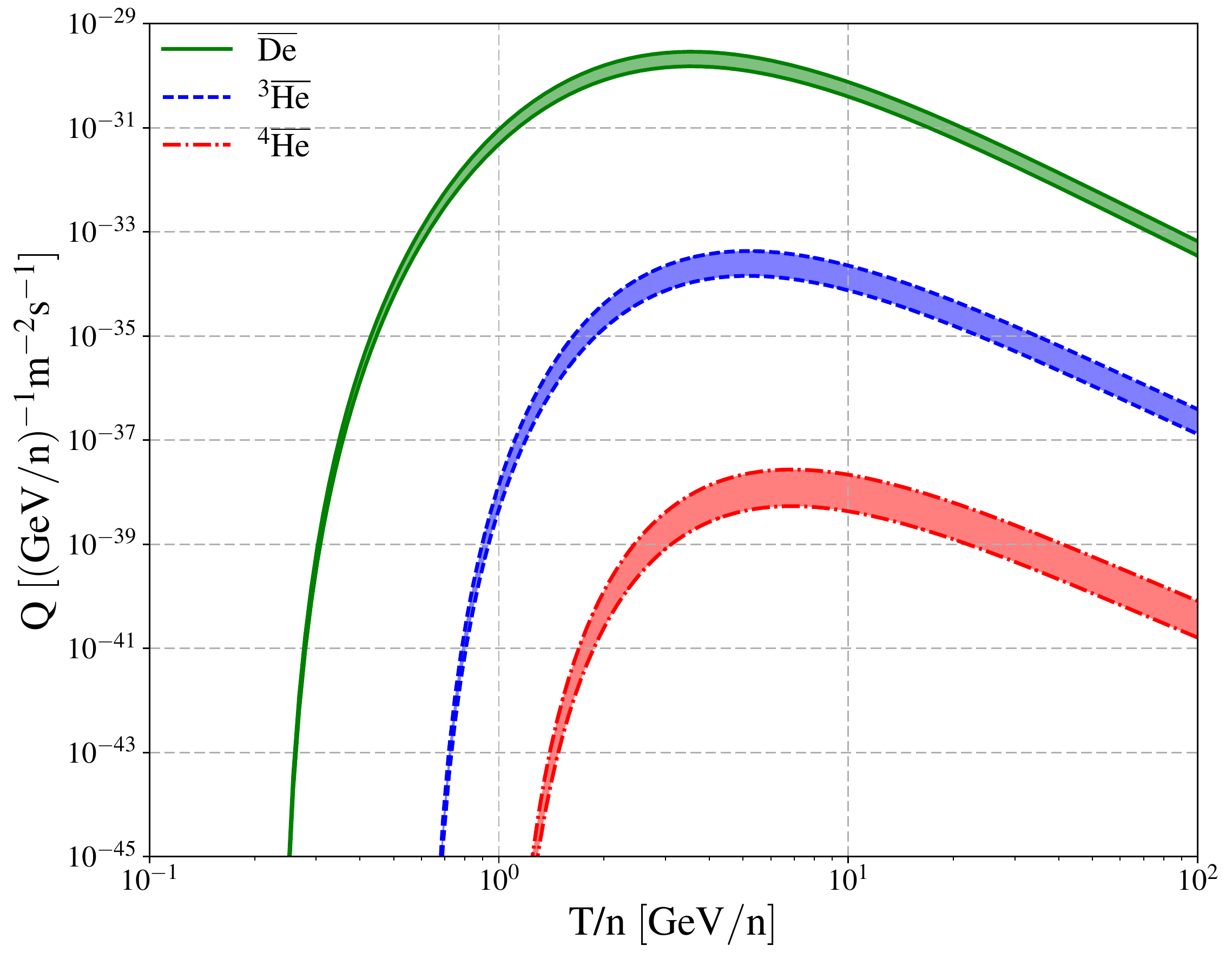}
\caption{Local  source term  for the secondary production of $\bar{d}$,  $\bartHe$ and $\barqHe$. The width of the prediction represents the uncertainty on the coalescence parameter $B_A$.}
\label{fig:source_terms}
\end{figure}
%
Similarly, for the anti-helium-3 and -4 we evaluate cross-sections decreasing the available center of mass energy $\sqrt{s}$ by $2E^{*}_p$ for each subsequent produced anti-nucleon. This ansatz has the merit of imposing energy conservation, although others are possible (see the discussion in Ref.~\cite{Chardonnet:1997dv}). We checked that adopting the other prescription $\sqrt{s} - m_p - E_p$ ̄suggested in Ref.~\cite{Chardonnet:1997dv} does not affect our conclusions.
Another possibility to extend the coalescence analysis down to near-threshold collision energies is to introduce an interpolating factor $R$ in the RHS of eq.~(\ref{eq:coal_1}) as suggested, e.g., in \cite{Duperray:2002pj,Blum:2017qnn}.
The secondary source term can then be readily computed as:
\begin{equation}
Q_{\rm sec}^{ij} (E_A)= 4\pi n_{ j}\int_{E_{\rm th}}^{\infty}dE_i \, \phi_i(E_i) \, \frac{d\sigma^{ij}_A}{dE_A}(E_i,E_A)\,,
\end{equation}
with
\begin{equation}
\frac{d\sigma^{ij}_A}{dE_A} = 2\pi k_A\int_{-1}^1\bigg\{E_A \, \frac{d^3\sigma^{ij}_A}{d^3k_A}\bigg\}d(-cos\theta)\,.
\end{equation}

We assume the density of target hydrogen and helium in the ISM $n_{ j}$ to be $0.9$ g/cm$^3$ and  $0.1$ g/cm$^3$ respectively and make use of the demodulated flux of hydrogen and helium from {\it AMS-02} with Fisk potential $730$ MV.  To calculate the contribution of the main channel $p+p\to\bar{A}+X$, we make use of the recent  $p+p\to\bar{p}+X$  cross-section parameterization from ref.~\cite{diMauro:2014zea}. In order to incorporate other production channels (i.e. from spallation of and onto ${}^4$He), we make use of scaling relations derived in ref.~\cite{Norbury:2006hp} and multiply the $p+p\to\bar{p}+X$ cross section by $(A_TA_P)^{2.2/3}$ where $A_P$  and $A_T$ are the nucleon numbers of the projectile and target nuclei.
The result of our computation is plotted in Fig.~\ref{fig:source_terms}. A nice feature of the coalescence scenario is that it naturally predicts, for simple kinematic reasons, a hierarchical relation between the flux of $\overline{p}$, $\overline{d}$, $\bartHe$ and $\barqHe$, where each subsequent nucleus gets suppressed by a factor $10^{-4}-10^{-3}$. 

%
\begin{figure}
\centering
\includegraphics[scale=0.39]{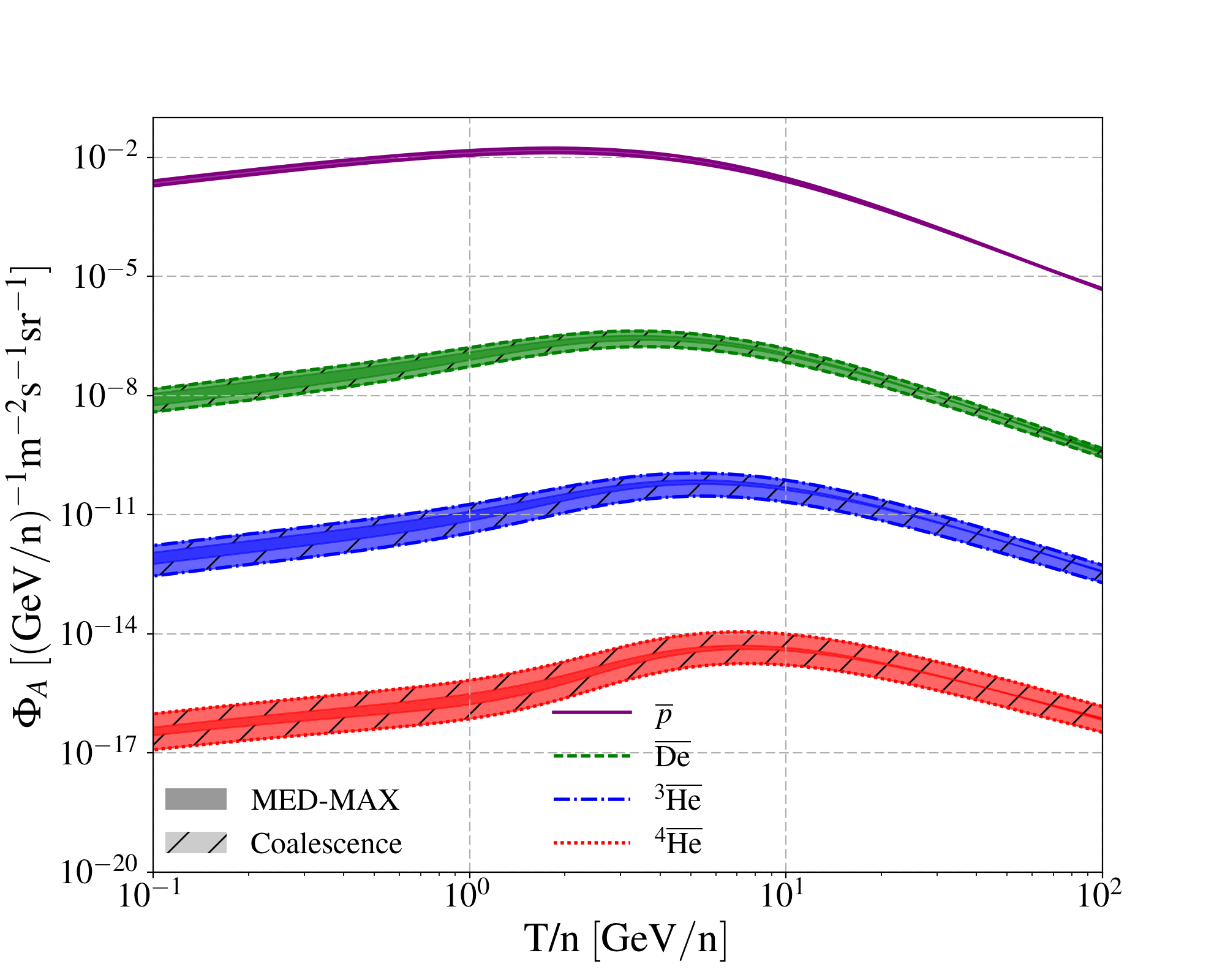}
\caption{Predicted secondary flux of $\bar{p}$, $\bar{d}$,  $\bartHe$ and $\barqHe$ showing the uncertainty associated to the propagation and the coalescence momentum.}
\label{fig:all_species}
\end{figure}
%
%
\begin{figure}
\centering
\includegraphics[scale=0.39]{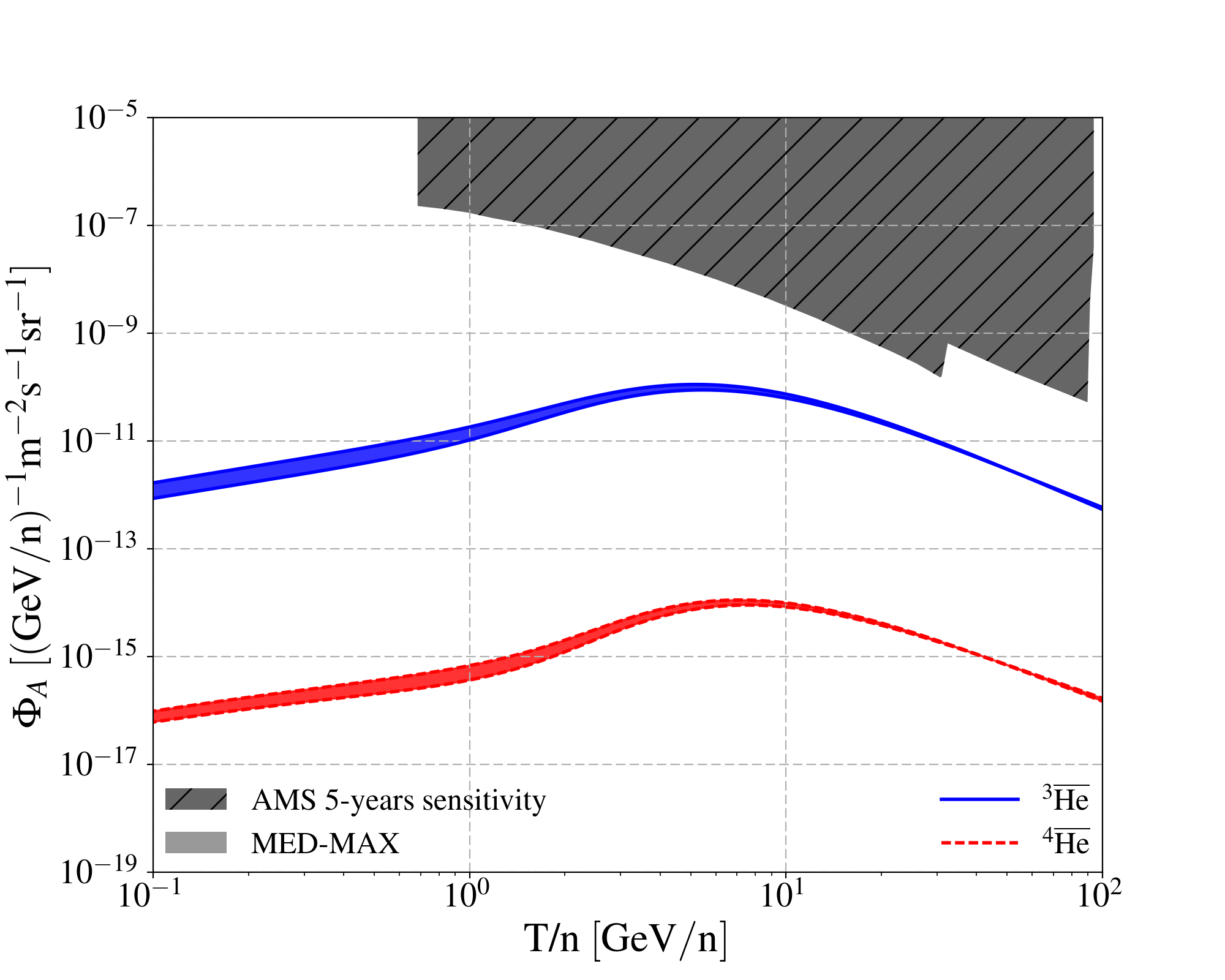}
\caption{Predicted secondary flux of ${}^3\overline{\rm He}$ and ${}^4\overline{\rm He}$ using the upper limit on the coalescence momentum deduced from the ALICE experiment and showing the uncertainty associated to the  MED to MAX propagation model from Ref.~\cite{Donato:2003xg}. We also show the expected sensitivity from {\it AMS-02} \cite{Kounine}.}
\label{fig:hebar}
\end{figure}
%

\subsection{Propagation in the Galaxy}
To deal with propagation, we adapt the code developed in Refs.~\cite{Boudaud:2014qra,Giesen:2015ufa}. We model the Galaxy as a thin disk embedded in a 2D cylindrical (turbulent) magnetic halo and solve semi-analytically the full transport equation for a charged particle. We include all relevant effects, namely diffusion, diffusive reacceleration, convection, energy losses, $\overline{\rm A}$ annihilation and tertiary production. The equation governing the evolution of the energy and spatial distribution function $f$ of any species  reads
\begin{eqnarray}\label{eq:propag}
& & \partial_t f +\partial_E\{b(E,\vec{x})f-K_{EE}(E)\partial_Ef\}+\partial_z\{{\rm sign}(z)fV_c\}\nonumber\\
& &-K(E)\nabla^2f=Q_{\rm II}+Q_{\rm III}-2h\delta(z)\Gamma_{\rm ann}f\,.
\end{eqnarray}

 We choose a homogeneous and isotropic diffusion coefficient $K(E)=\beta K_0(R(E)/{\rm 1 GV})^\delta$ where $\beta$ is the velocity of the particle and $R = p/(Ze)$ its rigidity, the ratio between the momentum $p=\sqrt{E^2-m_A^2}$ and electric charge $Ze$. The diffusive reacceleration coefficient is expressed as $K_{EE}(E)=(2v_a^2E^2\beta^4)/(9K(E))$ where $v_a$ is the drift - or Alfv\`en - velocity of the diffusion centers. The (subdominant) energy losses are taken only in the disk $b(e,\vec{x})=2h\delta(z)b(E)$ where $h=100~$pc is the half-height of the disk and include ionization, Coulomb and adiabatic losses. The gradient of $V_c$ represents the convective wind, pushing outwards CR nuclei with respect to the disk. Possible annihilations of anti-nuclei $A$ in the disk are encoded in the last term on the RHS of eq.~\ref{eq:propag}. The annilation rate takes the form $\Gamma_{\rm ann}=(n_{\rm H}+4^{2.2/3}n_{\rm He})v\sigma_{\rm ann}$, where $\sigma_{\rm ann}$ is the inelastic annihilation cross-section. To estimate the deuteron annihilation cross-section, we make use of the parameterization of the total cross-section of $\overline{\rm H}d$ from Ref.~\cite{Hikasa:1992je} from which we remove the non-annihilating contribution using a measurement presented in Ref.~\cite{Duperray:2005si}, that is $\sigma_{\rm no-ann}^{\bar{d}{\rm H}} = 4$ mb. This non-annihilation contribution is also used to calculate the tertiary source term $Q_{\rm III}$ following Ref.~\cite{Boudaud:2014qra}. Our prescription for annihilation and tertiary is in very good agreement with that presented in Ref.~\cite{Duperray:2005si}.  To calculate the annihilation and tertiary production of anti-helium-3 and 4, we re-scale all cross-sections by a factor $(A/2)^{2.2/3}$.
 We treat the solar modulation in the force field approximation, setting the Fisk potential to 0.730 GV, the average value over AMS02 data taking period \cite{Ghelfi:2016pcv}. Our secondary predictions of  anti-deuteron, anti-helium-3 and -4 fluxes $\phi=\beta c f / (4\pi)$ are plotted in Fig.~\ref{fig:all_species}. We also show the antiproton flux associated to the same cross-section and propagation parameters, in order to illustrate the relative amount of each anti-species from secondary production in our Galaxy. However, we stress that our secondary prediction for antiproton {\em is not} the most up-to-date one and can be within 50\% of the most recent calculation done in Ref.~\cite{Reinert:2017aga}.  We thus implemented the antiproton cross-section parameterization from Ref.~\cite{Reinert:2017aga} and checked that it does not affect our conclusions regarding anti-deuteron and anti-helium. We also checked that the impact of a break in the diffusion coefficient, as advocated in Ref.~\cite{Genolini:2017dfb,Reinert:2017aga} from an analysis of the recent {\it AMS-02} proton, helium and B/C data, is negligible in the energy range we are interested in.  Similarly, changing the value of the Fisk potential does not affect our prediction above a few GeV per nucleon. 
 
In Fig.~\ref{fig:hebar} we show the secondary prediction on anti-helium-3 and -4 compared to the advocated sensitivity of {\it AMS-02} after 5 years \cite{Kounine}. In principle, we should compare our prediction to the measured flux, but this one is not available. Still, we can deduce from the claimed ratio of $\overline{\rm He}/{\rm He}\sim10^{-8}$ that this flux is larger by a factor of $\sim 10$ than the advocated sensitivity of {\it AMS-02} after 5 years around 10 GeV. Hence, we confirm that it is very challenging to explain the potential AMS02 anti-He signal as a pure secondary component. The $\bartHe$  is typically one to two orders of magnitude below the sensitivity of {\it AMS-02} after 5 years, and the $\barqHe$ is roughly 5 orders of magnitude below {\it AMS-02} sensitivity. Our results are in very good agreement with Ref.~\cite{Korsmeier:2017xzj}, who also found that the secondary prediction is, at best, roughly an order of magnitude below the tentative detection. Ref.~\cite{Blum:2017qnn} on the other hand, concluded that a pure secondary explanation of the $\bartHe$ events was still viable. The main difference with Ref.~\cite{Blum:2017qnn} lies in the range of values considered for the coalescence momentum. As the analysis of Ref.~\cite{Blum:2017qnn} was completed, the Alice experiment had not yet published updated values for the coalescence factor of anti-helium $B_3$. Hence, the considered uncertainty range considered by Ref.~\cite{Blum:2017qnn} is much broader (up to $20\times10^{-4}$ GeV$^4$) than the one considered in this work. When considering similar values of $B_3$, our results are in good agreement, even though the propagation of cosmic rays is treated in very different manners.

\subsection{Boosting the production by spallation}
Given the uncertainty on the mass measurement, it is conceivable that  all of the anti-helium nuclei are actually $\bartHe$ isotopes.  The standard $\bartHe$ calculation  yields a flux that is a factor $\sim30-100$ below what is measured by {\it AMS-02}. While uncertainties in the propagation are unlikely to be responsible for such mismatch, one might argue that the production term from spallation is underestimated. In order to boost the secondary flux, one needs to increase the coalescence factor $B_3$ by the same amount. The ALICE experiment has reported a measurement of the coalescence factor from $pp$-collision with center of mass energies of $0.9$ to $7$ TeV instead of the few hundreds of GeV at which collisions occur in the ISM. It is conceivable that the $B_3$ factor differs at lower energies. However, there are several arguments going against a large increase of the coalescence factor at low energy:
\begin{enumerate}
\item[\textbullet] To commence, within the range of energies considered by ALICE (which spans an order of magnitude), the coalescence factor is very close to constant.
\item[\textbullet] Then, there exist measurements~\cite{Lemaire:1980qw} of the $B_2$ and $B_3$ factors from heavy ion collisions with beam energies between 0.4 and 2.1 GeV/nuc. Albeit probed at a much lower center-of-mass energy than in the case of ALICE, the coalescence momentum is found to lie in the range 0.173--0.304 GeV for deuterium and 0.130--0.187 GeV for tritium and helium-3. In the latter case, $p_{\rm coal}$ is smaller than what is found by ALICE at LHC energies. Our prediction for the production of $\bartHe$ in primary CR collisions onto the ISM tends to overestimate the actual rate. 
\item[\textbullet] Finally, from a theoretical perspective, we expect the rate of coalescence of nucleons to be higher at high energy than at low energy. Indeed, the collision of a high energy particle (having a large Lorentz boost)  will create a jet of particles whose opening angle is smaller than that of a low-energy collision. This in turn will increase the correlation of nucleons within the shower and thus the probability for nucleons to merge. Moreover, the production of many nucleons in low-energy collisions is strongly suppressed by the phase-space. This theoretical consideration is in good agreement with what has been found in recent Monte-Carlo simulation studies \cite{Gomez-Coral:2018yuk}. The coalescence momentum (and hence the coalescence factor) decreases with lower center of mass energy. Hence, using the value obtained from ALICE data leads to a conversative over-estimation of the anti-nuclei secondary fluxes.
\end{enumerate}
Alternatively, increasing the grammage\footnote{The grammage measures the column density of interstellar matter crossed by CR. In an homogeneous and isotropic propagation model, it is directly proportional to the interstellar secondary flux.}  seen by primary CRs along their journey towards Earth would enhance the yields of secondary nuclei. However such a scenario would result in all secondaries being affected in a similar way. Given the very good agreement (at the $\sim 20\%$ level) between the measurement of the $\bar{p}$ flux and its current best secondary estimate, a large increase in the grammage of our Galaxy is not realistic.

In conclusion, it seems to us extremely unlikely that a boosted production by spallation is responsible for such a large  $\bartHe$ flux. Naturally, the presence of $\barqHe$ goes as well against this scenario.

\subsection{A word on $\bartHe$  and $\barqHe$  from Dark Matter annihilation}
A recent reanalysis of the $\bartHe$ yield from Galactic DM annihilation has been presented in Refs.~\cite{Coogan:2017pwt,Korsmeier:2017xzj}. The formalism is very similar to that of production by spallation, and the estimate of the DM source term depends as well on the knowledge of the coalescence momentum $p_A$ previously introduced. Similarly to secondary production, one expects a hierarchical relation between the fluxes of $\bar{p}$, $\bar{d}$, $\bartHe$ and $\barqHe$. According to Refs.~\cite{Coogan:2017pwt,Korsmeier:2017xzj}, if DM is responsible for {\it AMS-02} events, it seems unlikely to observe $\bartHe$ without seeing a single $\overline{d}$ or overshooting $\overline{p}$ data. One caveat to this argument is that the sensitivity of {\it AMS-02} to $\overline{d}$ might be (much) smaller than that to $\bartHe$ in some energy range \cite{Kounine}. Still, the possible presence of $\barqHe$ events is at odds with the DM scenario.
 
\section{$\bartHe$ and $\barqHe$ as an indication for an anti-world}\label{sec:antidomain}
Motivated by the $\bartHe$ and $\barqHe$, in this section we
discuss the possibility that extended regions made of anti-matter
have survived in our Galactic environment. There are many scenarios discussed in the literature and we present a few 
possibilities in sec.~\ref{sec:discussion}. The two possible cases that regions of anti-matter are present in our Galactic environment, are\footnote{Additionally, compact objects might exist but would most likely not lead to the injection of high energy cosmic-rays and we therefore do not consider them here.} i) as ambient anti-matter mixed with regular matter in the ISM or in the form of anti-clouds; ii) in the form of anti-stars.  
The presence of $\barqHe$, if confirmed, would be a hint at the presence of such anti-regions. 
However, the fact that {\it AMS-02} measures more $\bartHe$ than $\barqHe$ (roughly 3:1) is also 
interesting. As we discuss below, the isotopic ratio of anti-helium can potentially carry information about 
the physical conditions (in particular anti-matter and matter densities) within these regions.

\subsection{Anti-clouds}
\label{sec:anticloud}
We argue that the presence of clouds of anti-matter in our local 
environment can be responsible for the {\it AMS-02} events. We discuss properties of these clouds and constraints that apply to this scenario (in particular from non-observation of $\gamma$-rays from matter-anti-matter annihilation).

\subsubsection{Exotic BBN as an explanation of the anti-helium isotopic ratio}
\begin{figure}
\includegraphics[scale=0.45]{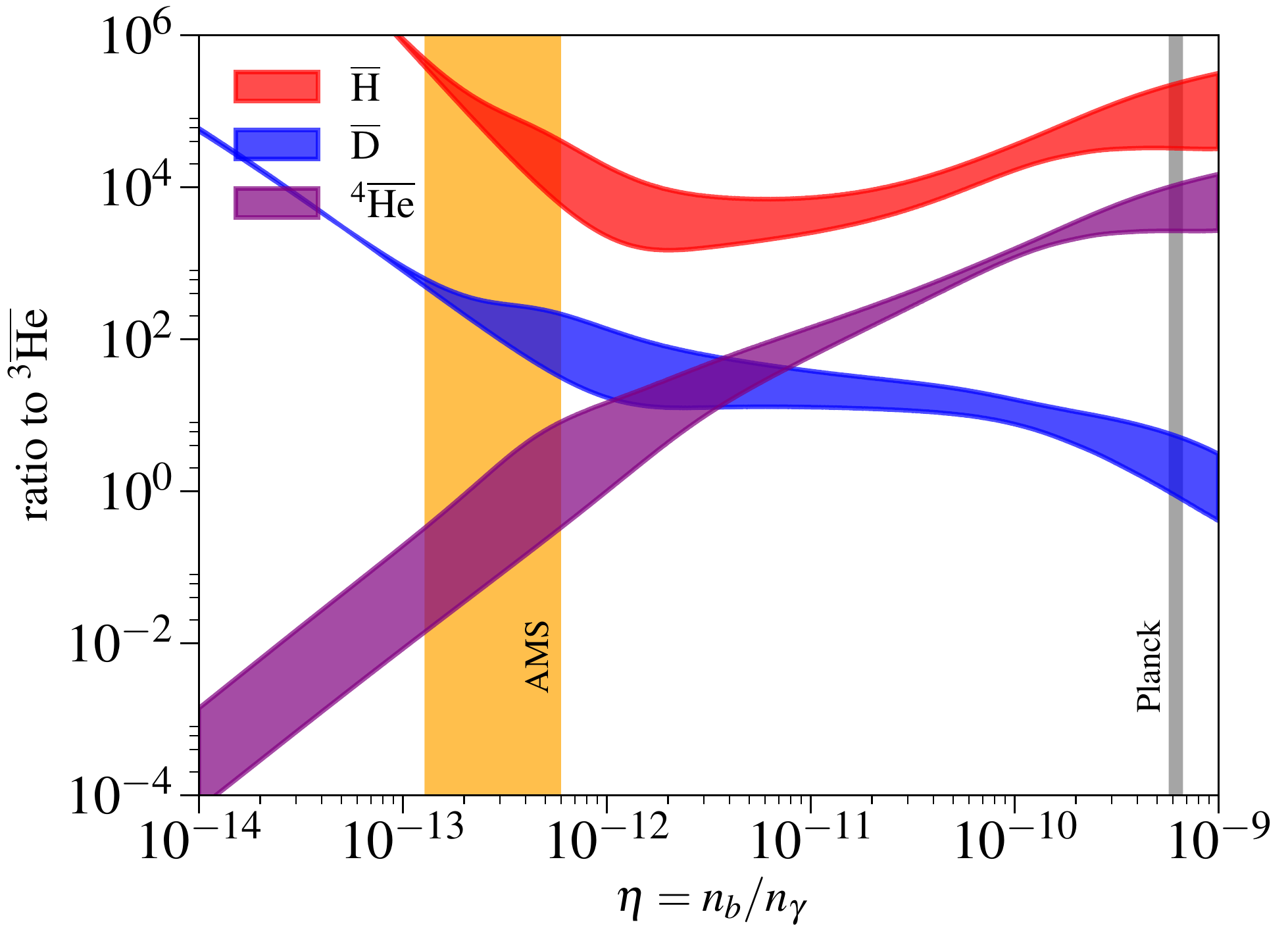}
\caption{Abundance of $\overline{\rm H}$, $\overline{\rm D}$ and $\barqHe$ with respect to that of 
$\bartHe$ as a function of the (anti-)baryon-to-photon ratio $\bar{\eta}$. The {\em Planck} value is represented 
by the grey band. The value required by the {\it AMS-02} experiment is shown by the orange band.}
\label{fig:BBN}
\end{figure}

Standard big-bang nucleosynthesis (BBN) predicts the presence of many more ${}^4$He events, 
compared to  ${}^3$He. For normal matter, the isotopic ratio of ${}^4$He:${}^3$He is roughly 
$10^4$:$1$. Within CRs, the isotopic ratio is higher since ${}^3$He  can be produced through spallation of ${}^4$He and, according to {\it PAMELA} \cite{Adriani:2015aps}, it reaches $\sim$ 5:1 at a few GeV/n. Still, this is much lower than the possible measurement from {\it AMS-02}. As we have argued previously, increasing spallation by an order of magnitude is not realistic as it would affect all secondary species equally and lead to an over-prediction of $\bar{p}$s\footnote{In sec.~\ref{sec:antistar}, we estimate that this could be realistic close to compact objects such as anti-stars.}.  Hence, inverting the isotopic ratio requires the presence of anisotropic BBN in
regions where the (anti)-baryon-to-photon ratio strongly differs from that measured by {\em Planck} 
\cite{Ade:2015xua}. We therefore re-calculated the BBN yields for a large number 
of $\eta$ values using the BBN-code \texttt{AlterBBN}{}\footnote{https://alterbbn.hepforge.org} \cite{Arbey:2011nf} assuming CP-invariance for simplicity (Similar results are obtained with the BBN public code \texttt{PRIMAT}{}\footnote{http://www2.iap.fr/users/pitrou/primat.htm} \cite{Pitrou:2018cgg}). We show in fig.~\ref{fig:BBN} the number density of $\overline{\rm H}$, $\overline{\rm D}$ and $\barqHe$ 
normalized to  $\bartHe$ as a function of the (anti-)baryon-to-photon ratio $\bar{\eta}$. The width of 
the band features the nuclear rate uncertainties \footnote{A caveat is that nuclear uncertainty 
correlations  are not provided. Hence, to calculate these bands,  we simply vary all rates in the 
same way (increase them all or reduce them all simultaneously), i.e. we assume that all nuclear 
uncertainties  are completely correlated, following the prescriptions implemented in \texttt{AlterBBN}{}. This leads to the smallest uncertainty on the ratio and 
therefore a broader range of $\bar{\eta}$ values might in fact be allowed. A detailed study is left to
future work.}. It is possible then to obtain the right isotopic ratio (i.e. roughly 
$\barqHe:\bartHe$ of 1:3) for  $\bar{\eta}\simeq1.3\!-\!6\times10^{-13}$. Interestingly, this also predicts the presence of non-negligible CR $\bar{p}$ and 
$\bar{d}$ fluxes that we comment on in sec.~\ref{sec:antiCR}. 
We also point out that while $\eta$ from {\em Planck} refers to an average over the whole 
observable Universe,  $\bar{\eta}$ is based on the isotopic ratio of $\barqHe:\bartHe$ and is therefore a local quantity. Depending on the object from which anti-helium events are originating, it is conceivable that this number varies from place to place even within our Galaxy such that {\em on average} the isotopic ratio of $\barqHe:\bartHe$ is as measured.
%
\subsubsection{Some properties of the anti-domains}
We can get some information about the anti-cloud regions from the ratio $\phi_{\overline{\rm He}}
/\phi_{\rm He}$ (integrated over all energies), assuming that it reflects the ratio of the abundance of $\overline{\rm He}$ to 
${\rm He}$ in the ISM (i.e. acceleration and propagation of CR are identical for matter and anti-matter), $N_{\overline{\rm He}}/N_{\rm He}$,
\begin{equation}
\frac{\phi_{\overline{\rm He}}}{\phi_{\rm He}} \simeq \frac{N_{\overline{\rm He}}}{N_{\rm He}} \simeq
\bigg(\frac{n_{\overline{\rm He}}}{n_{\bar{p}}}\bigg)\bigg(\frac{n_p}{n_{\rm He}}\bigg)
\bigg(\frac{n_{\overline{b}}}{n_b}\bigg)\bigg(\frac{V_{\overline {\rm M}}}{V_{\rm M}}\bigg)\,,
\end{equation}
where $V_{\overline {\rm M}}$ and $V_{\rm M}$ represent the total volume of the anti-matter 
and matter regions in our Galaxy.  
We have assumed here that $n_b = n_p$ which is correct at the 10$\%$ level and that
$n_{\overline{b}} = n_{\bar{p}}$ that, as shown in Fig.~\ref{fig:BBN} is also correct at better than the 10$\%$ level. The CR data can also tell us about some of these ratios: 
i)  $\phi_{\overline{\rm He}}/\phi_{\rm He}$ is of order $\sim10^{-8}$; ii) the ratio 
$n_p/n_{\rm He}$ is of order $\sim$10; iii) the ratio $n_{\overline{\rm He}}/n_{\overline{p}}$, 
from the BBN calculation motivated by the isotopic ratio ${}^4\overline{\rm He}$:${}^3\overline{\rm He}$, is of order 
$\sim 10^{-5.5}\!-\!10^{-4}$.  Hence we can get a constraint on the product of the total 
volume and density of these regions:
\begin{equation}
\label{eq:constraints_antimatter}
\bigg(\frac{n_{\overline{b}}}{n_b}\bigg)\bigg(\frac{V_{\overline {\rm M}}}{V_{\rm M}}\bigg)
\sim 10^{-5}\!-\!10^{-3.5}\,.
\end{equation}
It is possible to go one step further since the typical density of matter in the ISM is $n_b= 1$ cm$^{-3}$ and $V_{\rm M} = 2h\pi R_{\rm gal}^2\sim6\times10^{10}$ pc$^3$. Moreover, if we assume that anti-matter forms spherical anti-clouds of radius $r_{\overline{\rm{c}}}$, we get $V_{\overline {\rm M}} = N_{\overline{\rm{c}}} (4\pi/3r_{\overline{\rm{c}}}^3)$ and derive
\begin{equation}
\label{eq:nbar}
n_{\overline{b}}\simeq10^{5}-10^{6.5}N_{\overline{\mathrm{c}}}^{-1}\bigg(\frac{n_b}{1~\mathrm{cm}^{-3}}\bigg)\bigg(\frac{r_{\overline{c}}}{1~{\rm pc}}\bigg)^{-3}\mathrm{cm}^{-3}\,.
\end{equation}
This key relation mostly relies on {\it AMS-02} data and knowledge about Galactic properties. The only theoretical assumption so far is that the isotopic ratio of anti-helium is derived from BBN.  From this we have additionally derived that at the time of BBN, $\bar{\eta}/\eta=n_{\bar{b}}/n_b\sim10^{-3.5}-10^{-3}$.   If this ratio still holds today, it would imply that there are $N_{\overline{\rm{c}}}\sim 10^8-10^{10}(r_{\overline{c}}/1{\rm pc})^{-3}$ anti-clouds in our Galaxy.  The higher end of $N_{\overline{\mathrm{c}}}$ is close to the situation where the anti-clouds are connected in the ISM. However, this probably strongly overestimates the number of such objects, as cosmological evolution can affect these regions (and in particular the ratio $n_{\bar{b}}/n_b$) compared to primordial conditions. More realistically, AMS measured events would originate from a few highly-dense clouds.

\subsubsection{Survival time of anti-matter in the Milky Way and the early Universe}

We can gain information about the properties of the anti-matter regions and in particular constrain the amount of {\em normal} matter within them by estimating the typical lifetime of anti-matter in our Galaxy. The lifetime depends on the relative velocity between matter and anti-matter particles, as the annihilation cross-section can be strongly enhanced at low-velocity. For our estimates, we will follow the parameterization suggested in Ref.~\cite{Steigman:1976ev} and split the cross-section in three regimes; a high-energy regime where the cross-section scales with the inverse of the velocity; a Sommerfeld enhanced-regime where the cross-section scales with the inverse of the {\em square} of the velocity; a saturation limit once the cross-section reaches the size of an atom\footnote{We note that this parameterization has a discontinuity around $10^4$ K, therefore the constraints obtained around that energy should be taken with a grain of salt. Fortunately, most of the constraining power comes from the regime where the cross-section is well-behaved.}. In practice, we use
\begin{equation}
\label{eq:cross-section}
\langle\sigma_{p\bar{p}\,}v\rangle\simeq \left\{ \begin{array}{lr}1.5\times10^{-15}~{\rm cm}^{3}/{\rm s}& T>10^{10}~{\rm K}\,,\\
10^{-10}\left(\frac{{\rm K}}{T}\right)^{1/2}~{\rm cm}^{3}/{\rm s}\quad&10^{10}~{\rm K}> T>10^4~{\rm K}\,,\\
10^{-10}~{\rm cm}^{3}/{\rm s}&10^4~{\rm K}> T\,.\end{array}\right.
\end{equation}
The survival rate depends on whether anti-matter is in the form of cold clouds, where $T\sim{\cal O}(30)$ K, or in hot ionized clouds, where $T\sim{\cal O}(10^6)$ K. 
In the former scenario, the lifetime $\tau_{\rm ann}^ {\rm cold}$ is roughly
\begin{eqnarray}
 \label{eq:AnnLifetime}
\tau_{\rm ann}^ {\rm cold} &=& (n_p\langle\sigma_{p\overline{p}}v\rangle)^{-1} \simeq  10^{10}\bigg(\frac{n_{p}}{1~{\rm cm}^{-3}}\bigg)^{-1}\,~{\rm s}\,,
\end{eqnarray}
which is to be compared with the (much longer) age of our Galaxy $t_{\rm gal}\simeq2.8\times10^{17}$s\,. Hence, this requires the hydrogen density within cold anti-matter clouds to verify 
\begin{equation}
n_p^{\rm cold}<3.5\times10^{-8}~{\rm cm}^{-3}\,,
\end{equation}
for such anti-clouds to survive in our Galaxy.
The same calculation in hot ionized cloud yields
\begin{equation}
\tau_{\rm ann}^ {\rm hot} \simeq 1.7\times 10^{13}\bigg(\frac{n_{p}}{1~{\rm cm}^{-3}}\bigg)^{-1}\!\!{\rm s} \,\Rightarrow \,n_p^{\rm hot}<6.1\times10^{-5}~{\rm cm}^{-3}\,.
\end{equation}
Note that these numbers are independent of the size and density of anti-matter regions and agrees well with Refs.~\cite{Steigman:1976ev,Bambi:2007cc}. 
We conclude from this short analysis that anti-matter would survive 
in our Galaxy only if there is some separation between the species, in which case it could be a 
viable candidate to explain the anti-helium events.  However, diffuse anti-matter occupying all the volume of our Galaxy would not survive over the life span of our Galaxy.

Additionally, we can perform the same calculation in the early Universe, splitting between three periods depending on the annihilation regime. Before BBN, annihilations happen in the relativistic regime, and we can deduce
\begin{eqnarray}
\tau_{\rm ann}({z>z_{\rm BBN}})&\simeq& \frac{3.3\times10^{21}}{(1+z)^{3}}\,
\frac{n_p^{\rm cosmo}(z)}{n_p^{\rm local}(z)}~{\rm s}\,,\nonumber
\end{eqnarray}
where $n_p^{\rm local}(z)$ and $n_p^{\rm cosmo}(z)$ respectively stand for the local and cosmological proton densities at redshift $z$ while $z_{\rm BBN}\simeq3.5\times10^9$. Comparing to the Hubble time at that epoch $t_H\simeq 5\times10^{19}(1+z)^{-2}$ s, we find that the hydrogen density inside anti-matter regions must satisfy
\begin{equation}
\frac{n_p^{\rm local}}{n_p^{\rm cosmo}}({z>{z_{\rm BBN}}})<\bigg(\frac{67}{1+z}\bigg)\,,
\end{equation}
which means that, in the most optimistic scenario where such regions were formed right before BBN, the local proton density satisfies
$n_p^{\rm local} < 1.9\times10^{-8}\,n_p^{\rm cosmo}$\,. After BBN, and roughly until matter-radiation equality, the constraint becomes
\begin{eqnarray}
& \tau_{\rm ann}&({z_{\rm eq}<z<z_{\rm BBN}})\simeq\frac{8.3\times10^{16}}{(1+z)^{5/2}} \, \frac{n_p^{\rm cosmo}(z)}{n_p^{\rm local}(z)}~{\rm s}\nonumber\\
& & \Rightarrow \frac{n_p^{\rm local}}{n_p^{\rm cosmo}}(z_{\rm eq}<z<z_{\rm BBN})<\bigg(\frac{1.7\times10^{-3}}{\sqrt{1+z}}\bigg)\,.\nonumber\\
\end{eqnarray}
Finally, deep in the matter-dominated regime when anti-matter is non-relativistic and $t_H\simeq8\times10^{17}(1+z)^{-3/2}$ we get
\begin{eqnarray}
& \tau_{\rm ann}&(z<z_{\rm eq})\simeq \frac{5\times10^{16}}{(1+z)^{3}} \, \frac{n_p^{\rm cosmo}(z)}{n_p^{\rm local}(z)}~{\rm s}\nonumber\\
& & \Rightarrow \frac{n_p^{\rm local}}{n_p^{\rm cosmo}}(z<z_{\rm eq}) < \frac{6.3\times10^{-2}}{(1+z)^{3/2}} \,.\nonumber\\
\end{eqnarray}
This confirms that anti-matter must have formed 
in regions where the density of protons was much lower than the cosmological average (at least ${\cal O}(10^{-8})$ if these regions form just at the start BBN), such that annihilations only occur at the border of the anti-matter dominated domains. It is conceivable that today these regions have 
survived in their pristine form, i.e., with little annihilation taking place inside them, although galaxy formation will likely have mixed up partially the species. Hence, this would imply the existence of some exotic segregation mechanism which makes the existence of such anti-clouds rather improbable. 

\subsubsection{Constraints from the CMB}
\label{sec:CMB}
For over a decade, observations of the CMB have been used to constrain scenarios leading to exotic energy injection, in particular dark matter annihilations \cite{Padman05,Hooper09,Cirelli09,Huetsi:2009ex,Slatyer09,Natarajan08,Natarajan09,Natarajan10,Valdes:2009cq,Evoli:2012zz,Galli13,Finkbeiner11,Hutsi:2011vx,Slatyer12,Giesen,Slatyer13,Slatyer15-1,Lopez-Honorez:2013lcm,Poulin:2015pna,Liu:2016cnk,Stocker:2018avm}. Interestingly, it is possible to recast constraints from these analyses onto the case of anti-matter annihilations\footnote{The non-observation of CMB spectral distortions can also be used to set constraints, but these are usually much weaker than that coming from anistropy power spectra analysis except if the cross-section is boosted at high-velocities (e.g. p-wave) \cite{Chluba:2013wsa}.}. Since constraints on DM annihilation assume that the DM in our Universe is homogeneously distributed, they are only stricly applicable for the case of well mixed matter and anti-matter regions. A full analysis of the case where energy is injected in an inhomogeneous manner is left to future work.
To translate CMB bounds, we start by writing the energy injection rate from DM annihilation:
\begin{equation}
\label{eq:injection_anni}
\frac{\mathrm{d}^2 E}{\mathrm{d} V \mathrm{d}t}\bigg|_{\rm DM} = \kappa \rho^2_c c^2 \Omega_\mathrm{CDM}^2 \left( 1+z \right)^6\frac{{\langle\sigma_{\textrm{ann}} v\rangle}}{m_\mathrm{DM}}\,.
\end{equation}
where $\kappa=1$ for Majorana particles and $\kappa=1/2$ for Dirac particles. The DM density today is well known from CMB data and the prefactor $\rho^2_c c^2 \Omega_\mathrm{CDM}^2\simeq4.5\times10^{-37}$kg$^2$ s$^{-2}$ m$^{-4}$.
In CMB analyses that constrain DM annihilation, the parameter $p_{\rm ann}\equiv{\langle\sigma_{\textrm{ann}} v\rangle}/m_\mathrm{DM}$ is often introduced. Recently, the {\em Planck} collaboration has derived \cite{Aghanim:2018eyx}  $p_{\rm ann}< 1.8 \times 10^6 {\rm m}^3 {\rm s}^{-1} {\rm kg}^{-1}$, assuming a constant thermally-averaged annihilation cross-section times velocity (hereafter dubbed ``the cross-section'' for simplicity)\footnote{Additionnally, we assume here that all the energy is efficiently absorbed by the plasma for simplicity. This can over-estimate the bound by up to an order of magnitude compared to more accurate analyses \cite{Slatyer09,Finkbeiner11,Slatyer:2016qyl,Poulin2016}.}. Under this hypothesis, we can thus constrain the amount of energy injection from annihilation to be
\begin{equation}
\label{eq:planck_constraints}
\frac{\mathrm{d}^2 E}{\mathrm{d} V \mathrm{d}t}\bigg|_{\rm ann} < 8.1\times10^{-31} \left( 1+z \right)^6{\rm J}~{\rm m}^{-3}~{\rm s}^{-1}\,.
\end{equation}
This can be applied to the specific case of non-relativistic anti-matter annihilation
\begin{equation}
\frac{\mathrm{d}^2 E}{\mathrm{d} V \mathrm{d}t}\bigg|_{b\bar{b}-{\rm ann}}={\langle\sigma_{p\bar{p}}v\rangle}n_pn_{\bar{p}}2 m_pc^2\,
\end{equation}
leading to 
\begin{equation}
n_{\bar{p}}^0 \, n_p^0\lesssim 2.7 \times10^{-5}\bigg(\frac{10^{-16}{\rm m}^3/{\rm s}}{\langle\sigma_{p\bar{p}}v\rangle}\bigg){\rm m}^{-6}\,.
\end{equation}
Hence, we find $n_{\bar{p}}^0<1.35\times10^{-10}$cm$^{-3}$ on cosmological scales, which is not in tension with {\it AMS-02} requirement (although strictly speaking the latter is valid within our Galaxy), as given by Eq.~\ref{eq:constraints_antimatter}. We stress that the constraint derived here is very rough as CMB analyses rely on hypothesis of homogeneity and  constant cross-section, and thus deserves a thorough investigation in a separate paper if AMS measurement was confirmed.

\subsubsection{\textit{Using Gamma Ray observations to place limits}}
Alternatively to searches using early Universe cosmology, gamma-ray observations are routinely used to place constraints on exotic physics including dark 
matter annihilation or decay. There are three types of searches that have provided strong constraints on these scenarios: i)  searches for
distinctive spectral features as would be the case for a gamma-ray line \cite{Weniger:2012tx, Bringmann:2012ez,
Ackermann:2013uma,Ackermann:2015lka}; ii) searches for 
morphological features localized on the sky, either from extended sources or from point 
sources on the sky (i.e. of angular size smaller than the point spread function of the instrument) 
\cite{Abdo:2010ex, Aleksic:2011jx, GeringerSameth:2011iw, Aliu:2012ga, Cholis:2012am, Ackermann:2013yva, 
Fermi-LAT:2016uux}; iii) searches for a continuous spectrum of gamma-rays extending over 
a large area on the sky as for instance from the extragalactic gamma-ray 
background \cite{Golubkov:2000xz,Abdo:2010nz, Ackermann:2014usa,}. In the following we present limits for the cases i) and ii) as they provide the strongest constraints on anti-matter regions.
Additionally, let us mention that, while we focus on annihilations of antiprotons, the requirement of overal neutrality of anti-regions implies that there are as many positrons whose annihilations can also be searched for. For instance, in the case of annihilations (almost) at rest, we expect photons in the MeV range, extending down to the 511 keV line. These can be looked for in  \textit{INTEGRAL} data and will also lead to strong constraints on the presence of anti-clouds, a task we leave to future work.
\subsubsection*{Annihilations at rest and $\gamma$-ray line limit at 0.93 GeV}
We start by discussing constraints of type i), which arise from non-relativistic protons annihilating 
within or on the borders of the anti-matter clouds/regions and resulting in gamma-rays. These 
gamma-rays will come from the channels that produce a neutral meson and a gamma-ray, as 
would be the case for $p \bar{p} \rightarrow \pi^{0} \gamma$, $\eta \gamma$, $\omega \gamma$, 
$\eta' \gamma$, $\phi \gamma$, $\gamma \gamma$. All these channels produce lines in the rest
frame of the $p\,\bar{p}$-pair with energies between 0.66 and 0.938 GeV.  The dominant reaction is the $\pi^{0} \gamma$ channel at 0.933 GeV with a branching ratio ($BR$) of 
$4.4 \times 10^{-5}$. Many more gamma-rays will come from the decays of neutral mesons 
produced from the $p \bar{p}$ annihilations, but these would result in a continuous spectrum 
below 0.938 GeV (see \cite{Amsler:1997up} for a full discussion on the $p \bar{p}$ annihilation products). 
 A dedicated analysis accounting for all the annihilation channels would lead to stronger limits, but is beyond the scope of this paper. However, there are limits from the Fermi-LAT observations on gamma-ray line features at these energies 
 \cite{Ackermann:2015lka} which we use as a ``proof-of-principle'' that these studies can already severely 
 constrain anti-clouds. We make use of the constraints derived from the ``R180'' region ~\cite{Ackermann:2015lka} that covers the entire sky apart from a thin stripe along most of the Galactic plane and thus probes the averaged annihilation rate in a large part of the Milky Way around
our location. Since most of the disk is excluded from the analysis, these constraints are to be taken with a grain of salt: it is conceivable that there are only a few highly dense clouds in our Galaxy contributing to the {\it AMS-02} flux, which would escape the analysis in Ref.~\cite{Ackermann:2015lka}. Assuming on the other hand that anti-regions are rather numerous and homogenously distributed in the Galactic disk, these constraints would be on the conservative side.
We use the 95$\%$ 
upper limit flux of $\Phi_{\gamma \gamma}^{0.947} = 6.8 \times 10^{-7}$  $\#\gamma\,\textrm{cm}^{-2}
\textrm{s}^{-1}$, where $\Phi_{\gamma \gamma}$ just refers to the emission of two gamma-rays 
per annihilation event within an energy bin centered at 0.947 GeV and having a width of 
$\simeq 0.03$ GeV.


Assuming anti-matter to form cold clouds, we estimate  the $\gamma$-production per unit volume to be
\begin{eqnarray}
\rho_{\pi^{0} \gamma}^{\rm MW} &=&  {\rm Br}_{\pi^0\gamma}\bigg(\frac{V_{\rm \overline{M}} \, n_{\bar{p}}}{V_{\rm M}}\bigg)n_p\langle\sigma_{p\bar{p}}v\rangle\\
&\simeq& 4.4 \times 10^{-15} (10^{-5}-10^{-3.5}) \bigg(\frac{n_{b}}{1~{\rm cm}^{-3}}\bigg)\nonumber\\
& & \times \bigg(\frac{n_{p}^{\rm local}}{1~{\rm cm}^{-3}}\bigg)\, \#\gamma \, \textrm{cm}^{-3} \textrm{s}^{-1}\,,
\nonumber
\end{eqnarray} 
where the ratio $V_{\rm \overline{M}} \, n_{\bar{p}}/V_{\rm M}$ is given by Eq.~(\ref{eq:constraints_antimatter}) and we make the distinction between the average baryon number density in the Galaxy $n_{b}$ and the density of proton within anti-clouds $n_{p}^{\rm local}$.
We assume that this rate is homogeneous in the Galactic disk (as would arise from a scenario with numerous clouds) but 
drops as we move perpendicularly away from the Galactic plane, following a Gaussian\footnote{This choice is arbitrary and just ensures that the gas and anti-gas density drops abruptly above and below the Galactic plane. We checked that using a more sharply dropping hyperbolic tangent 
gives similar result.} with a width 
$\sigma_{z}$ of 0.1 kpc. Integrating along the line of sight and averaging 
over all relevant directions for R180 of \cite{Ackermann:2015lka}, the gamma-ray line flux is
\begin{eqnarray}
 \Phi_{\pi^{0} \gamma}^{m_{p}} &=& \frac{\int^{R180} \; d\ell \; d\Omega \; \rho_{\pi^{0} \gamma}^{\rm MW}}
 {\int^{R180} \; d\Omega} \\
 & = & 2.42 \times(10^{2}-10^{3.5})\bigg(\frac{n_{b}}{1~{\rm cm}^{-3}}\bigg)\nonumber\\
 & &  \bigg(\frac{n_{p}^{\rm local}}{1~{\rm cm}^{-3}}\bigg)\#\gamma\,\textrm{cm}^{-2} \textrm{s}^{-1}.
 \nonumber
\end{eqnarray} 
This is a factor of  $4.4 \times 10^{8}- 1.1 \times 10^{10}$ larger than the reported limit (we recall that the uncertainty range comes from the uncertainty on $\bar{\eta}$). 
We point out that while this result is 
an approximation that relies on certain assumptions on the anti-matter distribution properties, as well as on the level of overlap of matter and anti-matter,
such a strong tension can not be easily circumvented. In fact we consider these limits to be 
very constraining of such a possibility unless matter and anti-matter regions overlap only by ${\cal O}(10^{-10})$, such that the density of matter within anti-region is constrained to be
\begin{equation}
\label{eq:AnnConstraints}
n_{p}^{\rm local} \; \lesssim 10^{-10}  \, - \, \;2 \times 10^{-9}    \; \textrm{cm}^{-3}.
\end{equation}
In the case of hot clouds, note that this constraint can relax by a factor of ${\cal O}(2000)$.
We recall that those limits are calculated based on an optimization of the region of interest (``R180'' in this case) that was chosen for a possible signal of DM decay all over the DM halo, and in 
fact are not optimal for searching a gamma-ray line signal from ambient anti-matter 
or anti-matter clouds in the Galactic disk. They would apply -- and are conservative -- if anti-clouds are numerous and distributed following the Galactic disk profile, while they vanish if there are only a few very dense anti-clouds in our Galaxy.

\subsubsection*{ $\gamma$-rays from CR annihilations in close-by anti-clouds}

Even though anti-clouds are devoid of matter so that the above mentioned constraints are satisfied, nothing prevents CR protons to penetrate into these clouds where they annihilate. Contrary to a spectral feature arising from annihilation at rest, accelerated particles such as CR can yield a strong annihilation signal appearing as a continuous emission.  Moreover, localized features on the sky (case ii) in previous discussion) can arise if anti-matter regions are well localized in space. Of the 
{\em regular} matter clouds, the densest are the molecular clouds that have sizes from tenths of a
parsec up to ${\cal O}(10)$ pc, with the smallest ones in size having number densities as high as $10^{6}$ cm$^{-3}$ \cite{Ferriere:2001rg}. Instead the cold and warm atomic Hydrogen is 
more diffuse but large clouds can have sizes of $\sim 100$ pc with densities of 0.2-50 cm$^{-3}$
\cite{Ferriere:2001rg}. Finally ionized gas clouds have densities of $7\times10^{-3} - 0.5$ cm$^{-3}$
with a maximum size of ${\cal O}(100)$ pc. 

As we have discussed previously, {\it AMS-02} does not give a precise measurement of the density of anti-matter, but rather constrains the product of the total volume and density of these regions (see Eq.~(\ref{eq:constraints_antimatter})). To avoid making exact 
assumptions on the size and density of these clouds (since these two parameters vary observationally 
by many orders of magnitude) we will assume that anti-matter clouds have a typical mass of 
$M_{\overline{\mathrm{c}}}\equiv V_{\bar{c}}m_{\bar{b}} n_{\bar{b}} = 10^{3} M_{\odot}$ \footnote{From observations on matter clouds that 
quantity can also vary by at least a couple of orders of magnitude either towards larger or smaller 
mass values.}. In that case there are,
\begin{eqnarray}
N_{\overline{\mathrm{c}}} &=& \frac{V_{\rm M} n_{b}m_p}{M_{\overline{\mathrm{c}}}} 
(10^{-5}-10^{-3.5}) \\
&\simeq& 1.6\times10^{4} (10^{-1.5}-1) \left(\frac{M_{b}^{\rm MW}}{5 \times 10^{10} M_{\odot}} \right)
 \left(\frac{M_{\overline{\mathrm{c}}}}{10^{3} M_{\odot}}\right)^{-1}\,, \nonumber
\end{eqnarray}
where $M_{b}^{\rm MW}$ is the total mass of baryons in the Milky Way and where, as usual, we made use of Eq.~(\ref{eq:constraints_antimatter}).
We can estimate the typical distance $D_{\overline{\mathrm{c}}} $ separating these objects (and therefore the Earth from them), assuming that within the Galactic disk the anti-matter 
clouds are homogeneously distributed. We get
\begin{eqnarray}
\label{eq:distance}
D_{\overline{\mathrm{c}}} &\simeq&\frac{1}{2}\left(\frac{V_{\rm M}}{N_{\overline{\mathrm{c}}}}\right)^{1/3} \\
&\simeq& 80 \; \textrm{pc} \times (1-10^{0.5}) \left(\frac{V_{\rm M}}{6 \times 10^{10} \textrm{pc}^{3}}\right)^{1/3} 
\nonumber \\
&&\times  \left(\frac{M_{b}^{\rm MW}}{5 \times 10^{10} M_{\odot}} \right)^{-1/3}
\left(\frac{M_{\overline{\mathrm{c}}}}{10^{3} M_{\odot}} \right)^{1/3}\,. \nonumber
\end{eqnarray}

If such a cloud is  of a size $< 1$ pc its angular extension is $\simeq 0.7^{\circ}$. Hence,
even the closest anti-matter cloud would appear as a point source at gamma-ray energies of 1 GeV.  The local CR proton spectrum is $dN/dE \simeq 10^{3} (E/1 \textrm{GeV})^{-2.8}$ m$^{-2}$s$^{-1}$
sr$^{-1}$GeV$^{-1}$ \cite{Adriani:2011cu}. These protons colliding with antiprotons would give 
relativistic neutral mesons that after decaying would result in 
a similar gamma-ray spectrum above $\sim 1$ GeV. At gamma-ray energies between 1-3 GeV the 
flux is,
\begin{eqnarray}
\label{eq:PhiGammaDC}
\Phi_{\gamma} &=&\frac{L_{\gamma}}{4 \pi D_{\overline{\mathrm{c}}}^{2}} = \frac{1}{4 \pi D_{\overline{\mathrm{c}}}^{2}}
\int dV_{\overline{\mathrm{c}}}n_{\bar{p}}n_{p}\sigma_{p\bar{p}}v \\
&=& 1.2 \times 10^{-9} \; (10^{-1}-1)  \left(\frac{V_{\rm M}}{6 \times 10^{10} \textrm{pc}^{3}}\right)^{-2/3} \nonumber \\
&&\times \left(\frac{M_{b}^{\rm MW}}{5 \times 10^{10} M_{\odot}} \right)^{2/3}
\left(\frac{M_{\overline{\mathrm{c}}}}{10^{3} M_{\odot}} \right)^{1/3} \#\gamma\,\textrm{cm}^{-2}\textrm{s}^{-1}. \nonumber
\end{eqnarray}
We have assumed here four photons per annihilation coming from the average $\pi^{0}$ multiplicity 
of $\simeq 2$ from $p\,\bar{p}$ annihilations \cite{Amsler:1997up}. We took the branching ratio to neutral mesons to be 4$\%$ 
and have integrated $dN/dE$ between 3 and 10 GeV for the CR protons, in order
to be able to directly compare to the Fermi-LAT point source sensitivity at the same energy range 
as reported in Ref.~\cite{Acero:2015hja}. Note that at such energies, we can use the high-energy limit of Eq.~\ref{eq:cross-section} for the annihilation cross-section. For this energy range the sensitivity is $\Phi_{\gamma} \simeq 10^{-10}
\#\gamma\,\textrm{cm}^{-2} \textrm{s}^{-1}$. 
Using Eq.~(\ref{eq:PhiGammaDC}), we deduce that such annihilations would be detectable up to a distance of $0.1-0.3$ kpc. From Eq.~(\ref{eq:distance}), we conclude that roughly $2-40$  point sources could be
detectable. Interestingly, a number of these could contribute to the 334 3FGL unassociated point sources in the Galactic plane\footnote{We note that over the entire sky, there are 992 unindentified point sources.} (within Galactic latitude $| b | < 5^{\circ}$).  Alternatively, the non-detection of anti-clouds by the Fermi LAT allows to constrain the number (and in turn the mass) of these objects. Note that this constraint does not depend on the amount of matter within anti-matter domains; as CR propagate they would travel through anti-regions even if these are poor in matter originally. Hence, this estimate is fairly robust to conditions occurring within anti-domains. 
Also, a careful analysis of the spectrum of the unassociated sources would be necessary to assess whether these are anti-cloud regions. For instance, we anticipate that when annihilations between antiprotons and protons occur nearly at rest (as is in most cases), then a continuous spectrum in gamma-rays with a cut-off at $\simeq$1 GeV should always be produced. A proper population analysis should also take into account the variation in the luminosity of these new type of sources depending on their size and their distance from the Earth.  

Finally, we point out that one could also do a dedicated search for extended continuous spectrum features 
on the Galactic sky; i.e. what we described as type iii). These could be coming from very close-by anti-cloud 
regions or from the combined emission of a very large number of them along the Galactic-disk plane. Given 
the uncertainties on their distribution and that one would also need to account for the many charged pions 
produced by the $p\,\bar{p}$ annihilations leading to $e^{\pm}$ pairs which in turn if relativistic would give 
inverse Compton scattering as they propagate though the Milky-Way, this possible search 
channel is beyond the scope of this paper. At low-energies,
we know from \textit{INTEGRAL} observations that $e^{\pm}$ pairs pervade the inner parts of the Galaxy \cite{Knodlseder:2005yq, 
Weidenspointner:2007rs, 2008Natur.451..159W}. From \textit{Fermi}-LAT studies, we also know of the GeV 
Galactic Center excess in gamma-rays \cite{Hooper:2010mq, Hooper:2011ti, Abazajian:2012pn, Daylan:2014rsa, 
Calore:2014xka, TheFermi-LAT:2015kwa}. Yet, neither the Galactic Center excess, nor the 511-keV line are strongly
 correlated to known gas structures. Moreover, the Galactic Center excess has a high-energy tail that would demand 
 highly boosted anti-matter that has survived in a dense matter environment. In conclusion, we 
 consider it very unlikely that these excesses could be associated to anti-cloud regions.

\subsubsection{\textit{Using Cosmic Ray anti-matter observations to look for anti-regions}}\label{sec:antiCR}
One of the important questions associated to anti-clouds is the acceleration of anti-matter 
within the cloud. Supernovae shock waves, following the explosion of a massive star\footnote{This 
acceleration mechanism could also arise from the explosion of a massive anti-star, if such 
regions with that massive stars still exist.}, accelerate the material in the ISM. Any anti-matter 
particle within -- or close to -- these environments can also be accelerated by these waves.
The spectra of the injected particles would be very similar to that of normal matter, i.e., following 
a power law in energy with index $\sim 2.2-2.8$. The index values of $\simeq 2.7-2.8$ come 
naturally for CR protons and He at $\sim 10-200$ GeV, as a result 
of several SNRs in the Milky-Way, i.e., the averaged CR nuclei spectra. A harder index value of 
2.4 or 2.2 would instead arise if a near-by (of $\sim$ kpc or $\sim$100pc distance) SNR was the 
dominant contributor of CR antiparticles observed by {\it AMS-02}.

Interestingly, from our BBN range of $\overline{\eta}$-values (and given the uncertainties on the injection index 
of these CRs) we can calculate what fluxes of $\bar{p}$ and $\bar{d}$ should be expected in {\it AMS-02} data. 
Antiproton data from {\it AMS-02} are already available and can alternatively be used to place constraints on 
the $\bar{p}$ primary CR flux component (coming from the acceleration of the ISM $\bar{p}$). 

We calculate first the $\bar{p}$ flux from the primary component associated with the $\bartHe$ and 
$\barqHe$ events. We evaluate first 3-$\sigma$ upper limits from the $\bar{p}/p$ ratio \cite{Aguilar:2016kjl}.
To set the normalisation, we take into account that eight $\bartHe$ and $\barqHe$ events have been observed with at 
least one with $E_{\textrm{kin}}/n$ between 6 and 10 GeV. Moreover, we account for all 
the relevant uncertainties. These are associated with,
\begin{enumerate}
\item[\textbullet] The injection and propagation through the ISM of the matter CRs, mainly protons
and Helium nuclei, that through the inelastic collisions with the ISM gas lead to the production of 
the conventional secondary $\bar{p}$s. The part of the ISM uncertainties affects also the propagation 
of the secondary $\bar{p}$s. 
\item[\textbullet] The antiproton production cross-section from these collisions, affecting the spectrum 
and overall flux of the secondary $\bar{p}$ component.
\item[\textbullet] The matter gas in the local ISM affecting the overall normalization of the secondary 
$\bar{p}$ component.
\item[\textbullet] The Solar modulation of CRs as they propagate through the Heliosphere before getting
detected by {\it AMS-02}. These uncertainties affect both the secondary and primary $\bar{p}$ components as
well as the heavier anti-nuclei fluxes.  
\item[\textbullet] The primary $\bar{p}$ flux index $n$ range of 2.2-2.8 that is associated with the 
uncertainties of their propagation through the ISM, i.e. their locality of origin or not.
\item[\textbullet] The $\bar{\eta}$-range of $1.3-6 \times 10^{-13}$, affecting the ratios of primary anti-nuclei fluxes. 
\end{enumerate}

To account for the first four of the above mentioned uncertainties we marginalize over them following the prescription of Ref.~\cite{Cholis:2017qlb} based on results of Refs.~\cite{diMauro:2014zea, Cholis:2015gna}. 
For the latter two we just take a few extreme cases of $(n,\bar{\eta})\, = \, (2.2, 1.3\times10^{-13})$, $(2.4, 1.3\times10^{-13})$
$(2.8, 1.3\times10^{-13})$, $(2.2, 6\times10^{-13})$, $(2.4, 6\times10^{-13})$ and $(2.8, 6\times10^{-13})$.  
The primary $\bar{p}$ flux are described by,
\begin{equation}
\label{eq:PrimCRpbar}
\frac{dN^{\bar{p}}}{dE_{\textrm{kin}}} = \textrm{Norm}^{\bar{p}} \left( \frac{E_{\textrm{kin}}}{1 \, \textrm{GeV}} 
\right)^{-n} \; \; (\textrm{GeV}^{-1} \textrm{m}^{-2} \textrm{s}^{-1} \textrm{sr}^{-1}), 
\end{equation}
and in turn the $\bar{d}$ and $\overline{{}{\rm He}}$ (: $\bartHe$ \& $\barqHe$) primary fluxes are,
\begin{eqnarray}
\label{eq:PrimCRantinucl}
\frac{dN^{\bar{d}}}{dE_{\textrm{kin}}} &=& \frac{\bar{d}}{\bar{p}}(\bar{\eta}) \frac{dN^{\bar{p}}}
{dE_{\textrm{kin}}} \; \textrm{and} \\
\frac{dN^{\overline{{}{\rm He}}}}{dE_{\textrm{kin}}} &=& \frac{\overline{{}{\rm He}}}{\bar{p}}
(\bar{\eta}) \frac{dN^{\bar{p}}}{dE_{\textrm{kin}}},
\end{eqnarray}
where $E_{\textrm{kin}}$ is the per nucleon kinetic energy. For $\bar{\eta} = 1.3 \times 10^{-13}$, 
$\bar{d}/\bar{p} \simeq 10^{-2.5}$, $\overline{{}{\rm He}}/\bar{p} \simeq 10^{-5.5}$, while for 
$\bar{\eta} = 6 \times 10^{-13}$, $\bar{d}/\bar{p} \simeq 10^{-2}$, $\overline{{}{\rm He}}/\bar{p} 
\simeq 10^{-4}$.

In general, we find that anti-clouds can leave significant traces in the $\bar{p}/p$ ratio. In fact, depending on the propagation configuration, it could even lead to an excess of antiprotons. For instance, for an injection index of $n = 2.2 (2.4)$ and a given propagation model (model E of Ref.~\cite{Cholis:2015gna}), we find that the $\bar{p}/p$ ratio of {\it AMS-02} \cite{Aguilar:2016kjl} can restrict $\bar{\eta} \geq 1.3 (2.0) \times 10^{-13}$ at 3-$\sigma$ (the proportion of $\bar{p}$ decreases as $\bar{\eta}$ increases). If we saturate this limit, we predict  $5.1 (2.7) \times 
10^{4}$ primary CR $\bar{p}$ detected events by {\it AMS-02} after 6 years of data collection and $0.1<$ $\bar{d}$ events in the same period.
If instead we assume $n=2.8$, we find that $\bar{\eta} \geq 4 \times 10^{-13}$, and predict $8.4 \times 10^{3}$ primary CR $\bar{p}$ in {\it AMS-02} data and again only $\simeq 0.1$ $\bar{d}$s. 

In conclusion, CRs provide a strong probe of the anti-cloud scenarios. Interestingly, a number of $\bar{p}$ events detected by {\it AMS-02} might originate from anti-regions. The {\it GAPS} experiment, sensitive to $\bar{d}$ at lower energies than {\it AMS-02}, could detect a few events. A study of the implication of this finding in light of the recent claims of an excess of antiprotons at ten's of GeV energies \cite{Cuoco:2016eej} would be worthwile. 

%

\subsection{Anti-stars in a dense environment}
\label{sec:antistar}
\subsubsection{Properties of anti-stars from {\it AMS-02} measurement}
An alternative possibility is that the anti-matter is in the form of stars. This is likely more realistic, since anti-stars would naturally be free of matter at their heart, and annihilation are limited to their surface.  In that case, the isotopic ratio measured by {\it AMS-02} can inform us about the stellar population. Taken at face value, the presence of a high number of $\bartHe$ is also difficult to explain in this scenario.  One possibility is that anti-stars are relatively light. Indeed, by analogy with normal matter, the main material within an anti-star with $M_{\bar{*}}\lesssim 0.6 M_\odot$ (but higher than $0.08 M\odot$ such as to initiate hydrogen fusion into deuterium) could be $\bartHe$. This would however require the presence of low density regions so that the primordial material from which the star has formed is poor in anti-helium-4, and this scenario is thus affected by the same difficulty as the anti-cloud one.

A more realistic case, already suggested in Ref.~\cite{Dolgov:1992pu}, and more recently in 
 Ref.~\cite{Blinnikov:2014nea} is that the anti-star has  formed from a very dense clump within an anti-matter domain, which could have survived since the early Universe. BBN in a very dense medium would result in the creation of very large amounts of $\barqHe$, so that the anti-star could be largely dominated by $\barqHe$. Difficulties associated to this scenario are two fold: i) a mechanism responsible for the acceleration of $\barqHe$ up to 50 GeV energy is required; ii) the isotopic ratio $\barqHe:\bartHe$ must be inverted during propagation close to the source.  

Depending on the answer to point i), the estimate of the number of such objects can largely vary. A single close-by anti-star might be responsible for the entire anti-helium flux seen by {\it AMS-02}. A possible acceleration mechanism is that large chunks of normal matter, e.g. asteroid-mass clumps,  hit the anti-star, resulting in a powerful annihilation reaction which would eject and accelerate $\barqHe$ nuclei from within the anti-star. Impacts of asteroid-mass clumps with neutron stars are, for example, key elements of a current model for fast radio bursts as in Ref.\cite{Mottez:2014awa} and have been used to constrain the possible presence of anti-matter in our Galaxy \cite{Golubkov:2000xz}.
 As a more relevant  example, we estimate that an object of the size of the Earth annihilating onto the surface of such an anti-star could liberate an energy of order $\sim 10^{49}$ ergs. This is enough for a shell of anti-matter with mass $\sim 0.01$ M$_{\odot}$ to be expelled in outer space with a velocity of $10^{4}$ km/s. This coincides with half the rotational energy of the Crab pulsar which is a well-known potential source of high-energy positrons and electrons. To quantify, if a fraction $f_{\rm acc}$ of a {\em single} anti-star experienced such an event, the total amount of anti-helium ejected in the Galaxy would be approximately
\begin{eqnarray}
\Phi_{\overline{\rm He}} & = & \bigg(\frac{c}{V_{\rm gal}}\bigg)\bigg(\frac{f_{\overline{\rm He}}M_{\bar{*}}}{m_{\overline{\rm He}}}\bigg)f_{\rm acc}\\
& \simeq & 10^{-9} \bigg(\frac{(4\pi/3)(10~{\rm kpc})^3}{V_{\rm gal}}\bigg)\bigg(\frac{M_{\bar{*}}}{M_\odot}\bigg)\nonumber\\
& &  \bigg(\frac{f_{\rm acc}}{10^{-8}}\bigg) \bigg(\frac{f_{\overline{\rm He}}}{1}\bigg)~\#\overline{\rm He}~{\rm cm}^{-2}{\rm s}^{-1}\nonumber\,,
\end{eqnarray}
where $f_{\overline{\rm He}}$ represents the fraction of anti-helium-4 within the anti-star. Interestingly, for $f_{\overline{\rm He}}=1$ and $f_{\rm acc}=10^{-8}$, this is in good agreement with the measured {\it AMS-02} flux in the GeV range.
However, given that CR nuclei stay confined within the magnetic halo over a timescale ranging from $\sim 10^7$ to $3 \times 10^8$ yr, which is short compared to the $\sim 10^{10}$ yr of existence of our Galaxy, the probability that such an event occured nowadays is smaller than $3\%$, and it is therefore more likely that there exists a population of such stars. If anti-stars are formed in star clusters, more conventional acceleration mechanisms (e.g. SN shock-waves, jets, outflows) can also be responsible for CRs anti-helium at such energies.  We note that massive stars leading to SN explosions are short-lived, and therefore primordial anti-stars would most likely not survive over the course of the Universe. This acceleration mechanism would  require to form anti-stars from the gas at a much later time. Given the strong constraints on the anti-clouds scenario, this case seems disfavored. However, one of these other routes to anti-matter CR acceleration from anti-stars is the case where a binary of anti-matter white dwarfs would merge giving an anti-matter type Ia supernova. Regular white-dwarf mergers occur at a rate per unit stellar mass of $1.4 \times 10^{-13}$ yr$^{-1}$ M$_{\odot}^{-1}$ \cite{Badenes:2012ak}. Requiring that at least one binary of anti-matter white dwarfs merges over a typical CR diffusion timescale translates into a minimal stellar population of anti-stars of $\sim 2.4 \times 10^{4}$ to $7 \times 10^{5}$ M$_{\odot}$ within 10~kpc from the Earth. This is very small compared to the Galactic stellar population which amounts to $\sim 6 \times 10^{10}$ M$_{\odot}$.
In order to achieve point ii), spallation around the source needs to be efficient enough such as to convert a large amount of $\barqHe$ into $\bartHe$.   Given the total cross-section for $\bar{p}\,^4$He interactions as well as the fraction of events going into $^3$He$+X$ measured by the Lear collaboration \cite{Balestra:1985wk}, we estimate that a grammage of order ~$20$ g/cm$^2$ would be enough to generate an isotopic ratio $\bartHe$:$\barqHe$ of roughly 3:1. A similar estimate can be calculated from the measurement of the isotopic ratio of ${}^4$He:${}^3$He by {\it PAMELA} \cite{Adriani:2015aps}, that is  $\sim$ 5:1 aroAund a few GeV/n, and from the fact that the grammage in our Galaxy below 100 GeV is $\sim~3$ g/cm$^2$ (deduced from B/C analysis \cite{DAngelo:2015cfw}).  The grammage required for anti-helium is reasonable as it corresponds to a layer 200 m thick with density $10^{-3}$ g/cm$^{3}$, i.e., 1/50th of our atmosphere. If true, the origin of this grammage woud most likely be related to the origin of the anti-star itself. Indeed, we expect anti-stars to be surrounded by much denser material than that around normal stars, as the former are born from large over-densities at a much earlier time.

\subsubsection{Constraints on anti-stars}
Given that a single anti-star could explain {\it AMS-02} data, there is no strong constrain on the presence of such objects in our Galaxy.  Indeed, even if all of the anti-helium-4 is converted to antiprotons, it would only lead to a handful of events that can easily be hidden within the $\sim10^5$ $\overline{p}$  events observed by {\it AMS-02} \cite{Aguilar:2016kjl}. We can however constrain the presence of such object in the vicinity of the Sun. Assuming spherical (Bondi) accretion and making use of unidentified source in the 2FGL Fermi-LAT catalog, Ref.~\cite{vonBallmoos:2014zza} constrained the local environment, within 150 pc from the Sun, to have $N_{\bar{*}}< 4\times10^{-5}N_{*}$. 
The brightest unassociated source from the 3FGL catalog emits $2\times10^{-8}\#\gamma$ cm$^{-2}$s$^{-1}$ above 1 GeV \cite{Acero:2015hja}. From this, we can estimate the distance of the closest anti-star assuming that its luminosity is sourced by annihilation at its surface. The luminosity associated to the emission is \begin{eqnarray}
\label{eq:Lum}
L_{\bar{*}} & = & 8\pi R_{\bar{*}}^2vn_p \\
& \simeq & 10^{31}\bigg( \frac{R_{\bar{*}}}{10^{11}~{\rm cm}} \bigg)^2\bigg( \frac{v}{300 {\rm km}~{\rm s}^{-1}} \bigg)\bigg( \frac{n_p}{1 {\rm cm}^{-3}} \bigg){\#\gamma}~{\rm  s}^{-1}\,, \nonumber
\end{eqnarray}
where we assumed that the dominant channel for prompt photon emission is through $\pi_0$ production (whose average multiplicity is 2 per annihilation at rest \cite{Amsler:1997up}). 
The minimal distance of such an object is obtained by requiring
\begin{equation}
\frac{L_{\bar{*}}}{4\pi d_{\bar{*}}^2}\leq2\times10^{-8}\#\gamma\,{\rm cm}^{-2}{\rm s}^{-1}\,,
\end{equation}
which yields 
\begin{equation}
d_{\bar{*}}\geq 6\times10^{18}\sqrt{\bigg( \frac{R_{\bar{*}}}{10^{11}~{\rm cm}} \bigg)\bigg( \frac{v}{300 {\rm km}~{\rm s}^{-1}} \bigg)\bigg( \frac{n_p}{1 {\rm cm}^{-3}} \bigg)}~{\rm cm}\,.
\end{equation}
Hence, it is possible that an anti-star whose main source of emission is annihilation at its surface lies in a close-by environment $\sim{\cal O}(1 ~{\rm pc})$ away from the Sun.

Although constraints in our Galaxy are weak, bounds on the scenario can potentially be derived from annihilations and energy injection in the early Universe. Any realistic scenario would lead to the creation of a population of such objects that in turn could lead to spectral distortions of the CMB and modify the CMB anisotropy power spectra.  We have calculated in sec.~\ref{sec:CMB} the specific case of homogeneously distributed anti-matter domains. A similar calculation can be done to get a rough constraint on the number density of anti-stars from CMB data. We can calculate the energy injection rate from annihilation at the surface of an anti-star moving in the photon-baryon plasma at a velocity $v\sim 30$km/s (the typical relative velocity between baryons and CDM-like component at early times \cite{Tseliakhovich:2010bj}):
\begin{eqnarray}
\frac{\mathrm{d}^2 E}{\mathrm{d} V \mathrm{d}t}\bigg|_{\bar{\star}}
& = & 8\pi R_{\bar{*}}^2vn_pm_pc^2n_{\bar{\star}}\\
& \simeq& 10^{13} n_{\bar{\star}}~{\rm J ~s}^{-1}\nonumber\\
& \times &\bigg( \frac{R_{\bar{*}}}{10^{11}~{\rm cm}} \bigg)\bigg( \frac{v}{30 {\rm km}~{\rm s}^{-1}} \bigg)\bigg( \frac{n_p^0}{2\times10^{-7} {\rm cm}^{-3}} \bigg)\,.\nonumber
\end{eqnarray}
Applying the constraints from {\em Planck} given by Eq.~(\ref{eq:planck_constraints}), we can derive that on cosmological scales
\begin{equation}
n_{\bar{\star}} \lesssim 10^{24}(1+z)^3{\rm Mpc}^{-3}\,,
\end{equation}
which trivially satisfies AMS measurements. 
We stress that this very weak constraint assumes that the main source of ionizing radiation is annihilation at the surface of anti-stars. A more accurate constraint would also take into account radiation coming from nuclear processes at play within anti-stars, which would require additional assumptions about these objects.

\section{Discussion and conclusions}
\label{sec:discussion}
In this work, we have studied the implications of the potential discovery of anti-helium-3 and -4 nuclei by the {\it AMS-02} experiment. Using up-to-date semi-analytical tools, we have shown that it is impossible to explain these events as secondaries, i.e., from the spallation of CR protons and helium nuclei onto the ISM.
The $\bartHe$  is typically one to two orders of magnitude below the sensitivity of {\it AMS-02} after 5 years, and the $\barqHe$ is roughly 5 orders of magnitude below {\it AMS-02} reach. It is conceivable that  $\bartHe$ has been misidentified for $\barqHe$. Still, we have argued that the pure secondary explanation would require a large increase of the coalescence momentum at low energies, a behavior that goes against theoretical considerations and experimental results. The DM scenario suffers the same difficulties. Hence, we have discussed how this detection, if confirmed, would indicate the existence of an anti-world, in the form of anti-stars or anti-clouds.
We summarize what we have learned about the properties of anti-matter regions:
\begin{itemize}
\item[\textbullet] Taken at face value the isotopic ratio of anti-helium nuclei is puzzling. We have shown that it can be explained by anisotropic BBN in regions where $\bar{\eta}\sim1.3-6\times10^{-13}$.
\item[\textbullet] The density, size and number of anti-matter domains is constrained by {\it AMS-02} observations and our knowledge of Galactic properties to verify Eq.~(\ref{eq:nbar}). The only theoretical assumption behind is that the isotopic ratio measured by {\it AMS-02} comes from BBN. Interestingly, a few highly dense clouds are sufficient to explain {\it AMS-02} measurements.
\item[\textbullet] The annihilation rate of anti-matter in our Galaxy requires anti-domains to be poor in normal matter (typically a tenth or less of the normal matter density). Considering the annihilation rate in the early Universe leads to even stronger requirements, which would imply the existence of some exotic mechanism allowing segregation of matter and anti-matter domains all along cosmic evolution that makes the existence of such anti-clouds quite improbable. 
\item[\textbullet] Additionally, gamma rays can provide strong constraints on this scenario. Non-observation of spectral features in the form of lines with energies close to the proton mass strongly constrains the proton density in anti-matter domain, as given by Eq.~(\ref{eq:AnnConstraints}). However, this constraints apply only if anti-matter domains are numerous and homogeneously distributed within the Galactic disk. We anticipate that very competitive constraints can be obtained from non-observation of positron annihilations and/or pion decays.
\item[\textbullet] Anti-clouds could produce a measurable flux of $\bar{p}$ and $\bar{d}$. Most of the parameter space evades current $\bar{p}$ constraints but could be probed by GAPS.

\item[\textbullet] Alternatively (and more likely), these anti-helium events could originate from anti-star(s) whose main material is anti-helium-4, converted into anti-helium-3 via spallation in the dense environment surrounding the anti-star(s). 

\item[\textbullet] Part of the 3FGL unassociated point sources can be anti-clouds experiencing annihilations due to CRs propagating through them. They can also be anti-stars which experience annihilations as they propagate in the ISM.

\item[\textbullet]  Depending on the (unknown) acceleration mechanism, it is conceivable that a single near-by anti-star (whose distance to the Earth must be larger than $\sim$1 pc) contributes to the {\it AMS-02} observation.
\end{itemize}

All these hints can be used to build a scenario for their formation in the early Universe. Needless to say, the successful creation and survival of such objects within a coherent cosmological model is far from obvious. Here we just mention that there are many scenarios discussed in the literature \cite{Bambi:2007cc,Blinnikov:2014nea,Khlopov:1998uy}, including the Affleck-Dine mechanism \cite{Affleck:1984fy}, which would  lead to the formation of ``bubbles'' of matter and anti-matter with arbitrarily large values of the baryon-asymmetry locally. Depending on the relation between their mass and the corresponding Jeans mass, these bubbles can then lead to the formation  of anti-star-like objects, either through specific inflation scenarios with large density contrast \cite{Carr:1974nx,Carr:1975qj}
 on scales re-entering the horizon around the QCD phase-transition, i.e., $T\sim{\cal O}(100~{\rm MeV})$, or from peculiar dynamics of the plasma within the bubble, as described for instance in Ref.~\cite{Dolgov:1992pu}. In the latter scenario, the {\em negative} pressure perturbation inside the bubble leads to the collapse of baryons within this region. If the value of the baryon-asymmetry in the bubble is very large, it is even possible that different expansion rate (due to more non-relativistic matter inside the bubble) naturally leads to the growth of density perturbations much earlier than outside of these regions. 
Given the strong implications of the discovery of a single anti-helium-4 nucleus for cosmology, important theoretical and experimental efforts must be undertaken in order to assess whether the reported events could be explained by a more mundane source, such as interactions within the detector, or another source of yet unkown systematic error. Still, this potential discovery would represent an important probe of conditions prevailing in the very early Universe and should be investigated further in future work. 
\appendix

\acknowledgments
We thank Kim Boddy, Kfir Blum and Robert K. Schaefer for very interesting discussions. We thank Annika Reinert and Martin Winkler for clarifications about the antiproton cross-section parameterization. We thank Alexandre Arbey for his help with the \texttt{AlterBBN}{} code, as well as Elisabeth Vangioni, Alain Coc, Cyril Pitrou and Jean-Philippe Uzan for helping us check our results with the \texttt{PRIMAT} code. We thank Pasquale D. Serpico, Philip von Doetinchem and Julien Lavalle for their critical and insightful comments on an earlier version of this draft.   This work was partly supported at Johns Hopkins by NSF Grant No.\ 0244990, NASA NNX17AK38G, and the Simons Foundation. P.S. would like to thank Institut Universitaire de France for its support.This research project was conducted using computational resources at the Maryland Advanced Research Computing Center (MARCC).

\bibliography{biblio}

\begin{thebibliography}{122}%
\makeatletter
\providecommand \@ifxundefined [1]{%
 \@ifx{#1\undefined}
}%
\providecommand \@ifnum [1]{%
 \ifnum #1\expandafter \@firstoftwo
 \else \expandafter \@secondoftwo
 \fi
}%
\providecommand \@ifx [1]{%
 \ifx #1\expandafter \@firstoftwo
 \else \expandafter \@secondoftwo
 \fi
}%
\providecommand \natexlab [1]{#1}%
\providecommand \enquote  [1]{``#1''}%
\providecommand \bibnamefont  [1]{#1}%
\providecommand \bibfnamefont [1]{#1}%
\providecommand \citenamefont [1]{#1}%
\providecommand \href@noop [0]{\@secondoftwo}%
\providecommand \href [0]{\begingroup \@sanitize@url \@href}%
\providecommand \@href[1]{\@@startlink{#1}\@@href}%
\providecommand \@@href[1]{\endgroup#1\@@endlink}%
\providecommand \@sanitize@url [0]{\catcode `\\12\catcode `\$12\catcode
  `\&12\catcode `\#12\catcode `\^12\catcode `\_12\catcode `\%12\relax}%
\providecommand \@@startlink[1]{}%
\providecommand \@@endlink[0]{}%
\providecommand \url  [0]{\begingroup\@sanitize@url \@url }%
\providecommand \@url [1]{\endgroup\@href {#1}{\urlprefix }}%
\providecommand \urlprefix  [0]{URL }%
\providecommand \Eprint [0]{\href }%
\providecommand \doibase [0]{http://dx.doi.org/}%
\providecommand \selectlanguage [0]{\@gobble}%
\providecommand \bibinfo  [0]{\@secondoftwo}%
\providecommand \bibfield  [0]{\@secondoftwo}%
\providecommand \translation [1]{[#1]}%
\providecommand \BibitemOpen [0]{}%
\providecommand \bibitemStop [0]{}%
\providecommand \bibitemNoStop [0]{.\EOS\space}%
\providecommand \EOS [0]{\spacefactor3000\relax}%
\providecommand \BibitemShut  [1]{\csname bibitem#1\endcsname}%
\let\auto@bib@innerbib\@empty
\bibitem [{\citenamefont {Adriani}\ \emph {et~al.}(2010)\citenamefont {Adriani}
  \emph {et~al.}}]{Adriani:2010rc}%
  \BibitemOpen
  \bibfield  {author} {\bibinfo {author} {\bibfnamefont {O.}~\bibnamefont
  {Adriani}} \emph {et~al.} (\bibinfo {collaboration} {PAMELA}),\ }\href
  {\doibase 10.1103/PhysRevLett.105.121101} {\bibfield  {journal} {\bibinfo
  {journal} {Phys. Rev. Lett.}\ }\textbf {\bibinfo {volume} {105}},\ \bibinfo
  {pages} {121101} (\bibinfo {year} {2010})},\ \Eprint
  {http://arxiv.org/abs/1007.0821} {arXiv:1007.0821 [astro-ph.HE]} \BibitemShut
  {NoStop}%
\bibitem [{\citenamefont {Bergstrom}\ \emph {et~al.}(2008)\citenamefont
  {Bergstrom}, \citenamefont {Bringmann},\ and\ \citenamefont
  {Edsjo}}]{Bergstrom:2008gr}%
  \BibitemOpen
  \bibfield  {author} {\bibinfo {author} {\bibfnamefont {L.}~\bibnamefont
  {Bergstrom}}, \bibinfo {author} {\bibfnamefont {T.}~\bibnamefont
  {Bringmann}}, \ and\ \bibinfo {author} {\bibfnamefont {J.}~\bibnamefont
  {Edsjo}},\ }\href {\doibase 10.1103/PhysRevD.78.103520} {\bibfield  {journal}
  {\bibinfo  {journal} {Phys. Rev.}\ }\textbf {\bibinfo {volume} {D78}},\
  \bibinfo {pages} {103520} (\bibinfo {year} {2008})},\ \Eprint
  {http://arxiv.org/abs/0808.3725} {arXiv:0808.3725 [astro-ph]} \BibitemShut
  {NoStop}%
\bibitem [{\citenamefont {Cirelli}\ and\ \citenamefont
  {Strumia}(2008)}]{Cirelli:2008jk}%
  \BibitemOpen
  \bibfield  {author} {\bibinfo {author} {\bibfnamefont {M.}~\bibnamefont
  {Cirelli}}\ and\ \bibinfo {author} {\bibfnamefont {A.}~\bibnamefont
  {Strumia}},\ }\bibfield  {booktitle} {\emph {\bibinfo {booktitle}
  {{Proceedings, 7th International Workshop on the Identification of Dark
  Matter (IDM 2008): Stockholm, Sweden, August 18-22, 2008}}},\ }\href@noop {}
  {\bibfield  {journal} {\bibinfo  {journal} {PoS}\ }\textbf {\bibinfo {volume}
  {IDM2008}},\ \bibinfo {pages} {089} (\bibinfo {year} {2008})},\ \Eprint
  {http://arxiv.org/abs/0808.3867} {arXiv:0808.3867 [astro-ph]} \BibitemShut
  {NoStop}%
\bibitem [{\citenamefont {Cirelli}\ \emph
  {et~al.}(2009{\natexlab{a}})\citenamefont {Cirelli}, \citenamefont
  {Kadastik}, \citenamefont {Raidal},\ and\ \citenamefont
  {Strumia}}]{Cirelli:2008pk}%
  \BibitemOpen
  \bibfield  {author} {\bibinfo {author} {\bibfnamefont {M.}~\bibnamefont
  {Cirelli}}, \bibinfo {author} {\bibfnamefont {M.}~\bibnamefont {Kadastik}},
  \bibinfo {author} {\bibfnamefont {M.}~\bibnamefont {Raidal}}, \ and\ \bibinfo
  {author} {\bibfnamefont {A.}~\bibnamefont {Strumia}},\ }\href {\doibase
  10.1016/j.nuclphysb.2013.05.002, 10.1016/j.nuclphysb.2008.11.031} {\bibfield
  {journal} {\bibinfo  {journal} {Nucl. Phys.}\ }\textbf {\bibinfo {volume}
  {B813}},\ \bibinfo {pages} {1} (\bibinfo {year} {2009}{\natexlab{a}})},\
  \bibinfo {note} {[Addendum: Nucl. Phys.B873,530(2013)]},\ \Eprint
  {http://arxiv.org/abs/0809.2409} {arXiv:0809.2409 [hep-ph]} \BibitemShut
  {NoStop}%
\bibitem [{\citenamefont {Nelson}\ and\ \citenamefont
  {Spitzer}(2010)}]{Nelson:2008hj}%
  \BibitemOpen
  \bibfield  {author} {\bibinfo {author} {\bibfnamefont {A.~E.}\ \bibnamefont
  {Nelson}}\ and\ \bibinfo {author} {\bibfnamefont {C.}~\bibnamefont
  {Spitzer}},\ }\href {\doibase 10.1007/JHEP10(2010)066} {\bibfield  {journal}
  {\bibinfo  {journal} {JHEP}\ }\textbf {\bibinfo {volume} {10}},\ \bibinfo
  {pages} {066} (\bibinfo {year} {2010})},\ \Eprint
  {http://arxiv.org/abs/0810.5167} {arXiv:0810.5167 [hep-ph]} \BibitemShut
  {NoStop}%
\bibitem [{\citenamefont {Arkani-Hamed}\ \emph {et~al.}(2009)\citenamefont
  {Arkani-Hamed}, \citenamefont {Finkbeiner}, \citenamefont {Slatyer},\ and\
  \citenamefont {Weiner}}]{ArkaniHamed:2008qn}%
  \BibitemOpen
  \bibfield  {author} {\bibinfo {author} {\bibfnamefont {N.}~\bibnamefont
  {Arkani-Hamed}}, \bibinfo {author} {\bibfnamefont {D.~P.}\ \bibnamefont
  {Finkbeiner}}, \bibinfo {author} {\bibfnamefont {T.~R.}\ \bibnamefont
  {Slatyer}}, \ and\ \bibinfo {author} {\bibfnamefont {N.}~\bibnamefont
  {Weiner}},\ }\href {\doibase 10.1103/PhysRevD.79.015014} {\bibfield
  {journal} {\bibinfo  {journal} {Phys. Rev.}\ }\textbf {\bibinfo {volume}
  {D79}},\ \bibinfo {pages} {015014} (\bibinfo {year} {2009})},\ \Eprint
  {http://arxiv.org/abs/0810.0713} {arXiv:0810.0713 [hep-ph]} \BibitemShut
  {NoStop}%
\bibitem [{\citenamefont {Harnik}\ and\ \citenamefont
  {Kribs}(2009)}]{Harnik:2008uu}%
  \BibitemOpen
  \bibfield  {author} {\bibinfo {author} {\bibfnamefont {R.}~\bibnamefont
  {Harnik}}\ and\ \bibinfo {author} {\bibfnamefont {G.~D.}\ \bibnamefont
  {Kribs}},\ }\href {\doibase 10.1103/PhysRevD.79.095007} {\bibfield  {journal}
  {\bibinfo  {journal} {Phys. Rev.}\ }\textbf {\bibinfo {volume} {D79}},\
  \bibinfo {pages} {095007} (\bibinfo {year} {2009})},\ \Eprint
  {http://arxiv.org/abs/0810.5557} {arXiv:0810.5557 [hep-ph]} \BibitemShut
  {NoStop}%
\bibitem [{\citenamefont {Fox}\ and\ \citenamefont
  {Poppitz}(2009)}]{Fox:2008kb}%
  \BibitemOpen
  \bibfield  {author} {\bibinfo {author} {\bibfnamefont {P.~J.}\ \bibnamefont
  {Fox}}\ and\ \bibinfo {author} {\bibfnamefont {E.}~\bibnamefont {Poppitz}},\
  }\href {\doibase 10.1103/PhysRevD.79.083528} {\bibfield  {journal} {\bibinfo
  {journal} {Phys. Rev.}\ }\textbf {\bibinfo {volume} {D79}},\ \bibinfo {pages}
  {083528} (\bibinfo {year} {2009})},\ \Eprint {http://arxiv.org/abs/0811.0399}
  {arXiv:0811.0399 [hep-ph]} \BibitemShut {NoStop}%
\bibitem [{\citenamefont {Pospelov}\ and\ \citenamefont
  {Ritz}(2009)}]{Pospelov:2008jd}%
  \BibitemOpen
  \bibfield  {author} {\bibinfo {author} {\bibfnamefont {M.}~\bibnamefont
  {Pospelov}}\ and\ \bibinfo {author} {\bibfnamefont {A.}~\bibnamefont
  {Ritz}},\ }\href {\doibase 10.1016/j.physletb.2008.12.012} {\bibfield
  {journal} {\bibinfo  {journal} {Phys. Lett.}\ }\textbf {\bibinfo {volume}
  {B671}},\ \bibinfo {pages} {391} (\bibinfo {year} {2009})},\ \Eprint
  {http://arxiv.org/abs/0810.1502} {arXiv:0810.1502 [hep-ph]} \BibitemShut
  {NoStop}%
\bibitem [{\citenamefont {March-Russell}\ and\ \citenamefont
  {West}(2009)}]{MarchRussell:2008tu}%
  \BibitemOpen
  \bibfield  {author} {\bibinfo {author} {\bibfnamefont {J.~D.}\ \bibnamefont
  {March-Russell}}\ and\ \bibinfo {author} {\bibfnamefont {S.~M.}\ \bibnamefont
  {West}},\ }\href {\doibase 10.1016/j.physletb.2009.04.010} {\bibfield
  {journal} {\bibinfo  {journal} {Phys. Lett.}\ }\textbf {\bibinfo {volume}
  {B676}},\ \bibinfo {pages} {133} (\bibinfo {year} {2009})},\ \Eprint
  {http://arxiv.org/abs/0812.0559} {arXiv:0812.0559 [astro-ph]} \BibitemShut
  {NoStop}%
\bibitem [{\citenamefont {Dienes}\ \emph {et~al.}(2013)\citenamefont {Dienes},
  \citenamefont {Kumar},\ and\ \citenamefont {Thomas}}]{Dienes:2013xff}%
  \BibitemOpen
  \bibfield  {author} {\bibinfo {author} {\bibfnamefont {K.~R.}\ \bibnamefont
  {Dienes}}, \bibinfo {author} {\bibfnamefont {J.}~\bibnamefont {Kumar}}, \
  and\ \bibinfo {author} {\bibfnamefont {B.}~\bibnamefont {Thomas}},\ }\href
  {\doibase 10.1103/PhysRevD.88.103509} {\bibfield  {journal} {\bibinfo
  {journal} {Phys. Rev.}\ }\textbf {\bibinfo {volume} {D88}},\ \bibinfo {pages}
  {103509} (\bibinfo {year} {2013})},\ \Eprint {http://arxiv.org/abs/1306.2959}
  {arXiv:1306.2959 [hep-ph]} \BibitemShut {NoStop}%
\bibitem [{\citenamefont {Kopp}(2013)}]{Kopp:2013eka}%
  \BibitemOpen
  \bibfield  {author} {\bibinfo {author} {\bibfnamefont {J.}~\bibnamefont
  {Kopp}},\ }\href {\doibase 10.1103/PhysRevD.88.076013} {\bibfield  {journal}
  {\bibinfo  {journal} {Phys. Rev.}\ }\textbf {\bibinfo {volume} {D88}},\
  \bibinfo {pages} {076013} (\bibinfo {year} {2013})},\ \Eprint
  {http://arxiv.org/abs/1304.1184} {arXiv:1304.1184 [hep-ph]} \BibitemShut
  {NoStop}%
\bibitem [{\citenamefont {Yuksel}\ \emph {et~al.}(2009)\citenamefont {Yuksel},
  \citenamefont {Kistler},\ and\ \citenamefont {Stanev}}]{Yuksel:2008rf}%
  \BibitemOpen
  \bibfield  {author} {\bibinfo {author} {\bibfnamefont {H.}~\bibnamefont
  {Yuksel}}, \bibinfo {author} {\bibfnamefont {M.~D.}\ \bibnamefont {Kistler}},
  \ and\ \bibinfo {author} {\bibfnamefont {T.}~\bibnamefont {Stanev}},\ }\href
  {\doibase 10.1103/PhysRevLett.103.051101} {\bibfield  {journal} {\bibinfo
  {journal} {Phys. Rev. Lett.}\ }\textbf {\bibinfo {volume} {103}},\ \bibinfo
  {pages} {051101} (\bibinfo {year} {2009})},\ \Eprint
  {http://arxiv.org/abs/0810.2784} {arXiv:0810.2784 [astro-ph]} \BibitemShut
  {NoStop}%
\bibitem [{\citenamefont {Profumo}(2011)}]{Profumo:2008ms}%
  \BibitemOpen
  \bibfield  {author} {\bibinfo {author} {\bibfnamefont {S.}~\bibnamefont
  {Profumo}},\ }\href {\doibase 10.2478/s11534-011-0099-z} {\bibfield
  {journal} {\bibinfo  {journal} {Central Eur. J. Phys.}\ }\textbf {\bibinfo
  {volume} {10}},\ \bibinfo {pages} {1} (\bibinfo {year} {2011})},\ \Eprint
  {http://arxiv.org/abs/0812.4457} {arXiv:0812.4457 [astro-ph]} \BibitemShut
  {NoStop}%
\bibitem [{\citenamefont {Kawanaka}\ \emph {et~al.}(2010)\citenamefont
  {Kawanaka}, \citenamefont {Ioka},\ and\ \citenamefont
  {Nojiri}}]{Kawanaka:2009dk}%
  \BibitemOpen
  \bibfield  {author} {\bibinfo {author} {\bibfnamefont {N.}~\bibnamefont
  {Kawanaka}}, \bibinfo {author} {\bibfnamefont {K.}~\bibnamefont {Ioka}}, \
  and\ \bibinfo {author} {\bibfnamefont {M.~M.}\ \bibnamefont {Nojiri}},\
  }\href {\doibase 10.1088/0004-637X/710/2/958} {\bibfield  {journal} {\bibinfo
   {journal} {Astrophys. J.}\ }\textbf {\bibinfo {volume} {710}},\ \bibinfo
  {pages} {958} (\bibinfo {year} {2010})},\ \Eprint
  {http://arxiv.org/abs/0903.3782} {arXiv:0903.3782 [astro-ph.HE]} \BibitemShut
  {NoStop}%
\bibitem [{\citenamefont {Yuan}\ \emph {et~al.}(2015)\citenamefont {Yuan},
  \citenamefont {Bi}, \citenamefont {Chen}, \citenamefont {Guo}, \citenamefont
  {Lin},\ and\ \citenamefont {Zhang}}]{Yuan:2013eja}%
  \BibitemOpen
  \bibfield  {author} {\bibinfo {author} {\bibfnamefont {Q.}~\bibnamefont
  {Yuan}}, \bibinfo {author} {\bibfnamefont {X.-J.}\ \bibnamefont {Bi}},
  \bibinfo {author} {\bibfnamefont {G.-M.}\ \bibnamefont {Chen}}, \bibinfo
  {author} {\bibfnamefont {Y.-Q.}\ \bibnamefont {Guo}}, \bibinfo {author}
  {\bibfnamefont {S.-J.}\ \bibnamefont {Lin}}, \ and\ \bibinfo {author}
  {\bibfnamefont {X.}~\bibnamefont {Zhang}},\ }\href {\doibase
  10.1016/j.astropartphys.2014.05.005} {\bibfield  {journal} {\bibinfo
  {journal} {Astropart. Phys.}\ }\textbf {\bibinfo {volume} {60}},\ \bibinfo
  {pages} {1} (\bibinfo {year} {2015})},\ \Eprint
  {http://arxiv.org/abs/1304.1482} {arXiv:1304.1482 [astro-ph.HE]} \BibitemShut
  {NoStop}%
\bibitem [{\citenamefont {Yin}\ \emph {et~al.}(2013)\citenamefont {Yin},
  \citenamefont {Yu}, \citenamefont {Yuan},\ and\ \citenamefont
  {Bi}}]{Yin:2013vaa}%
  \BibitemOpen
  \bibfield  {author} {\bibinfo {author} {\bibfnamefont {P.-F.}\ \bibnamefont
  {Yin}}, \bibinfo {author} {\bibfnamefont {Z.-H.}\ \bibnamefont {Yu}},
  \bibinfo {author} {\bibfnamefont {Q.}~\bibnamefont {Yuan}}, \ and\ \bibinfo
  {author} {\bibfnamefont {X.-J.}\ \bibnamefont {Bi}},\ }\href {\doibase
  10.1103/PhysRevD.88.023001} {\bibfield  {journal} {\bibinfo  {journal} {Phys.
  Rev.}\ }\textbf {\bibinfo {volume} {D88}},\ \bibinfo {pages} {023001}
  (\bibinfo {year} {2013})},\ \Eprint {http://arxiv.org/abs/1304.4128}
  {arXiv:1304.4128 [astro-ph.HE]} \BibitemShut {NoStop}%
\bibitem [{\citenamefont {Hooper}\ \emph {et~al.}(2009)\citenamefont {Hooper},
  \citenamefont {Blasi},\ and\ \citenamefont {Serpico}}]{Hooper:2008kg}%
  \BibitemOpen
  \bibfield  {author} {\bibinfo {author} {\bibfnamefont {D.}~\bibnamefont
  {Hooper}}, \bibinfo {author} {\bibfnamefont {P.}~\bibnamefont {Blasi}}, \
  and\ \bibinfo {author} {\bibfnamefont {P.~D.}\ \bibnamefont {Serpico}},\
  }\href {\doibase 10.1088/1475-7516/2009/01/025} {\bibfield  {journal}
  {\bibinfo  {journal} {JCAP}\ }\textbf {\bibinfo {volume} {0901}},\ \bibinfo
  {pages} {025} (\bibinfo {year} {2009})},\ \Eprint
  {http://arxiv.org/abs/0810.1527} {arXiv:0810.1527 [astro-ph]} \BibitemShut
  {NoStop}%
\bibitem [{\citenamefont {Cholis}\ \emph
  {et~al.}(2009{\natexlab{a}})\citenamefont {Cholis}, \citenamefont
  {Finkbeiner}, \citenamefont {Goodenough},\ and\ \citenamefont
  {Weiner}}]{Cholis:2008qq}%
  \BibitemOpen
  \bibfield  {author} {\bibinfo {author} {\bibfnamefont {I.}~\bibnamefont
  {Cholis}}, \bibinfo {author} {\bibfnamefont {D.~P.}\ \bibnamefont
  {Finkbeiner}}, \bibinfo {author} {\bibfnamefont {L.}~\bibnamefont
  {Goodenough}}, \ and\ \bibinfo {author} {\bibfnamefont {N.}~\bibnamefont
  {Weiner}},\ }\href {\doibase 10.1088/1475-7516/2009/12/007} {\bibfield
  {journal} {\bibinfo  {journal} {JCAP}\ }\textbf {\bibinfo {volume} {0912}},\
  \bibinfo {pages} {007} (\bibinfo {year} {2009}{\natexlab{a}})},\ \Eprint
  {http://arxiv.org/abs/0810.5344} {arXiv:0810.5344 [astro-ph]} \BibitemShut
  {NoStop}%
\bibitem [{\citenamefont {Cholis}\ \emph
  {et~al.}(2009{\natexlab{b}})\citenamefont {Cholis}, \citenamefont
  {Goodenough}, \citenamefont {Hooper}, \citenamefont {Simet},\ and\
  \citenamefont {Weiner}}]{Cholis:2008hb}%
  \BibitemOpen
  \bibfield  {author} {\bibinfo {author} {\bibfnamefont {I.}~\bibnamefont
  {Cholis}}, \bibinfo {author} {\bibfnamefont {L.}~\bibnamefont {Goodenough}},
  \bibinfo {author} {\bibfnamefont {D.}~\bibnamefont {Hooper}}, \bibinfo
  {author} {\bibfnamefont {M.}~\bibnamefont {Simet}}, \ and\ \bibinfo {author}
  {\bibfnamefont {N.}~\bibnamefont {Weiner}},\ }\href {\doibase
  10.1103/PhysRevD.80.123511} {\bibfield  {journal} {\bibinfo  {journal} {Phys.
  Rev.}\ }\textbf {\bibinfo {volume} {D80}},\ \bibinfo {pages} {123511}
  (\bibinfo {year} {2009}{\natexlab{b}})},\ \Eprint
  {http://arxiv.org/abs/0809.1683} {arXiv:0809.1683 [hep-ph]} \BibitemShut
  {NoStop}%
\bibitem [{\citenamefont {Cholis}\ and\ \citenamefont
  {Hooper}(2013)}]{Cholis:2013psa}%
  \BibitemOpen
  \bibfield  {author} {\bibinfo {author} {\bibfnamefont {I.}~\bibnamefont
  {Cholis}}\ and\ \bibinfo {author} {\bibfnamefont {D.}~\bibnamefont
  {Hooper}},\ }\href {\doibase 10.1103/PhysRevD.88.023013} {\bibfield
  {journal} {\bibinfo  {journal} {Phys. Rev.}\ }\textbf {\bibinfo {volume}
  {D88}},\ \bibinfo {pages} {023013} (\bibinfo {year} {2013})},\ \Eprint
  {http://arxiv.org/abs/1304.1840} {arXiv:1304.1840 [astro-ph.HE]} \BibitemShut
  {NoStop}%
\bibitem [{\citenamefont {Boudaud}\ \emph
  {et~al.}(2015{\natexlab{a}})\citenamefont {Boudaud} \emph
  {et~al.}}]{Boudaud:2014dta}%
  \BibitemOpen
  \bibfield  {author} {\bibinfo {author} {\bibfnamefont {M.}~\bibnamefont
  {Boudaud}} \emph {et~al.},\ }\href {\doibase 10.1051/0004-6361/201425197}
  {\bibfield  {journal} {\bibinfo  {journal} {Astron. Astrophys.}\ }\textbf
  {\bibinfo {volume} {575}},\ \bibinfo {pages} {A67} (\bibinfo {year}
  {2015}{\natexlab{a}})},\ \Eprint {http://arxiv.org/abs/1410.3799}
  {arXiv:1410.3799 [astro-ph.HE]} \BibitemShut {NoStop}%
\bibitem [{\citenamefont {Boudaud}\ \emph {et~al.}(2017)\citenamefont
  {Boudaud}, \citenamefont {Bueno}, \citenamefont {Caroff}, \citenamefont
  {Genolini}, \citenamefont {Poulin}, \citenamefont {Poireau}, \citenamefont
  {Putze}, \citenamefont {Rosier}, \citenamefont {Salati},\ and\ \citenamefont
  {Vecchi}}]{Boudaud:2016jvj}%
  \BibitemOpen
  \bibfield  {author} {\bibinfo {author} {\bibfnamefont {M.}~\bibnamefont
  {Boudaud}}, \bibinfo {author} {\bibfnamefont {E.~F.}\ \bibnamefont {Bueno}},
  \bibinfo {author} {\bibfnamefont {S.}~\bibnamefont {Caroff}}, \bibinfo
  {author} {\bibfnamefont {Y.}~\bibnamefont {Genolini}}, \bibinfo {author}
  {\bibfnamefont {V.}~\bibnamefont {Poulin}}, \bibinfo {author} {\bibfnamefont
  {V.}~\bibnamefont {Poireau}}, \bibinfo {author} {\bibfnamefont
  {A.}~\bibnamefont {Putze}}, \bibinfo {author} {\bibfnamefont
  {S.}~\bibnamefont {Rosier}}, \bibinfo {author} {\bibfnamefont
  {P.}~\bibnamefont {Salati}}, \ and\ \bibinfo {author} {\bibfnamefont
  {M.}~\bibnamefont {Vecchi}},\ }\href {\doibase 10.1051/0004-6361/201630321}
  {\bibfield  {journal} {\bibinfo  {journal} {Astron. Astrophys.}\ }\textbf
  {\bibinfo {volume} {605}},\ \bibinfo {pages} {A17} (\bibinfo {year}
  {2017})},\ \Eprint {http://arxiv.org/abs/1612.03924} {arXiv:1612.03924
  [astro-ph.HE]} \BibitemShut {NoStop}%
\bibitem [{\citenamefont {Hooper}\ \emph {et~al.}(2017)\citenamefont {Hooper},
  \citenamefont {Cholis}, \citenamefont {Linden},\ and\ \citenamefont
  {Fang}}]{Hooper:2017gtd}%
  \BibitemOpen
  \bibfield  {author} {\bibinfo {author} {\bibfnamefont {D.}~\bibnamefont
  {Hooper}}, \bibinfo {author} {\bibfnamefont {I.}~\bibnamefont {Cholis}},
  \bibinfo {author} {\bibfnamefont {T.}~\bibnamefont {Linden}}, \ and\ \bibinfo
  {author} {\bibfnamefont {K.}~\bibnamefont {Fang}},\ }\href {\doibase
  10.1103/PhysRevD.96.103013} {\bibfield  {journal} {\bibinfo  {journal} {Phys.
  Rev.}\ }\textbf {\bibinfo {volume} {D96}},\ \bibinfo {pages} {103013}
  (\bibinfo {year} {2017})},\ \Eprint {http://arxiv.org/abs/1702.08436}
  {arXiv:1702.08436 [astro-ph.HE]} \BibitemShut {NoStop}%
\bibitem [{\citenamefont {Cholis}\ \emph
  {et~al.}(2018{\natexlab{a}})\citenamefont {Cholis}, \citenamefont {Karwal},\
  and\ \citenamefont {Kamionkowski}}]{Cholis:2017ccs}%
  \BibitemOpen
  \bibfield  {author} {\bibinfo {author} {\bibfnamefont {I.}~\bibnamefont
  {Cholis}}, \bibinfo {author} {\bibfnamefont {T.}~\bibnamefont {Karwal}}, \
  and\ \bibinfo {author} {\bibfnamefont {M.}~\bibnamefont {Kamionkowski}},\
  }\href {\doibase 10.1103/PhysRevD.97.123011} {\bibfield  {journal} {\bibinfo
  {journal} {Phys. Rev.}\ }\textbf {\bibinfo {volume} {D97}},\ \bibinfo {pages}
  {123011} (\bibinfo {year} {2018}{\natexlab{a}})},\ \Eprint
  {http://arxiv.org/abs/1712.00011} {arXiv:1712.00011 [astro-ph.HE]}
  \BibitemShut {NoStop}%
\bibitem [{\citenamefont {Cholis}\ \emph
  {et~al.}(2018{\natexlab{b}})\citenamefont {Cholis}, \citenamefont {Karwal},\
  and\ \citenamefont {Kamionkowski}}]{Cholis:2018izy}%
  \BibitemOpen
  \bibfield  {author} {\bibinfo {author} {\bibfnamefont {I.}~\bibnamefont
  {Cholis}}, \bibinfo {author} {\bibfnamefont {T.}~\bibnamefont {Karwal}}, \
  and\ \bibinfo {author} {\bibfnamefont {M.}~\bibnamefont {Kamionkowski}},\
  }\href@noop {} {\  (\bibinfo {year} {2018}{\natexlab{b}})},\ \Eprint
  {http://arxiv.org/abs/1807.05230} {arXiv:1807.05230 [astro-ph.HE]}
  \BibitemShut {NoStop}%
\bibitem [{\citenamefont {Cuoco}\ \emph {et~al.}(2017)\citenamefont {Cuoco},
  \citenamefont {Krämer},\ and\ \citenamefont {Korsmeier}}]{Cuoco:2016eej}%
  \BibitemOpen
  \bibfield  {author} {\bibinfo {author} {\bibfnamefont {A.}~\bibnamefont
  {Cuoco}}, \bibinfo {author} {\bibfnamefont {M.}~\bibnamefont {Krämer}}, \
  and\ \bibinfo {author} {\bibfnamefont {M.}~\bibnamefont {Korsmeier}},\ }\href
  {\doibase 10.1103/PhysRevLett.118.191102} {\bibfield  {journal} {\bibinfo
  {journal} {Phys. Rev. Lett.}\ }\textbf {\bibinfo {volume} {118}},\ \bibinfo
  {pages} {191102} (\bibinfo {year} {2017})},\ \Eprint
  {http://arxiv.org/abs/1610.03071} {arXiv:1610.03071 [astro-ph.HE]}
  \BibitemShut {NoStop}%
\bibitem [{\citenamefont {Reinert}\ and\ \citenamefont
  {Winkler}(2017)}]{Reinert:2017aga}%
  \BibitemOpen
  \bibfield  {author} {\bibinfo {author} {\bibfnamefont {A.}~\bibnamefont
  {Reinert}}\ and\ \bibinfo {author} {\bibfnamefont {M.~W.}\ \bibnamefont
  {Winkler}},\ }\href@noop {} {\  (\bibinfo {year} {2017})},\ \Eprint
  {http://arxiv.org/abs/1712.00002} {arXiv:1712.00002 [astro-ph.HE]}
  \BibitemShut {NoStop}%
\bibitem [{\citenamefont {Chardonnet}\ \emph {et~al.}(1997)\citenamefont
  {Chardonnet}, \citenamefont {Orloff},\ and\ \citenamefont
  {Salati}}]{Chardonnet:1997dv}%
  \BibitemOpen
  \bibfield  {author} {\bibinfo {author} {\bibfnamefont {P.}~\bibnamefont
  {Chardonnet}}, \bibinfo {author} {\bibfnamefont {J.}~\bibnamefont {Orloff}},
  \ and\ \bibinfo {author} {\bibfnamefont {P.}~\bibnamefont {Salati}},\ }\href
  {\doibase 10.1016/S0370-2693(97)00870-8} {\bibfield  {journal} {\bibinfo
  {journal} {Phys. Lett.}\ }\textbf {\bibinfo {volume} {B409}},\ \bibinfo
  {pages} {313} (\bibinfo {year} {1997})},\ \Eprint
  {http://arxiv.org/abs/astro-ph/9705110} {arXiv:astro-ph/9705110 [astro-ph]}
  \BibitemShut {NoStop}%
\bibitem [{\citenamefont {Donato}\ \emph {et~al.}(2000)\citenamefont {Donato},
  \citenamefont {Fornengo},\ and\ \citenamefont {Salati}}]{Donato:1999gy}%
  \BibitemOpen
  \bibfield  {author} {\bibinfo {author} {\bibfnamefont {F.}~\bibnamefont
  {Donato}}, \bibinfo {author} {\bibfnamefont {N.}~\bibnamefont {Fornengo}}, \
  and\ \bibinfo {author} {\bibfnamefont {P.}~\bibnamefont {Salati}},\ }\href
  {\doibase 10.1103/PhysRevD.62.043003} {\bibfield  {journal} {\bibinfo
  {journal} {Phys. Rev.}\ }\textbf {\bibinfo {volume} {D62}},\ \bibinfo {pages}
  {043003} (\bibinfo {year} {2000})},\ \Eprint
  {http://arxiv.org/abs/hep-ph/9904481} {arXiv:hep-ph/9904481 [hep-ph]}
  \BibitemShut {NoStop}%
\bibitem [{\citenamefont {Duperray}\ \emph {et~al.}(2005)\citenamefont
  {Duperray}, \citenamefont {Baret}, \citenamefont {Maurin}, \citenamefont
  {Boudoul}, \citenamefont {Barrau}, \citenamefont {Derome}, \citenamefont
  {Protasov},\ and\ \citenamefont {Buenerd}}]{Duperray:2005si}%
  \BibitemOpen
  \bibfield  {author} {\bibinfo {author} {\bibfnamefont {R.}~\bibnamefont
  {Duperray}}, \bibinfo {author} {\bibfnamefont {B.}~\bibnamefont {Baret}},
  \bibinfo {author} {\bibfnamefont {D.}~\bibnamefont {Maurin}}, \bibinfo
  {author} {\bibfnamefont {G.}~\bibnamefont {Boudoul}}, \bibinfo {author}
  {\bibfnamefont {A.}~\bibnamefont {Barrau}}, \bibinfo {author} {\bibfnamefont
  {L.}~\bibnamefont {Derome}}, \bibinfo {author} {\bibfnamefont
  {K.}~\bibnamefont {Protasov}}, \ and\ \bibinfo {author} {\bibfnamefont
  {M.}~\bibnamefont {Buenerd}},\ }\href {\doibase 10.1103/PhysRevD.71.083013}
  {\bibfield  {journal} {\bibinfo  {journal} {Phys. Rev.}\ }\textbf {\bibinfo
  {volume} {D71}},\ \bibinfo {pages} {083013} (\bibinfo {year} {2005})},\
  \Eprint {http://arxiv.org/abs/astro-ph/0503544} {arXiv:astro-ph/0503544
  [astro-ph]} \BibitemShut {NoStop}%
\bibitem [{\citenamefont {von Doetinchem}\ \emph {et~al.}(2016)\citenamefont
  {von Doetinchem} \emph {et~al.}}]{vonDoetinchem:2015yva}%
  \BibitemOpen
  \bibfield  {author} {\bibinfo {author} {\bibfnamefont {P.}~\bibnamefont {von
  Doetinchem}} \emph {et~al.},\ }\bibfield  {booktitle} {\emph {\bibinfo
  {booktitle} {{Proceedings, 34th International Cosmic Ray Conference (ICRC
  2015): The Hague, The Netherlands, July 30-August 6, 2015}}},\ }\href@noop {}
  {\bibfield  {journal} {\bibinfo  {journal} {PoS}\ }\textbf {\bibinfo {volume}
  {ICRC2015}},\ \bibinfo {pages} {1218} (\bibinfo {year} {2016})},\ \Eprint
  {http://arxiv.org/abs/1507.02712} {arXiv:1507.02712 [hep-ph]} \BibitemShut
  {NoStop}%
\bibitem [{\citenamefont {Carlson}\ \emph {et~al.}(2014)\citenamefont
  {Carlson}, \citenamefont {Coogan}, \citenamefont {Linden}, \citenamefont
  {Profumo}, \citenamefont {Ibarra},\ and\ \citenamefont
  {Wild}}]{Carlson:2014ssa}%
  \BibitemOpen
  \bibfield  {author} {\bibinfo {author} {\bibfnamefont {E.}~\bibnamefont
  {Carlson}}, \bibinfo {author} {\bibfnamefont {A.}~\bibnamefont {Coogan}},
  \bibinfo {author} {\bibfnamefont {T.}~\bibnamefont {Linden}}, \bibinfo
  {author} {\bibfnamefont {S.}~\bibnamefont {Profumo}}, \bibinfo {author}
  {\bibfnamefont {A.}~\bibnamefont {Ibarra}}, \ and\ \bibinfo {author}
  {\bibfnamefont {S.}~\bibnamefont {Wild}},\ }\href {\doibase
  10.1103/PhysRevD.89.076005} {\bibfield  {journal} {\bibinfo  {journal} {Phys.
  Rev.}\ }\textbf {\bibinfo {volume} {D89}},\ \bibinfo {pages} {076005}
  (\bibinfo {year} {2014})},\ \Eprint {http://arxiv.org/abs/1401.2461}
  {arXiv:1401.2461 [hep-ph]} \BibitemShut {NoStop}%
\bibitem [{\citenamefont {Cirelli}\ \emph {et~al.}(2014)\citenamefont
  {Cirelli}, \citenamefont {Fornengo}, \citenamefont {Taoso},\ and\
  \citenamefont {Vittino}}]{Cirelli:2014qia}%
  \BibitemOpen
  \bibfield  {author} {\bibinfo {author} {\bibfnamefont {M.}~\bibnamefont
  {Cirelli}}, \bibinfo {author} {\bibfnamefont {N.}~\bibnamefont {Fornengo}},
  \bibinfo {author} {\bibfnamefont {M.}~\bibnamefont {Taoso}}, \ and\ \bibinfo
  {author} {\bibfnamefont {A.}~\bibnamefont {Vittino}},\ }\href {\doibase
  10.1007/JHEP08(2014)009} {\bibfield  {journal} {\bibinfo  {journal} {JHEP}\
  }\textbf {\bibinfo {volume} {08}},\ \bibinfo {pages} {009} (\bibinfo {year}
  {2014})},\ \Eprint {http://arxiv.org/abs/1401.4017} {arXiv:1401.4017
  [hep-ph]} \BibitemShut {NoStop}%
\bibitem [{\citenamefont {Coogan}\ and\ \citenamefont
  {Profumo}(2017)}]{Coogan:2017pwt}%
  \BibitemOpen
  \bibfield  {author} {\bibinfo {author} {\bibfnamefont {A.}~\bibnamefont
  {Coogan}}\ and\ \bibinfo {author} {\bibfnamefont {S.}~\bibnamefont
  {Profumo}},\ }\href@noop {} {\  (\bibinfo {year} {2017})},\ \Eprint
  {http://arxiv.org/abs/1705.09664} {arXiv:1705.09664 [astro-ph.HE]}
  \BibitemShut {NoStop}%
\bibitem [{\citenamefont {Steigman}(1976)}]{Steigman:1976ev}%
  \BibitemOpen
  \bibfield  {author} {\bibinfo {author} {\bibfnamefont {G.}~\bibnamefont
  {Steigman}},\ }\href {\doibase 10.1146/annurev.aa.14.090176.002011}
  {\bibfield  {journal} {\bibinfo  {journal} {Ann. Rev. Astron. Astrophys.}\
  }\textbf {\bibinfo {volume} {14}},\ \bibinfo {pages} {339} (\bibinfo {year}
  {1976})}\BibitemShut {NoStop}%
\bibitem [{\citenamefont {Bambi}\ and\ \citenamefont
  {Dolgov}(2007)}]{Bambi:2007cc}%
  \BibitemOpen
  \bibfield  {author} {\bibinfo {author} {\bibfnamefont {C.}~\bibnamefont
  {Bambi}}\ and\ \bibinfo {author} {\bibfnamefont {A.~D.}\ \bibnamefont
  {Dolgov}},\ }\href {\doibase 10.1016/j.nuclphysb.2007.06.010} {\bibfield
  {journal} {\bibinfo  {journal} {Nucl. Phys.}\ }\textbf {\bibinfo {volume}
  {B784}},\ \bibinfo {pages} {132} (\bibinfo {year} {2007})},\ \Eprint
  {http://arxiv.org/abs/astro-ph/0702350} {arXiv:astro-ph/0702350 [astro-ph]}
  \BibitemShut {NoStop}%
\bibitem [{\citenamefont {Choutko}(2018)}]{Choutko}%
  \BibitemOpen
  \bibfield  {author} {\bibinfo {author} {\bibfnamefont {V.~A.}\ \bibnamefont
  {Choutko}},\ }\href@noop {} {\bibfield  {journal} {\bibinfo  {journal} {AMS
  days at la Palma, Spain}\ } (\bibinfo {year} {2018})}\BibitemShut {NoStop}%
\bibitem [{\citenamefont {Ting}(2016)}]{TingTalk}%
  \BibitemOpen
  \bibfield  {author} {\bibinfo {author} {\bibfnamefont {S.}~\bibnamefont
  {Ting}},\ }\href {https://indico.cern.ch/event/592392/} {\bibfield  {journal}
  {\bibinfo  {journal} {The First Five Years of the Alpha Magnetic Spectrometer
  on the ISS}\ } (\bibinfo {year} {2016})}\BibitemShut {NoStop}%
\bibitem [{\citenamefont {Acharya}\ \emph {et~al.}(2017)\citenamefont {Acharya}
  \emph {et~al.}}]{Acharya:2017fvb}%
  \BibitemOpen
  \bibfield  {author} {\bibinfo {author} {\bibfnamefont {S.}~\bibnamefont
  {Acharya}} \emph {et~al.} (\bibinfo {collaboration} {ALICE}),\ }\href@noop {}
  {\  (\bibinfo {year} {2017})},\ \Eprint {http://arxiv.org/abs/1709.08522}
  {arXiv:1709.08522 [nucl-ex]} \BibitemShut {NoStop}%
\bibitem [{\citenamefont {di~Mauro}\ \emph {et~al.}(2014)\citenamefont
  {di~Mauro}, \citenamefont {Donato}, \citenamefont {Goudelis},\ and\
  \citenamefont {Serpico}}]{diMauro:2014zea}%
  \BibitemOpen
  \bibfield  {author} {\bibinfo {author} {\bibfnamefont {M.}~\bibnamefont
  {di~Mauro}}, \bibinfo {author} {\bibfnamefont {F.}~\bibnamefont {Donato}},
  \bibinfo {author} {\bibfnamefont {A.}~\bibnamefont {Goudelis}}, \ and\
  \bibinfo {author} {\bibfnamefont {P.~D.}\ \bibnamefont {Serpico}},\ }\href
  {\doibase 10.1103/PhysRevD.90.085017} {\bibfield  {journal} {\bibinfo
  {journal} {Phys. Rev.}\ }\textbf {\bibinfo {volume} {D90}},\ \bibinfo {pages}
  {085017} (\bibinfo {year} {2014})},\ \Eprint {http://arxiv.org/abs/1408.0288}
  {arXiv:1408.0288 [hep-ph]} \BibitemShut {NoStop}%
\bibitem [{\citenamefont {Duperray}\ \emph {et~al.}(2003)\citenamefont
  {Duperray}, \citenamefont {Protasov},\ and\ \citenamefont
  {Voronin}}]{Duperray:2002pj}%
  \BibitemOpen
  \bibfield  {author} {\bibinfo {author} {\bibfnamefont {R.~P.}\ \bibnamefont
  {Duperray}}, \bibinfo {author} {\bibfnamefont {K.~V.}\ \bibnamefont
  {Protasov}}, \ and\ \bibinfo {author} {\bibfnamefont {A.~{\relax Yu}.}\
  \bibnamefont {Voronin}},\ }\href {\doibase 10.1140/epja/i2002-10074-0}
  {\bibfield  {journal} {\bibinfo  {journal} {Eur. Phys. J.}\ }\textbf
  {\bibinfo {volume} {A16}},\ \bibinfo {pages} {27} (\bibinfo {year} {2003})},\
  \Eprint {http://arxiv.org/abs/nucl-th/0209078} {arXiv:nucl-th/0209078
  [nucl-th]} \BibitemShut {NoStop}%
\bibitem [{\citenamefont {Blum}\ \emph {et~al.}(2017)\citenamefont {Blum},
  \citenamefont {Ng}, \citenamefont {Sato},\ and\ \citenamefont
  {Takimoto}}]{Blum:2017qnn}%
  \BibitemOpen
  \bibfield  {author} {\bibinfo {author} {\bibfnamefont {K.}~\bibnamefont
  {Blum}}, \bibinfo {author} {\bibfnamefont {K.~C.~Y.}\ \bibnamefont {Ng}},
  \bibinfo {author} {\bibfnamefont {R.}~\bibnamefont {Sato}}, \ and\ \bibinfo
  {author} {\bibfnamefont {M.}~\bibnamefont {Takimoto}},\ }\href@noop {} {\
  (\bibinfo {year} {2017})},\ \Eprint {http://arxiv.org/abs/1704.05431}
  {arXiv:1704.05431 [astro-ph.HE]} \BibitemShut {NoStop}%
\bibitem [{\citenamefont {Norbury}\ and\ \citenamefont
  {Townsend}(2007)}]{Norbury:2006hp}%
  \BibitemOpen
  \bibfield  {author} {\bibinfo {author} {\bibfnamefont {J.~W.}\ \bibnamefont
  {Norbury}}\ and\ \bibinfo {author} {\bibfnamefont {L.~W.}\ \bibnamefont
  {Townsend}},\ }\href {\doibase 10.1016/j.nimb.2006.11.054} {\bibfield
  {journal} {\bibinfo  {journal} {Nucl. Instrum. Meth.}\ }\textbf {\bibinfo
  {volume} {B254}},\ \bibinfo {pages} {187} (\bibinfo {year} {2007})},\ \Eprint
  {http://arxiv.org/abs/nucl-th/0612081} {arXiv:nucl-th/0612081 [nucl-th]}
  \BibitemShut {NoStop}%
\bibitem [{\citenamefont {Donato}\ \emph {et~al.}(2004)\citenamefont {Donato},
  \citenamefont {Fornengo}, \citenamefont {Maurin},\ and\ \citenamefont
  {Salati}}]{Donato:2003xg}%
  \BibitemOpen
  \bibfield  {author} {\bibinfo {author} {\bibfnamefont {F.}~\bibnamefont
  {Donato}}, \bibinfo {author} {\bibfnamefont {N.}~\bibnamefont {Fornengo}},
  \bibinfo {author} {\bibfnamefont {D.}~\bibnamefont {Maurin}}, \ and\ \bibinfo
  {author} {\bibfnamefont {P.}~\bibnamefont {Salati}},\ }\href {\doibase
  10.1103/PhysRevD.69.063501} {\bibfield  {journal} {\bibinfo  {journal} {Phys.
  Rev.}\ }\textbf {\bibinfo {volume} {D69}},\ \bibinfo {pages} {063501}
  (\bibinfo {year} {2004})},\ \Eprint {http://arxiv.org/abs/astro-ph/0306207}
  {arXiv:astro-ph/0306207 [astro-ph]} \BibitemShut {NoStop}%
\bibitem [{\citenamefont {Kounine}(2011)}]{Kounine}%
  \BibitemOpen
  \bibfield  {author} {\bibinfo {author} {\bibfnamefont {A.~A.}\ \bibnamefont
  {Kounine}},\ }\href@noop {} {\bibfield  {journal} {\bibinfo  {journal}
  {Proceedings, 32nd ICRC 2011}\ }\textbf {\bibinfo {volume} {c}},\ \bibinfo
  {pages} {5} (\bibinfo {year} {2011})}\BibitemShut {NoStop}%
\bibitem [{\citenamefont {Boudaud}\ \emph
  {et~al.}(2015{\natexlab{b}})\citenamefont {Boudaud}, \citenamefont {Cirelli},
  \citenamefont {Giesen},\ and\ \citenamefont {Salati}}]{Boudaud:2014qra}%
  \BibitemOpen
  \bibfield  {author} {\bibinfo {author} {\bibfnamefont {M.}~\bibnamefont
  {Boudaud}}, \bibinfo {author} {\bibfnamefont {M.}~\bibnamefont {Cirelli}},
  \bibinfo {author} {\bibfnamefont {G.}~\bibnamefont {Giesen}}, \ and\ \bibinfo
  {author} {\bibfnamefont {P.}~\bibnamefont {Salati}},\ }\href {\doibase
  10.1088/1475-7516/2015/05/013} {\bibfield  {journal} {\bibinfo  {journal}
  {JCAP}\ }\textbf {\bibinfo {volume} {1505}},\ \bibinfo {pages} {013}
  (\bibinfo {year} {2015}{\natexlab{b}})},\ \Eprint
  {http://arxiv.org/abs/1412.5696} {arXiv:1412.5696 [astro-ph.HE]} \BibitemShut
  {NoStop}%
\bibitem [{\citenamefont {Giesen}\ \emph {et~al.}(2015)\citenamefont {Giesen},
  \citenamefont {Boudaud}, \citenamefont {Génolini}, \citenamefont {Poulin},
  \citenamefont {Cirelli}, \citenamefont {Salati},\ and\ \citenamefont
  {Serpico}}]{Giesen:2015ufa}%
  \BibitemOpen
  \bibfield  {author} {\bibinfo {author} {\bibfnamefont {G.}~\bibnamefont
  {Giesen}}, \bibinfo {author} {\bibfnamefont {M.}~\bibnamefont {Boudaud}},
  \bibinfo {author} {\bibfnamefont {Y.}~\bibnamefont {Génolini}}, \bibinfo
  {author} {\bibfnamefont {V.}~\bibnamefont {Poulin}}, \bibinfo {author}
  {\bibfnamefont {M.}~\bibnamefont {Cirelli}}, \bibinfo {author} {\bibfnamefont
  {P.}~\bibnamefont {Salati}}, \ and\ \bibinfo {author} {\bibfnamefont {P.~D.}\
  \bibnamefont {Serpico}},\ }\href {\doibase 10.1088/1475-7516/2015/09/023,
  10.1088/1475-7516/2015/9/023} {\bibfield  {journal} {\bibinfo  {journal}
  {JCAP}\ }\textbf {\bibinfo {volume} {1509}},\ \bibinfo {pages} {023}
  (\bibinfo {year} {2015})},\ \Eprint {http://arxiv.org/abs/1504.04276}
  {arXiv:1504.04276 [astro-ph.HE]} \BibitemShut {NoStop}%
\bibitem [{\citenamefont {Hikasa}\ \emph {et~al.}(1992)\citenamefont {Hikasa}
  \emph {et~al.}}]{Hikasa:1992je}%
  \BibitemOpen
  \bibfield  {author} {\bibinfo {author} {\bibfnamefont {K.}~\bibnamefont
  {Hikasa}} \emph {et~al.} (\bibinfo {collaboration} {Particle Data Group}),\
  }\href {\doibase 10.1103/PhysRevD.46.5210, 10.1103/PhysRevD.45.S1} {\bibfield
   {journal} {\bibinfo  {journal} {Phys. Rev.}\ }\textbf {\bibinfo {volume}
  {D45}},\ \bibinfo {pages} {S1} (\bibinfo {year} {1992})},\ \bibinfo {note}
  {[Erratum: Phys. Rev.D46,5210(1992)]}\BibitemShut {NoStop}%
\bibitem [{\citenamefont {Ghelfi}\ \emph {et~al.}(2017)\citenamefont {Ghelfi},
  \citenamefont {Maurin}, \citenamefont {Cheminet}, \citenamefont {Derome},
  \citenamefont {Hubert},\ and\ \citenamefont {Melot}}]{Ghelfi:2016pcv}%
  \BibitemOpen
  \bibfield  {author} {\bibinfo {author} {\bibfnamefont {A.}~\bibnamefont
  {Ghelfi}}, \bibinfo {author} {\bibfnamefont {D.}~\bibnamefont {Maurin}},
  \bibinfo {author} {\bibfnamefont {A.}~\bibnamefont {Cheminet}}, \bibinfo
  {author} {\bibfnamefont {L.}~\bibnamefont {Derome}}, \bibinfo {author}
  {\bibfnamefont {G.}~\bibnamefont {Hubert}}, \ and\ \bibinfo {author}
  {\bibfnamefont {F.}~\bibnamefont {Melot}},\ }\href {\doibase
  10.1016/j.asr.2016.06.027} {\bibfield  {journal} {\bibinfo  {journal} {Adv.
  Space Res.}\ }\textbf {\bibinfo {volume} {60}},\ \bibinfo {pages} {833}
  (\bibinfo {year} {2017})},\ \Eprint {http://arxiv.org/abs/1607.01976}
  {arXiv:1607.01976 [astro-ph.HE]} \BibitemShut {NoStop}%
\bibitem [{\citenamefont {Génolini}\ \emph {et~al.}(2017)\citenamefont
  {Génolini} \emph {et~al.}}]{Genolini:2017dfb}%
  \BibitemOpen
  \bibfield  {author} {\bibinfo {author} {\bibfnamefont {Y.}~\bibnamefont
  {Génolini}} \emph {et~al.},\ }\href {\doibase
  10.1103/PhysRevLett.119.241101} {\bibfield  {journal} {\bibinfo  {journal}
  {Phys. Rev. Lett.}\ }\textbf {\bibinfo {volume} {119}},\ \bibinfo {pages}
  {241101} (\bibinfo {year} {2017})},\ \Eprint
  {http://arxiv.org/abs/1706.09812} {arXiv:1706.09812 [astro-ph.HE]}
  \BibitemShut {NoStop}%
\bibitem [{\citenamefont {Korsmeier}\ \emph {et~al.}(2017)\citenamefont
  {Korsmeier}, \citenamefont {Donato},\ and\ \citenamefont
  {Fornengo}}]{Korsmeier:2017xzj}%
  \BibitemOpen
  \bibfield  {author} {\bibinfo {author} {\bibfnamefont {M.}~\bibnamefont
  {Korsmeier}}, \bibinfo {author} {\bibfnamefont {F.}~\bibnamefont {Donato}}, \
  and\ \bibinfo {author} {\bibfnamefont {N.}~\bibnamefont {Fornengo}},\
  }\href@noop {} {\  (\bibinfo {year} {2017})},\ \Eprint
  {http://arxiv.org/abs/1711.08465} {arXiv:1711.08465 [astro-ph.HE]}
  \BibitemShut {NoStop}%
\bibitem [{\citenamefont {Lemaire}\ \emph {et~al.}(1979)\citenamefont
  {Lemaire}, \citenamefont {Nagamiya}, \citenamefont {Schnetzer}, \citenamefont
  {Steiner},\ and\ \citenamefont {Tanihata}}]{Lemaire:1980qw}%
  \BibitemOpen
  \bibfield  {author} {\bibinfo {author} {\bibfnamefont {M.~C.}\ \bibnamefont
  {Lemaire}}, \bibinfo {author} {\bibfnamefont {S.}~\bibnamefont {Nagamiya}},
  \bibinfo {author} {\bibfnamefont {S.}~\bibnamefont {Schnetzer}}, \bibinfo
  {author} {\bibfnamefont {H.}~\bibnamefont {Steiner}}, \ and\ \bibinfo
  {author} {\bibfnamefont {I.}~\bibnamefont {Tanihata}},\ }\href {\doibase
  10.1016/0370-2693(79)90772-X} {\bibfield  {journal} {\bibinfo  {journal}
  {Phys. Lett.}\ }\textbf {\bibinfo {volume} {85B}},\ \bibinfo {pages} {38}
  (\bibinfo {year} {1979})}\BibitemShut {NoStop}%
\bibitem [{\citenamefont {Gomez-Coral}\ \emph {et~al.}(2018)\citenamefont
  {Gomez-Coral}, \citenamefont {Rocha}, \citenamefont {Grabski}, \citenamefont
  {Datta}, \citenamefont {von Doetinchem},\ and\ \citenamefont
  {Shukla}}]{Gomez-Coral:2018yuk}%
  \BibitemOpen
  \bibfield  {author} {\bibinfo {author} {\bibfnamefont {D.-M.}\ \bibnamefont
  {Gomez-Coral}}, \bibinfo {author} {\bibfnamefont {A.~M.}\ \bibnamefont
  {Rocha}}, \bibinfo {author} {\bibfnamefont {V.}~\bibnamefont {Grabski}},
  \bibinfo {author} {\bibfnamefont {A.}~\bibnamefont {Datta}}, \bibinfo
  {author} {\bibfnamefont {P.}~\bibnamefont {von Doetinchem}}, \ and\ \bibinfo
  {author} {\bibfnamefont {A.}~\bibnamefont {Shukla}},\ }\href@noop {} {\
  (\bibinfo {year} {2018})},\ \Eprint {http://arxiv.org/abs/1806.09303}
  {arXiv:1806.09303 [astro-ph.HE]} \BibitemShut {NoStop}%
\bibitem [{\citenamefont {Adriani}\ \emph {et~al.}(2016)\citenamefont {Adriani}
  \emph {et~al.}}]{Adriani:2015aps}%
  \BibitemOpen
  \bibfield  {author} {\bibinfo {author} {\bibfnamefont {O.}~\bibnamefont
  {Adriani}} \emph {et~al.} (\bibinfo {collaboration} {PAMELA}),\ }\href
  {\doibase 10.3847/0004-637X/818/1/68} {\bibfield  {journal} {\bibinfo
  {journal} {Astrophys. J.}\ }\textbf {\bibinfo {volume} {818}},\ \bibinfo
  {pages} {68} (\bibinfo {year} {2016})},\ \Eprint
  {http://arxiv.org/abs/1512.06535} {arXiv:1512.06535 [astro-ph.HE]}
  \BibitemShut {NoStop}%
\bibitem [{\citenamefont {Ade}\ \emph {et~al.}(2016)\citenamefont {Ade} \emph
  {et~al.}}]{Ade:2015xua}%
  \BibitemOpen
  \bibfield  {author} {\bibinfo {author} {\bibfnamefont {P.~A.~R.}\
  \bibnamefont {Ade}} \emph {et~al.} (\bibinfo {collaboration} {Planck}),\
  }\href {\doibase 10.1051/0004-6361/201525830} {\bibfield  {journal} {\bibinfo
   {journal} {Astron. Astrophys.}\ }\textbf {\bibinfo {volume} {594}},\
  \bibinfo {pages} {A13} (\bibinfo {year} {2016})},\ \Eprint
  {http://arxiv.org/abs/1502.01589} {arXiv:1502.01589 [astro-ph.CO]}
  \BibitemShut {NoStop}%
\bibitem [{\citenamefont {Arbey}(2012)}]{Arbey:2011nf}%
  \BibitemOpen
  \bibfield  {author} {\bibinfo {author} {\bibfnamefont {A.}~\bibnamefont
  {Arbey}},\ }\href {\doibase 10.1016/j.cpc.2012.03.018} {\bibfield  {journal}
  {\bibinfo  {journal} {Comput. Phys. Commun.}\ }\textbf {\bibinfo {volume}
  {183}},\ \bibinfo {pages} {1822} (\bibinfo {year} {2012})},\ \Eprint
  {http://arxiv.org/abs/1106.1363} {arXiv:1106.1363 [astro-ph.CO]} \BibitemShut
  {NoStop}%
\bibitem [{\citenamefont {Pitrou}\ \emph {et~al.}(2018)\citenamefont {Pitrou},
  \citenamefont {Coc}, \citenamefont {Uzan},\ and\ \citenamefont
  {Vangioni}}]{Pitrou:2018cgg}%
  \BibitemOpen
  \bibfield  {author} {\bibinfo {author} {\bibfnamefont {C.}~\bibnamefont
  {Pitrou}}, \bibinfo {author} {\bibfnamefont {A.}~\bibnamefont {Coc}},
  \bibinfo {author} {\bibfnamefont {J.-P.}\ \bibnamefont {Uzan}}, \ and\
  \bibinfo {author} {\bibfnamefont {E.}~\bibnamefont {Vangioni}},\ }\href
  {\doibase 10.1016/j.physrep.2018.04.005} {\bibfield  {journal} {\bibinfo
  {journal} {Phys. Rept.}\ }\textbf {\bibinfo {volume} {04}},\ \bibinfo {pages}
  {005} (\bibinfo {year} {2018})},\ \Eprint {http://arxiv.org/abs/1801.08023}
  {arXiv:1801.08023 [astro-ph.CO]} \BibitemShut {NoStop}%
\bibitem [{\citenamefont {Padmanabhan}\ and\ \citenamefont
  {Finkbeiner}(2005)}]{Padman05}%
  \BibitemOpen
  \bibfield  {author} {\bibinfo {author} {\bibfnamefont {N.}~\bibnamefont
  {Padmanabhan}}\ and\ \bibinfo {author} {\bibfnamefont {D.~P.}\ \bibnamefont
  {Finkbeiner}},\ }\href {\doibase 10.1103/PhysRevD.72.023508} {\bibfield
  {journal} {\bibinfo  {journal} {Phys.Rev.}\ }\textbf {\bibinfo {volume}
  {D72}},\ \bibinfo {pages} {023508} (\bibinfo {year} {2005})},\ \Eprint
  {http://arxiv.org/abs/astro-ph/0503486} {arXiv:astro-ph/0503486 [astro-ph]}
  \BibitemShut {NoStop}%
\bibitem [{\citenamefont {Belikov}\ and\ \citenamefont
  {Hooper}(2009)}]{Hooper09}%
  \BibitemOpen
  \bibfield  {author} {\bibinfo {author} {\bibfnamefont {A.~V.}\ \bibnamefont
  {Belikov}}\ and\ \bibinfo {author} {\bibfnamefont {D.}~\bibnamefont
  {Hooper}},\ }\href {\doibase 10.1103/PhysRevD.80.035007} {\bibfield
  {journal} {\bibinfo  {journal} {Phys.Rev.}\ }\textbf {\bibinfo {volume}
  {D80}},\ \bibinfo {pages} {035007} (\bibinfo {year} {2009})},\ \Eprint
  {http://arxiv.org/abs/0904.1210} {arXiv:0904.1210 [hep-ph]} \BibitemShut
  {NoStop}%
\bibitem [{\citenamefont {Cirelli}\ \emph
  {et~al.}(2009{\natexlab{b}})\citenamefont {Cirelli}, \citenamefont {Iocco},\
  and\ \citenamefont {Panci}}]{Cirelli09}%
  \BibitemOpen
  \bibfield  {author} {\bibinfo {author} {\bibfnamefont {M.}~\bibnamefont
  {Cirelli}}, \bibinfo {author} {\bibfnamefont {F.}~\bibnamefont {Iocco}}, \
  and\ \bibinfo {author} {\bibfnamefont {P.}~\bibnamefont {Panci}},\ }\href
  {\doibase 10.1088/1475-7516/2009/10/009} {\bibfield  {journal} {\bibinfo
  {journal} {JCAP}\ }\textbf {\bibinfo {volume} {0910}},\ \bibinfo {pages}
  {009} (\bibinfo {year} {2009}{\natexlab{b}})},\ \Eprint
  {http://arxiv.org/abs/0907.0719} {arXiv:0907.0719 [astro-ph.CO]} \BibitemShut
  {NoStop}%
\bibitem [{\citenamefont {Huetsi}\ \emph {et~al.}(2009)\citenamefont {Huetsi},
  \citenamefont {Hektor},\ and\ \citenamefont {Raidal}}]{Huetsi:2009ex}%
  \BibitemOpen
  \bibfield  {author} {\bibinfo {author} {\bibfnamefont {G.}~\bibnamefont
  {Huetsi}}, \bibinfo {author} {\bibfnamefont {A.}~\bibnamefont {Hektor}}, \
  and\ \bibinfo {author} {\bibfnamefont {M.}~\bibnamefont {Raidal}},\ }\href
  {\doibase 10.1051/0004-6361/200912760} {\bibfield  {journal} {\bibinfo
  {journal} {Astron. Astrophys.}\ }\textbf {\bibinfo {volume} {505}},\ \bibinfo
  {pages} {999} (\bibinfo {year} {2009})},\ \Eprint
  {http://arxiv.org/abs/0906.4550} {arXiv:0906.4550 [astro-ph.CO]} \BibitemShut
  {NoStop}%
\bibitem [{\citenamefont {Slatyer}\ \emph {et~al.}(2009)\citenamefont
  {Slatyer}, \citenamefont {Padmanabhan},\ and\ \citenamefont
  {Finkbeiner}}]{Slatyer09}%
  \BibitemOpen
  \bibfield  {author} {\bibinfo {author} {\bibfnamefont {T.~R.}\ \bibnamefont
  {Slatyer}}, \bibinfo {author} {\bibfnamefont {N.}~\bibnamefont
  {Padmanabhan}}, \ and\ \bibinfo {author} {\bibfnamefont {D.~P.}\ \bibnamefont
  {Finkbeiner}},\ }\href {\doibase 10.1103/PhysRevD.80.043526} {\bibfield
  {journal} {\bibinfo  {journal} {Phys.Rev.}\ }\textbf {\bibinfo {volume}
  {D80}},\ \bibinfo {pages} {043526} (\bibinfo {year} {2009})},\ \Eprint
  {http://arxiv.org/abs/0906.1197} {arXiv:0906.1197 [astro-ph.CO]} \BibitemShut
  {NoStop}%
\bibitem [{\citenamefont {Natarajan}\ and\ \citenamefont
  {Schwarz}(2008)}]{Natarajan08}%
  \BibitemOpen
  \bibfield  {author} {\bibinfo {author} {\bibfnamefont {A.}~\bibnamefont
  {Natarajan}}\ and\ \bibinfo {author} {\bibfnamefont {D.~J.}\ \bibnamefont
  {Schwarz}},\ }\href {\doibase 10.1103/PhysRevD.78.103524,
  10.1103/PhysRevD.81.089905} {\bibfield  {journal} {\bibinfo  {journal}
  {Phys.Rev.}\ }\textbf {\bibinfo {volume} {D78}},\ \bibinfo {pages} {103524}
  (\bibinfo {year} {2008})},\ \Eprint {http://arxiv.org/abs/0805.3945}
  {arXiv:0805.3945 [astro-ph]} \BibitemShut {NoStop}%
\bibitem [{\citenamefont {Natarajan}\ and\ \citenamefont
  {Schwarz}(2009)}]{Natarajan09}%
  \BibitemOpen
  \bibfield  {author} {\bibinfo {author} {\bibfnamefont {A.}~\bibnamefont
  {Natarajan}}\ and\ \bibinfo {author} {\bibfnamefont {D.~J.}\ \bibnamefont
  {Schwarz}},\ }\href {\doibase 10.1103/PhysRevD.80.043529} {\bibfield
  {journal} {\bibinfo  {journal} {Phys.Rev.}\ }\textbf {\bibinfo {volume}
  {D80}},\ \bibinfo {pages} {043529} (\bibinfo {year} {2009})},\ \Eprint
  {http://arxiv.org/abs/0903.4485} {arXiv:0903.4485 [astro-ph.CO]} \BibitemShut
  {NoStop}%
\bibitem [{\citenamefont {Natarajan}\ and\ \citenamefont
  {Schwarz}(2010)}]{Natarajan10}%
  \BibitemOpen
  \bibfield  {author} {\bibinfo {author} {\bibfnamefont {A.}~\bibnamefont
  {Natarajan}}\ and\ \bibinfo {author} {\bibfnamefont {D.~J.}\ \bibnamefont
  {Schwarz}},\ }\href {\doibase 10.1103/PhysRevD.81.123510} {\bibfield
  {journal} {\bibinfo  {journal} {Phys.Rev.}\ }\textbf {\bibinfo {volume}
  {D81}},\ \bibinfo {pages} {123510} (\bibinfo {year} {2010})},\ \Eprint
  {http://arxiv.org/abs/1002.4405} {arXiv:1002.4405 [astro-ph.CO]} \BibitemShut
  {NoStop}%
\bibitem [{\citenamefont {Valdes}\ \emph {et~al.}(2010)\citenamefont {Valdes},
  \citenamefont {Evoli},\ and\ \citenamefont {Ferrara}}]{Valdes:2009cq}%
  \BibitemOpen
  \bibfield  {author} {\bibinfo {author} {\bibfnamefont {M.}~\bibnamefont
  {Valdes}}, \bibinfo {author} {\bibfnamefont {C.}~\bibnamefont {Evoli}}, \
  and\ \bibinfo {author} {\bibfnamefont {A.}~\bibnamefont {Ferrara}},\ }\href
  {\doibase 10.1111/j.1365-2966.2010.16387.x} {\bibfield  {journal} {\bibinfo
  {journal} {Mon. Not. Roy. Astron. Soc.}\ }\textbf {\bibinfo {volume} {404}},\
  \bibinfo {pages} {1569} (\bibinfo {year} {2010})},\ \Eprint
  {http://arxiv.org/abs/0911.1125} {arXiv:0911.1125 [astro-ph.CO]} \BibitemShut
  {NoStop}%
\bibitem [{\citenamefont {Evoli}\ \emph {et~al.}(2012)\citenamefont {Evoli},
  \citenamefont {Valdes}, \citenamefont {Ferrara},\ and\ \citenamefont
  {Yoshida}}]{Evoli:2012zz}%
  \BibitemOpen
  \bibfield  {author} {\bibinfo {author} {\bibfnamefont {C.}~\bibnamefont
  {Evoli}}, \bibinfo {author} {\bibfnamefont {M.}~\bibnamefont {Valdes}},
  \bibinfo {author} {\bibfnamefont {A.}~\bibnamefont {Ferrara}}, \ and\
  \bibinfo {author} {\bibfnamefont {N.}~\bibnamefont {Yoshida}},\ }\href
  {\doibase 10.1111/j.1365-2966.2012.20624.x} {\bibfield  {journal} {\bibinfo
  {journal} {Mon. Not. Roy. Astron. Soc.}\ }\textbf {\bibinfo {volume} {422}},\
  \bibinfo {pages} {420} (\bibinfo {year} {2012})}\BibitemShut {NoStop}%
\bibitem [{\citenamefont {Galli}\ \emph
  {et~al.}(2013{\natexlab{a}})\citenamefont {Galli}, \citenamefont {Slatyer},
  \citenamefont {Valdes},\ and\ \citenamefont {Iocco}}]{Galli13}%
  \BibitemOpen
  \bibfield  {author} {\bibinfo {author} {\bibfnamefont {S.}~\bibnamefont
  {Galli}}, \bibinfo {author} {\bibfnamefont {T.~R.}\ \bibnamefont {Slatyer}},
  \bibinfo {author} {\bibfnamefont {M.}~\bibnamefont {Valdes}}, \ and\ \bibinfo
  {author} {\bibfnamefont {F.}~\bibnamefont {Iocco}},\ }\href {\doibase
  10.1103/PhysRevD.88.063502} {\bibfield  {journal} {\bibinfo  {journal}
  {Phys.Rev.}\ }\textbf {\bibinfo {volume} {D88}},\ \bibinfo {pages} {063502}
  (\bibinfo {year} {2013}{\natexlab{a}})},\ \Eprint
  {http://arxiv.org/abs/1306.0563} {arXiv:1306.0563 [astro-ph.CO]} \BibitemShut
  {NoStop}%
\bibitem [{\citenamefont {Finkbeiner}\ \emph {et~al.}(2012)\citenamefont
  {Finkbeiner}, \citenamefont {Galli}, \citenamefont {Lin},\ and\ \citenamefont
  {Slatyer}}]{Finkbeiner11}%
  \BibitemOpen
  \bibfield  {author} {\bibinfo {author} {\bibfnamefont {D.~P.}\ \bibnamefont
  {Finkbeiner}}, \bibinfo {author} {\bibfnamefont {S.}~\bibnamefont {Galli}},
  \bibinfo {author} {\bibfnamefont {T.}~\bibnamefont {Lin}}, \ and\ \bibinfo
  {author} {\bibfnamefont {T.~R.}\ \bibnamefont {Slatyer}},\ }\href {\doibase
  10.1103/PhysRevD.85.043522} {\bibfield  {journal} {\bibinfo  {journal}
  {Phys.Rev.}\ }\textbf {\bibinfo {volume} {D85}},\ \bibinfo {pages} {043522}
  (\bibinfo {year} {2012})},\ \Eprint {http://arxiv.org/abs/1109.6322}
  {arXiv:1109.6322 [astro-ph.CO]} \BibitemShut {NoStop}%
\bibitem [{\citenamefont {Hutsi}\ \emph {et~al.}(2011)\citenamefont {Hutsi},
  \citenamefont {Chluba}, \citenamefont {Hektor},\ and\ \citenamefont
  {Raidal}}]{Hutsi:2011vx}%
  \BibitemOpen
  \bibfield  {author} {\bibinfo {author} {\bibfnamefont {G.}~\bibnamefont
  {Hutsi}}, \bibinfo {author} {\bibfnamefont {J.}~\bibnamefont {Chluba}},
  \bibinfo {author} {\bibfnamefont {A.}~\bibnamefont {Hektor}}, \ and\ \bibinfo
  {author} {\bibfnamefont {M.}~\bibnamefont {Raidal}},\ }\href {\doibase
  10.1051/0004-6361/201116914} {\bibfield  {journal} {\bibinfo  {journal}
  {Astron.Astrophys.}\ }\textbf {\bibinfo {volume} {535}},\ \bibinfo {pages}
  {A26} (\bibinfo {year} {2011})},\ \Eprint {http://arxiv.org/abs/1103.2766}
  {arXiv:1103.2766 [astro-ph.CO]} \BibitemShut {NoStop}%
\bibitem [{\citenamefont {Slatyer}(2013)}]{Slatyer12}%
  \BibitemOpen
  \bibfield  {author} {\bibinfo {author} {\bibfnamefont {T.~R.}\ \bibnamefont
  {Slatyer}},\ }\href {\doibase 10.1103/PhysRevD.87.123513} {\bibfield
  {journal} {\bibinfo  {journal} {Phys.Rev.}\ }\textbf {\bibinfo {volume}
  {D87}},\ \bibinfo {pages} {123513} (\bibinfo {year} {2013})},\ \Eprint
  {http://arxiv.org/abs/1211.0283} {arXiv:1211.0283 [astro-ph.CO]} \BibitemShut
  {NoStop}%
\bibitem [{\citenamefont {Giesen}\ \emph {et~al.}(2012)\citenamefont {Giesen},
  \citenamefont {Lesgourgues}, \citenamefont {Audren},\ and\ \citenamefont
  {Ali-Ha{\"i}moud}}]{Giesen}%
  \BibitemOpen
  \bibfield  {author} {\bibinfo {author} {\bibfnamefont {G.}~\bibnamefont
  {Giesen}}, \bibinfo {author} {\bibfnamefont {J.}~\bibnamefont {Lesgourgues}},
  \bibinfo {author} {\bibfnamefont {B.}~\bibnamefont {Audren}}, \ and\ \bibinfo
  {author} {\bibfnamefont {Y.}~\bibnamefont {Ali-Ha{\"i}moud}},\ }\href
  {\doibase 10.1088/1475-7516/2012/12/008} {\bibfield  {journal} {\bibinfo
  {journal} {JCAP}\ }\textbf {\bibinfo {volume} {1212}},\ \bibinfo {pages}
  {008} (\bibinfo {year} {2012})},\ \Eprint {http://arxiv.org/abs/1209.0247}
  {arXiv:1209.0247 [astro-ph.CO]} \BibitemShut {NoStop}%
\bibitem [{\citenamefont {Galli}\ \emph
  {et~al.}(2013{\natexlab{b}})\citenamefont {Galli}, \citenamefont {Slatyer},
  \citenamefont {Valdes},\ and\ \citenamefont {Iocco}}]{Slatyer13}%
  \BibitemOpen
  \bibfield  {author} {\bibinfo {author} {\bibfnamefont {S.}~\bibnamefont
  {Galli}}, \bibinfo {author} {\bibfnamefont {T.~R.}\ \bibnamefont {Slatyer}},
  \bibinfo {author} {\bibfnamefont {M.}~\bibnamefont {Valdes}}, \ and\ \bibinfo
  {author} {\bibfnamefont {F.}~\bibnamefont {Iocco}},\ }\href {\doibase
  10.1103/PhysRevD.88.063502} {\bibfield  {journal} {\bibinfo  {journal}
  {Phys.Rev.}\ }\textbf {\bibinfo {volume} {D88}},\ \bibinfo {pages} {063502}
  (\bibinfo {year} {2013}{\natexlab{b}})},\ \Eprint
  {http://arxiv.org/abs/1306.0563} {arXiv:1306.0563 [astro-ph.CO]} \BibitemShut
  {NoStop}%
\bibitem [{\citenamefont {Slatyer}(2015)}]{Slatyer15-1}%
  \BibitemOpen
  \bibfield  {author} {\bibinfo {author} {\bibfnamefont {T.~R.}\ \bibnamefont
  {Slatyer}},\ }\href@noop {} {\  (\bibinfo {year} {2015})},\ \Eprint
  {http://arxiv.org/abs/1506.03811} {arXiv:1506.03811 [hep-ph]} \BibitemShut
  {NoStop}%
\bibitem [{\citenamefont {Lopez-Honorez}\ \emph {et~al.}(2013)\citenamefont
  {Lopez-Honorez}, \citenamefont {Mena}, \citenamefont {Palomares-Ruiz},\ and\
  \citenamefont {Vincent}}]{Lopez-Honorez:2013lcm}%
  \BibitemOpen
  \bibfield  {author} {\bibinfo {author} {\bibfnamefont {L.}~\bibnamefont
  {Lopez-Honorez}}, \bibinfo {author} {\bibfnamefont {O.}~\bibnamefont {Mena}},
  \bibinfo {author} {\bibfnamefont {S.}~\bibnamefont {Palomares-Ruiz}}, \ and\
  \bibinfo {author} {\bibfnamefont {A.~C.}\ \bibnamefont {Vincent}},\ }\href
  {\doibase 10.1088/1475-7516/2013/07/046} {\bibfield  {journal} {\bibinfo
  {journal} {JCAP}\ }\textbf {\bibinfo {volume} {1307}},\ \bibinfo {pages}
  {046} (\bibinfo {year} {2013})},\ \Eprint {http://arxiv.org/abs/1303.5094}
  {arXiv:1303.5094 [astro-ph.CO]} \BibitemShut {NoStop}%
\bibitem [{\citenamefont {Poulin}\ \emph {et~al.}(2015)\citenamefont {Poulin},
  \citenamefont {Serpico},\ and\ \citenamefont {Lesgourgues}}]{Poulin:2015pna}%
  \BibitemOpen
  \bibfield  {author} {\bibinfo {author} {\bibfnamefont {V.}~\bibnamefont
  {Poulin}}, \bibinfo {author} {\bibfnamefont {P.~D.}\ \bibnamefont {Serpico}},
  \ and\ \bibinfo {author} {\bibfnamefont {J.}~\bibnamefont {Lesgourgues}},\
  }\href {\doibase 10.1088/1475-7516/2015/12/041} {\bibfield  {journal}
  {\bibinfo  {journal} {JCAP}\ }\textbf {\bibinfo {volume} {1512}},\ \bibinfo
  {pages} {041} (\bibinfo {year} {2015})},\ \Eprint
  {http://arxiv.org/abs/1508.01370} {arXiv:1508.01370 [astro-ph.CO]}
  \BibitemShut {NoStop}%
\bibitem [{\citenamefont {Liu}\ \emph {et~al.}(2016)\citenamefont {Liu},
  \citenamefont {Slatyer},\ and\ \citenamefont {Zavala}}]{Liu:2016cnk}%
  \BibitemOpen
  \bibfield  {author} {\bibinfo {author} {\bibfnamefont {H.}~\bibnamefont
  {Liu}}, \bibinfo {author} {\bibfnamefont {T.~R.}\ \bibnamefont {Slatyer}}, \
  and\ \bibinfo {author} {\bibfnamefont {J.}~\bibnamefont {Zavala}},\ }\href
  {\doibase 10.1103/PhysRevD.94.063507} {\bibfield  {journal} {\bibinfo
  {journal} {Phys. Rev.}\ }\textbf {\bibinfo {volume} {D94}},\ \bibinfo {pages}
  {063507} (\bibinfo {year} {2016})},\ \Eprint
  {http://arxiv.org/abs/1604.02457} {arXiv:1604.02457 [astro-ph.CO]}
  \BibitemShut {NoStop}%
\bibitem [{\citenamefont {St{\"o}cker}\ \emph {et~al.}(2018)\citenamefont
  {St{\"o}cker}, \citenamefont {Kr{\"a}mer}, \citenamefont {Lesgourgues},\ and\
  \citenamefont {Poulin}}]{Stocker:2018avm}%
  \BibitemOpen
  \bibfield  {author} {\bibinfo {author} {\bibfnamefont {P.}~\bibnamefont
  {St{\"o}cker}}, \bibinfo {author} {\bibfnamefont {M.}~\bibnamefont
  {Kr{\"a}mer}}, \bibinfo {author} {\bibfnamefont {J.}~\bibnamefont
  {Lesgourgues}}, \ and\ \bibinfo {author} {\bibfnamefont {V.}~\bibnamefont
  {Poulin}},\ }\href {\doibase 10.1088/1475-7516/2018/03/018} {\bibfield
  {journal} {\bibinfo  {journal} {JCAP}\ }\textbf {\bibinfo {volume} {1803}},\
  \bibinfo {pages} {018} (\bibinfo {year} {2018})},\ \Eprint
  {http://arxiv.org/abs/1801.01871} {arXiv:1801.01871 [astro-ph.CO]}
  \BibitemShut {NoStop}%
\bibitem [{\citenamefont {Chluba}(2013)}]{Chluba:2013wsa}%
  \BibitemOpen
  \bibfield  {author} {\bibinfo {author} {\bibfnamefont {J.}~\bibnamefont
  {Chluba}},\ }\href {\doibase 10.1093/mnras/stt1733} {\bibfield  {journal}
  {\bibinfo  {journal} {Mon. Not. Roy. Astron. Soc.}\ }\textbf {\bibinfo
  {volume} {436}},\ \bibinfo {pages} {2232} (\bibinfo {year} {2013})},\ \Eprint
  {http://arxiv.org/abs/1304.6121} {arXiv:1304.6121 [astro-ph.CO]} \BibitemShut
  {NoStop}%
\bibitem [{\citenamefont {Aghanim}\ \emph {et~al.}(2018)\citenamefont {Aghanim}
  \emph {et~al.}}]{Aghanim:2018eyx}%
  \BibitemOpen
  \bibfield  {author} {\bibinfo {author} {\bibfnamefont {N.}~\bibnamefont
  {Aghanim}} \emph {et~al.} (\bibinfo {collaboration} {Planck}),\ }\href@noop
  {} {\  (\bibinfo {year} {2018})},\ \Eprint {http://arxiv.org/abs/1807.06209}
  {arXiv:1807.06209 [astro-ph.CO]} \BibitemShut {NoStop}%
\bibitem [{\citenamefont {Slatyer}\ and\ \citenamefont
  {Wu}(2016)}]{Slatyer:2016qyl}%
  \BibitemOpen
  \bibfield  {author} {\bibinfo {author} {\bibfnamefont {T.~R.}\ \bibnamefont
  {Slatyer}}\ and\ \bibinfo {author} {\bibfnamefont {C.-L.}\ \bibnamefont
  {Wu}},\ }\href@noop {} {\  (\bibinfo {year} {2016})},\ \Eprint
  {http://arxiv.org/abs/1610.06933} {arXiv:1610.06933 [astro-ph.CO]}
  \BibitemShut {NoStop}%
\bibitem [{\citenamefont {Poulin}\ \emph {et~al.}(2017)\citenamefont {Poulin},
  \citenamefont {Lesgourgues},\ and\ \citenamefont {Serpico}}]{Poulin2016}%
  \BibitemOpen
  \bibfield  {author} {\bibinfo {author} {\bibfnamefont {V.}~\bibnamefont
  {Poulin}}, \bibinfo {author} {\bibfnamefont {J.}~\bibnamefont {Lesgourgues}},
  \ and\ \bibinfo {author} {\bibfnamefont {P.~D.}\ \bibnamefont {Serpico}},\
  }\href {\doibase 10.1088/1475-7516/2017/03/043} {\bibfield  {journal}
  {\bibinfo  {journal} {JCAP}\ }\textbf {\bibinfo {volume} {1703}},\ \bibinfo
  {pages} {043} (\bibinfo {year} {2017})},\ \Eprint
  {http://arxiv.org/abs/1610.10051} {arXiv:1610.10051 [astro-ph.CO]}
  \BibitemShut {NoStop}%
\bibitem [{\citenamefont {Weniger}(2012)}]{Weniger:2012tx}%
  \BibitemOpen
  \bibfield  {author} {\bibinfo {author} {\bibfnamefont {C.}~\bibnamefont
  {Weniger}},\ }\href {\doibase 10.1088/1475-7516/2012/08/007} {\bibfield
  {journal} {\bibinfo  {journal} {JCAP}\ }\textbf {\bibinfo {volume} {1208}},\
  \bibinfo {pages} {007} (\bibinfo {year} {2012})},\ \Eprint
  {http://arxiv.org/abs/1204.2797} {arXiv:1204.2797 [hep-ph]} \BibitemShut
  {NoStop}%
\bibitem [{\citenamefont {Bringmann}\ and\ \citenamefont
  {Weniger}(2012)}]{Bringmann:2012ez}%
  \BibitemOpen
  \bibfield  {author} {\bibinfo {author} {\bibfnamefont {T.}~\bibnamefont
  {Bringmann}}\ and\ \bibinfo {author} {\bibfnamefont {C.}~\bibnamefont
  {Weniger}},\ }\href {\doibase 10.1016/j.dark.2012.10.005} {\bibfield
  {journal} {\bibinfo  {journal} {Phys. Dark Univ.}\ }\textbf {\bibinfo
  {volume} {1}},\ \bibinfo {pages} {194} (\bibinfo {year} {2012})},\ \Eprint
  {http://arxiv.org/abs/1208.5481} {arXiv:1208.5481 [hep-ph]} \BibitemShut
  {NoStop}%
\bibitem [{\citenamefont {Ackermann}\ \emph {et~al.}(2013)\citenamefont
  {Ackermann} \emph {et~al.}}]{Ackermann:2013uma}%
  \BibitemOpen
  \bibfield  {author} {\bibinfo {author} {\bibfnamefont {M.}~\bibnamefont
  {Ackermann}} \emph {et~al.} (\bibinfo {collaboration} {Fermi-LAT}),\ }\href
  {\doibase 10.1103/PhysRevD.88.082002} {\bibfield  {journal} {\bibinfo
  {journal} {Phys. Rev.}\ }\textbf {\bibinfo {volume} {D88}},\ \bibinfo {pages}
  {082002} (\bibinfo {year} {2013})},\ \Eprint {http://arxiv.org/abs/1305.5597}
  {arXiv:1305.5597 [astro-ph.HE]} \BibitemShut {NoStop}%
\bibitem [{\citenamefont {Ackermann}\ \emph
  {et~al.}(2015{\natexlab{a}})\citenamefont {Ackermann} \emph
  {et~al.}}]{Ackermann:2015lka}%
  \BibitemOpen
  \bibfield  {author} {\bibinfo {author} {\bibfnamefont {M.}~\bibnamefont
  {Ackermann}} \emph {et~al.} (\bibinfo {collaboration} {Fermi-LAT}),\ }\href
  {\doibase 10.1103/PhysRevD.91.122002} {\bibfield  {journal} {\bibinfo
  {journal} {Phys. Rev.}\ }\textbf {\bibinfo {volume} {D91}},\ \bibinfo {pages}
  {122002} (\bibinfo {year} {2015}{\natexlab{a}})},\ \Eprint
  {http://arxiv.org/abs/1506.00013} {arXiv:1506.00013 [astro-ph.HE]}
  \BibitemShut {NoStop}%
\bibitem [{\citenamefont {Abdo}\ \emph
  {et~al.}(2010{\natexlab{a}})\citenamefont {Abdo} \emph
  {et~al.}}]{Abdo:2010ex}%
  \BibitemOpen
  \bibfield  {author} {\bibinfo {author} {\bibfnamefont {A.~A.}\ \bibnamefont
  {Abdo}} \emph {et~al.} (\bibinfo {collaboration} {Fermi-LAT}),\ }\href
  {\doibase 10.1088/0004-637X/712/1/147} {\bibfield  {journal} {\bibinfo
  {journal} {Astrophys. J.}\ }\textbf {\bibinfo {volume} {712}},\ \bibinfo
  {pages} {147} (\bibinfo {year} {2010}{\natexlab{a}})},\ \Eprint
  {http://arxiv.org/abs/1001.4531} {arXiv:1001.4531 [astro-ph.CO]} \BibitemShut
  {NoStop}%
\bibitem [{\citenamefont {Aleksic}\ \emph {et~al.}(2011)\citenamefont {Aleksic}
  \emph {et~al.}}]{Aleksic:2011jx}%
  \BibitemOpen
  \bibfield  {author} {\bibinfo {author} {\bibfnamefont {J.}~\bibnamefont
  {Aleksic}} \emph {et~al.} (\bibinfo {collaboration} {MAGIC}),\ }\href
  {\doibase 10.1088/1475-7516/2011/06/035} {\bibfield  {journal} {\bibinfo
  {journal} {JCAP}\ }\textbf {\bibinfo {volume} {1106}},\ \bibinfo {pages}
  {035} (\bibinfo {year} {2011})},\ \Eprint {http://arxiv.org/abs/1103.0477}
  {arXiv:1103.0477 [astro-ph.HE]} \BibitemShut {NoStop}%
\bibitem [{\citenamefont {Geringer-Sameth}\ and\ \citenamefont
  {Koushiappas}(2011)}]{GeringerSameth:2011iw}%
  \BibitemOpen
  \bibfield  {author} {\bibinfo {author} {\bibfnamefont {A.}~\bibnamefont
  {Geringer-Sameth}}\ and\ \bibinfo {author} {\bibfnamefont {S.~M.}\
  \bibnamefont {Koushiappas}},\ }\href {\doibase
  10.1103/PhysRevLett.107.241303} {\bibfield  {journal} {\bibinfo  {journal}
  {Phys. Rev. Lett.}\ }\textbf {\bibinfo {volume} {107}},\ \bibinfo {pages}
  {241303} (\bibinfo {year} {2011})},\ \Eprint {http://arxiv.org/abs/1108.2914}
  {arXiv:1108.2914 [astro-ph.CO]} \BibitemShut {NoStop}%
\bibitem [{\citenamefont {Aliu}\ \emph {et~al.}(2012)\citenamefont {Aliu} \emph
  {et~al.}}]{Aliu:2012ga}%
  \BibitemOpen
  \bibfield  {author} {\bibinfo {author} {\bibfnamefont {E.}~\bibnamefont
  {Aliu}} \emph {et~al.} (\bibinfo {collaboration} {VERITAS}),\ }\href
  {\doibase 10.1103/PhysRevD.85.062001, 10.1103/PhysRevD.91.129903} {\bibfield
  {journal} {\bibinfo  {journal} {Phys. Rev.}\ }\textbf {\bibinfo {volume}
  {D85}},\ \bibinfo {pages} {062001} (\bibinfo {year} {2012})},\ \bibinfo
  {note} {[Erratum: Phys. Rev.D91,no.12,129903(2015)]},\ \Eprint
  {http://arxiv.org/abs/1202.2144} {arXiv:1202.2144 [astro-ph.HE]} \BibitemShut
  {NoStop}%
\bibitem [{\citenamefont {Cholis}\ and\ \citenamefont
  {Salucci}(2012)}]{Cholis:2012am}%
  \BibitemOpen
  \bibfield  {author} {\bibinfo {author} {\bibfnamefont {I.}~\bibnamefont
  {Cholis}}\ and\ \bibinfo {author} {\bibfnamefont {P.}~\bibnamefont
  {Salucci}},\ }\href {\doibase 10.1103/PhysRevD.86.023528} {\bibfield
  {journal} {\bibinfo  {journal} {Phys. Rev.}\ }\textbf {\bibinfo {volume}
  {D86}},\ \bibinfo {pages} {023528} (\bibinfo {year} {2012})},\ \Eprint
  {http://arxiv.org/abs/1203.2954} {arXiv:1203.2954 [astro-ph.HE]} \BibitemShut
  {NoStop}%
\bibitem [{\citenamefont {Ackermann}\ \emph {et~al.}(2014)\citenamefont
  {Ackermann} \emph {et~al.}}]{Ackermann:2013yva}%
  \BibitemOpen
  \bibfield  {author} {\bibinfo {author} {\bibfnamefont {M.}~\bibnamefont
  {Ackermann}} \emph {et~al.} (\bibinfo {collaboration} {Fermi-LAT}),\ }\href
  {\doibase 10.1103/PhysRevD.89.042001} {\bibfield  {journal} {\bibinfo
  {journal} {Phys. Rev.}\ }\textbf {\bibinfo {volume} {D89}},\ \bibinfo {pages}
  {042001} (\bibinfo {year} {2014})},\ \Eprint {http://arxiv.org/abs/1310.0828}
  {arXiv:1310.0828 [astro-ph.HE]} \BibitemShut {NoStop}%
\bibitem [{\citenamefont {Albert}\ \emph {et~al.}(2017)\citenamefont {Albert}
  \emph {et~al.}}]{Fermi-LAT:2016uux}%
  \BibitemOpen
  \bibfield  {author} {\bibinfo {author} {\bibfnamefont {A.}~\bibnamefont
  {Albert}} \emph {et~al.} (\bibinfo {collaboration} {DES, Fermi-LAT}),\ }\href
  {\doibase 10.3847/1538-4357/834/2/110} {\bibfield  {journal} {\bibinfo
  {journal} {Astrophys. J.}\ }\textbf {\bibinfo {volume} {834}},\ \bibinfo
  {pages} {110} (\bibinfo {year} {2017})},\ \Eprint
  {http://arxiv.org/abs/1611.03184} {arXiv:1611.03184 [astro-ph.HE]}
  \BibitemShut {NoStop}%
\bibitem [{\citenamefont {Abdo}\ \emph
  {et~al.}(2010{\natexlab{b}})\citenamefont {Abdo} \emph
  {et~al.}}]{Abdo:2010nz}%
  \BibitemOpen
  \bibfield  {author} {\bibinfo {author} {\bibfnamefont {A.~A.}\ \bibnamefont
  {Abdo}} \emph {et~al.} (\bibinfo {collaboration} {Fermi-LAT}),\ }\href
  {\doibase 10.1103/PhysRevLett.104.101101} {\bibfield  {journal} {\bibinfo
  {journal} {Phys. Rev. Lett.}\ }\textbf {\bibinfo {volume} {104}},\ \bibinfo
  {pages} {101101} (\bibinfo {year} {2010}{\natexlab{b}})},\ \Eprint
  {http://arxiv.org/abs/1002.3603} {arXiv:1002.3603 [astro-ph.HE]} \BibitemShut
  {NoStop}%
\bibitem [{\citenamefont {Ackermann}\ \emph
  {et~al.}(2015{\natexlab{b}})\citenamefont {Ackermann} \emph
  {et~al.}}]{Ackermann:2014usa}%
  \BibitemOpen
  \bibfield  {author} {\bibinfo {author} {\bibfnamefont {M.}~\bibnamefont
  {Ackermann}} \emph {et~al.} (\bibinfo {collaboration} {Fermi-LAT}),\ }\href
  {\doibase 10.1088/0004-637X/799/1/86} {\bibfield  {journal} {\bibinfo
  {journal} {Astrophys. J.}\ }\textbf {\bibinfo {volume} {799}},\ \bibinfo
  {pages} {86} (\bibinfo {year} {2015}{\natexlab{b}})},\ \Eprint
  {http://arxiv.org/abs/1410.3696} {arXiv:1410.3696 [astro-ph.HE]} \BibitemShut
  {NoStop}%
\bibitem [{\citenamefont {Amsler}(1998)}]{Amsler:1997up}%
  \BibitemOpen
  \bibfield  {author} {\bibinfo {author} {\bibfnamefont {C.}~\bibnamefont
  {Amsler}},\ }\href {\doibase 10.1103/RevModPhys.70.1293} {\bibfield
  {journal} {\bibinfo  {journal} {Rev. Mod. Phys.}\ }\textbf {\bibinfo {volume}
  {70}},\ \bibinfo {pages} {1293} (\bibinfo {year} {1998})},\ \Eprint
  {http://arxiv.org/abs/hep-ex/9708025} {arXiv:hep-ex/9708025 [hep-ex]}
  \BibitemShut {NoStop}%
\bibitem [{\citenamefont {Ferriere}(2001)}]{Ferriere:2001rg}%
  \BibitemOpen
  \bibfield  {author} {\bibinfo {author} {\bibfnamefont {K.~M.}\ \bibnamefont
  {Ferriere}},\ }\href {\doibase 10.1103/RevModPhys.73.1031} {\bibfield
  {journal} {\bibinfo  {journal} {Rev. Mod. Phys.}\ }\textbf {\bibinfo {volume}
  {73}},\ \bibinfo {pages} {1031} (\bibinfo {year} {2001})},\ \Eprint
  {http://arxiv.org/abs/astro-ph/0106359} {arXiv:astro-ph/0106359 [astro-ph]}
  \BibitemShut {NoStop}%
\bibitem [{\citenamefont {Adriani}\ \emph {et~al.}(2011)\citenamefont {Adriani}
  \emph {et~al.}}]{Adriani:2011cu}%
  \BibitemOpen
  \bibfield  {author} {\bibinfo {author} {\bibfnamefont {O.}~\bibnamefont
  {Adriani}} \emph {et~al.} (\bibinfo {collaboration} {PAMELA}),\ }\href
  {\doibase 10.1126/science.1199172} {\bibfield  {journal} {\bibinfo  {journal}
  {Science}\ }\textbf {\bibinfo {volume} {332}},\ \bibinfo {pages} {69}
  (\bibinfo {year} {2011})},\ \Eprint {http://arxiv.org/abs/1103.4055}
  {arXiv:1103.4055 [astro-ph.HE]} \BibitemShut {NoStop}%
\bibitem [{\citenamefont {Acero}\ \emph {et~al.}(2015)\citenamefont {Acero}
  \emph {et~al.}}]{Acero:2015hja}%
  \BibitemOpen
  \bibfield  {author} {\bibinfo {author} {\bibfnamefont {F.}~\bibnamefont
  {Acero}} \emph {et~al.} (\bibinfo {collaboration} {Fermi-LAT}),\ }\href
  {\doibase 10.1088/0067-0049/218/2/23} {\bibfield  {journal} {\bibinfo
  {journal} {Astrophys. J. Suppl.}\ }\textbf {\bibinfo {volume} {218}},\
  \bibinfo {pages} {23} (\bibinfo {year} {2015})},\ \Eprint
  {http://arxiv.org/abs/1501.02003} {arXiv:1501.02003 [astro-ph.HE]}
  \BibitemShut {NoStop}%
\bibitem [{\citenamefont {Knodlseder}\ \emph {et~al.}(2005)\citenamefont
  {Knodlseder} \emph {et~al.}}]{Knodlseder:2005yq}%
  \BibitemOpen
  \bibfield  {author} {\bibinfo {author} {\bibfnamefont {J.}~\bibnamefont
  {Knodlseder}} \emph {et~al.},\ }\href {\doibase 10.1051/0004-6361:20042063}
  {\bibfield  {journal} {\bibinfo  {journal} {Astron. Astrophys.}\ }\textbf
  {\bibinfo {volume} {441}},\ \bibinfo {pages} {513} (\bibinfo {year}
  {2005})},\ \Eprint {http://arxiv.org/abs/astro-ph/0506026}
  {arXiv:astro-ph/0506026 [astro-ph]} \BibitemShut {NoStop}%
\bibitem [{\citenamefont {Weidenspointner}\ \emph {et~al.}(2007)\citenamefont
  {Weidenspointner} \emph {et~al.}}]{Weidenspointner:2007rs}%
  \BibitemOpen
  \bibfield  {author} {\bibinfo {author} {\bibfnamefont {G.}~\bibnamefont
  {Weidenspointner}} \emph {et~al.},\ }\bibfield  {booktitle} {\emph {\bibinfo
  {booktitle} {{SP-622 The Obscure Universe}}},\ }\href@noop {} {\  (\bibinfo
  {year} {2007})},\ \bibinfo {note} {[ESA Spec. Publ.622,25(2007)]},\ \Eprint
  {http://arxiv.org/abs/astro-ph/0702621} {arXiv:astro-ph/0702621 [ASTRO-PH]}
  \BibitemShut {NoStop}%
\bibitem [{\citenamefont {{Weidenspointner}}\ \emph {et~al.}(2008)\citenamefont
  {{Weidenspointner}}, \citenamefont {{Skinner}}, \citenamefont {{Jean}},
  \citenamefont {{Kn{\"o}dlseder}}, \citenamefont {{von Ballmoos}},
  \citenamefont {{Bignami}}, \citenamefont {{Diehl}}, \citenamefont {{Strong}},
  \citenamefont {{Cordier}}, \citenamefont {{Schanne}},\ and\ \citenamefont
  {{Winkler}}}]{2008Natur.451..159W}%
  \BibitemOpen
  \bibfield  {author} {\bibinfo {author} {\bibfnamefont {G.}~\bibnamefont
  {{Weidenspointner}}}, \bibinfo {author} {\bibfnamefont {G.}~\bibnamefont
  {{Skinner}}}, \bibinfo {author} {\bibfnamefont {P.}~\bibnamefont {{Jean}}},
  \bibinfo {author} {\bibfnamefont {J.}~\bibnamefont {{Kn{\"o}dlseder}}},
  \bibinfo {author} {\bibfnamefont {P.}~\bibnamefont {{von Ballmoos}}},
  \bibinfo {author} {\bibfnamefont {G.}~\bibnamefont {{Bignami}}}, \bibinfo
  {author} {\bibfnamefont {R.}~\bibnamefont {{Diehl}}}, \bibinfo {author}
  {\bibfnamefont {A.~W.}\ \bibnamefont {{Strong}}}, \bibinfo {author}
  {\bibfnamefont {B.}~\bibnamefont {{Cordier}}}, \bibinfo {author}
  {\bibfnamefont {S.}~\bibnamefont {{Schanne}}}, \ and\ \bibinfo {author}
  {\bibfnamefont {C.}~\bibnamefont {{Winkler}}},\ }\href {\doibase
  10.1038/nature06490} {\bibfield  {journal} {\bibinfo  {journal} {\nat}\
  }\textbf {\bibinfo {volume} {451}},\ \bibinfo {pages} {159} (\bibinfo {year}
  {2008})}\BibitemShut {NoStop}%
\bibitem [{\citenamefont {Hooper}\ and\ \citenamefont
  {Goodenough}(2011)}]{Hooper:2010mq}%
  \BibitemOpen
  \bibfield  {author} {\bibinfo {author} {\bibfnamefont {D.}~\bibnamefont
  {Hooper}}\ and\ \bibinfo {author} {\bibfnamefont {L.}~\bibnamefont
  {Goodenough}},\ }\href {\doibase 10.1016/j.physletb.2011.02.029} {\bibfield
  {journal} {\bibinfo  {journal} {Phys. Lett.}\ }\textbf {\bibinfo {volume}
  {B697}},\ \bibinfo {pages} {412} (\bibinfo {year} {2011})},\ \Eprint
  {http://arxiv.org/abs/1010.2752} {arXiv:1010.2752 [hep-ph]} \BibitemShut
  {NoStop}%
\bibitem [{\citenamefont {Hooper}\ and\ \citenamefont
  {Linden}(2011)}]{Hooper:2011ti}%
  \BibitemOpen
  \bibfield  {author} {\bibinfo {author} {\bibfnamefont {D.}~\bibnamefont
  {Hooper}}\ and\ \bibinfo {author} {\bibfnamefont {T.}~\bibnamefont
  {Linden}},\ }\href {\doibase 10.1103/PhysRevD.84.123005} {\bibfield
  {journal} {\bibinfo  {journal} {Phys. Rev.}\ }\textbf {\bibinfo {volume}
  {D84}},\ \bibinfo {pages} {123005} (\bibinfo {year} {2011})},\ \Eprint
  {http://arxiv.org/abs/1110.0006} {arXiv:1110.0006 [astro-ph.HE]} \BibitemShut
  {NoStop}%
\bibitem [{\citenamefont {Abazajian}\ and\ \citenamefont
  {Kaplinghat}(2012)}]{Abazajian:2012pn}%
  \BibitemOpen
  \bibfield  {author} {\bibinfo {author} {\bibfnamefont {K.~N.}\ \bibnamefont
  {Abazajian}}\ and\ \bibinfo {author} {\bibfnamefont {M.}~\bibnamefont
  {Kaplinghat}},\ }\href {\doibase 10.1103/PhysRevD.86.083511,
  10.1103/PhysRevD.87.129902} {\bibfield  {journal} {\bibinfo  {journal} {Phys.
  Rev.}\ }\textbf {\bibinfo {volume} {D86}},\ \bibinfo {pages} {083511}
  (\bibinfo {year} {2012})},\ \bibinfo {note} {[Erratum: Phys.
  Rev.D87,129902(2013)]},\ \Eprint {http://arxiv.org/abs/1207.6047}
  {arXiv:1207.6047 [astro-ph.HE]} \BibitemShut {NoStop}%
\bibitem [{\citenamefont {Daylan}\ \emph {et~al.}(2016)\citenamefont {Daylan},
  \citenamefont {Finkbeiner}, \citenamefont {Hooper}, \citenamefont {Linden},
  \citenamefont {Portillo}, \citenamefont {Rodd},\ and\ \citenamefont
  {Slatyer}}]{Daylan:2014rsa}%
  \BibitemOpen
  \bibfield  {author} {\bibinfo {author} {\bibfnamefont {T.}~\bibnamefont
  {Daylan}}, \bibinfo {author} {\bibfnamefont {D.~P.}\ \bibnamefont
  {Finkbeiner}}, \bibinfo {author} {\bibfnamefont {D.}~\bibnamefont {Hooper}},
  \bibinfo {author} {\bibfnamefont {T.}~\bibnamefont {Linden}}, \bibinfo
  {author} {\bibfnamefont {S.~K.~N.}\ \bibnamefont {Portillo}}, \bibinfo
  {author} {\bibfnamefont {N.~L.}\ \bibnamefont {Rodd}}, \ and\ \bibinfo
  {author} {\bibfnamefont {T.~R.}\ \bibnamefont {Slatyer}},\ }\href {\doibase
  10.1016/j.dark.2015.12.005} {\bibfield  {journal} {\bibinfo  {journal} {Phys.
  Dark Univ.}\ }\textbf {\bibinfo {volume} {12}},\ \bibinfo {pages} {1}
  (\bibinfo {year} {2016})},\ \Eprint {http://arxiv.org/abs/1402.6703}
  {arXiv:1402.6703 [astro-ph.HE]} \BibitemShut {NoStop}%
\bibitem [{\citenamefont {Calore}\ \emph {et~al.}(2015)\citenamefont {Calore},
  \citenamefont {Cholis},\ and\ \citenamefont {Weniger}}]{Calore:2014xka}%
  \BibitemOpen
  \bibfield  {author} {\bibinfo {author} {\bibfnamefont {F.}~\bibnamefont
  {Calore}}, \bibinfo {author} {\bibfnamefont {I.}~\bibnamefont {Cholis}}, \
  and\ \bibinfo {author} {\bibfnamefont {C.}~\bibnamefont {Weniger}},\ }\href
  {\doibase 10.1088/1475-7516/2015/03/038} {\bibfield  {journal} {\bibinfo
  {journal} {JCAP}\ }\textbf {\bibinfo {volume} {1503}},\ \bibinfo {pages}
  {038} (\bibinfo {year} {2015})},\ \Eprint {http://arxiv.org/abs/1409.0042}
  {arXiv:1409.0042 [astro-ph.CO]} \BibitemShut {NoStop}%
\bibitem [{\citenamefont {Ajello}\ \emph {et~al.}(2016)\citenamefont {Ajello}
  \emph {et~al.}}]{TheFermi-LAT:2015kwa}%
  \BibitemOpen
  \bibfield  {author} {\bibinfo {author} {\bibfnamefont {M.}~\bibnamefont
  {Ajello}} \emph {et~al.} (\bibinfo {collaboration} {Fermi-LAT}),\ }\href
  {\doibase 10.3847/0004-637X/819/1/44} {\bibfield  {journal} {\bibinfo
  {journal} {Astrophys. J.}\ }\textbf {\bibinfo {volume} {819}},\ \bibinfo
  {pages} {44} (\bibinfo {year} {2016})},\ \Eprint
  {http://arxiv.org/abs/1511.02938} {arXiv:1511.02938 [astro-ph.HE]}
  \BibitemShut {NoStop}%
\bibitem [{\citenamefont {Aguilar}\ \emph {et~al.}(2016)\citenamefont {Aguilar}
  \emph {et~al.}}]{Aguilar:2016kjl}%
  \BibitemOpen
  \bibfield  {author} {\bibinfo {author} {\bibfnamefont {M.}~\bibnamefont
  {Aguilar}} \emph {et~al.} (\bibinfo {collaboration} {AMS}),\ }\href {\doibase
  10.1103/PhysRevLett.117.091103} {\bibfield  {journal} {\bibinfo  {journal}
  {Phys. Rev. Lett.}\ }\textbf {\bibinfo {volume} {117}},\ \bibinfo {pages}
  {091103} (\bibinfo {year} {2016})}\BibitemShut {NoStop}%
\bibitem [{\citenamefont {Cholis}\ \emph {et~al.}(2017)\citenamefont {Cholis},
  \citenamefont {Hooper},\ and\ \citenamefont {Linden}}]{Cholis:2017qlb}%
  \BibitemOpen
  \bibfield  {author} {\bibinfo {author} {\bibfnamefont {I.}~\bibnamefont
  {Cholis}}, \bibinfo {author} {\bibfnamefont {D.}~\bibnamefont {Hooper}}, \
  and\ \bibinfo {author} {\bibfnamefont {T.}~\bibnamefont {Linden}},\ }\href
  {\doibase 10.1103/PhysRevD.95.123007} {\bibfield  {journal} {\bibinfo
  {journal} {Phys. Rev.}\ }\textbf {\bibinfo {volume} {D95}},\ \bibinfo {pages}
  {123007} (\bibinfo {year} {2017})},\ \Eprint
  {http://arxiv.org/abs/1701.04406} {arXiv:1701.04406 [astro-ph.HE]}
  \BibitemShut {NoStop}%
\bibitem [{\citenamefont {Cholis}\ \emph {et~al.}(2016)\citenamefont {Cholis},
  \citenamefont {Hooper},\ and\ \citenamefont {Linden}}]{Cholis:2015gna}%
  \BibitemOpen
  \bibfield  {author} {\bibinfo {author} {\bibfnamefont {I.}~\bibnamefont
  {Cholis}}, \bibinfo {author} {\bibfnamefont {D.}~\bibnamefont {Hooper}}, \
  and\ \bibinfo {author} {\bibfnamefont {T.}~\bibnamefont {Linden}},\ }\href
  {\doibase 10.1103/PhysRevD.93.043016} {\bibfield  {journal} {\bibinfo
  {journal} {Phys. Rev.}\ }\textbf {\bibinfo {volume} {D93}},\ \bibinfo {pages}
  {043016} (\bibinfo {year} {2016})},\ \Eprint
  {http://arxiv.org/abs/1511.01507} {arXiv:1511.01507 [astro-ph.SR]}
  \BibitemShut {NoStop}%
\bibitem [{\citenamefont {Dolgov}\ and\ \citenamefont
  {Silk}(1993)}]{Dolgov:1992pu}%
  \BibitemOpen
  \bibfield  {author} {\bibinfo {author} {\bibfnamefont {A.}~\bibnamefont
  {Dolgov}}\ and\ \bibinfo {author} {\bibfnamefont {J.}~\bibnamefont {Silk}},\
  }\href {\doibase 10.1103/PhysRevD.47.4244} {\bibfield  {journal} {\bibinfo
  {journal} {Phys. Rev.}\ }\textbf {\bibinfo {volume} {D47}},\ \bibinfo {pages}
  {4244} (\bibinfo {year} {1993})}\BibitemShut {NoStop}%
\bibitem [{\citenamefont {Blinnikov}\ \emph {et~al.}(2015)\citenamefont
  {Blinnikov}, \citenamefont {Dolgov},\ and\ \citenamefont
  {Postnov}}]{Blinnikov:2014nea}%
  \BibitemOpen
  \bibfield  {author} {\bibinfo {author} {\bibfnamefont {S.~I.}\ \bibnamefont
  {Blinnikov}}, \bibinfo {author} {\bibfnamefont {A.~D.}\ \bibnamefont
  {Dolgov}}, \ and\ \bibinfo {author} {\bibfnamefont {K.~A.}\ \bibnamefont
  {Postnov}},\ }\href {\doibase 10.1103/PhysRevD.92.023516} {\bibfield
  {journal} {\bibinfo  {journal} {Phys. Rev.}\ }\textbf {\bibinfo {volume}
  {D92}},\ \bibinfo {pages} {023516} (\bibinfo {year} {2015})},\ \Eprint
  {http://arxiv.org/abs/1409.5736} {arXiv:1409.5736 [astro-ph.HE]} \BibitemShut
  {NoStop}%
\bibitem [{\citenamefont {Mottez}\ and\ \citenamefont
  {Zarka}(2014)}]{Mottez:2014awa}%
  \BibitemOpen
  \bibfield  {author} {\bibinfo {author} {\bibfnamefont {F.}~\bibnamefont
  {Mottez}}\ and\ \bibinfo {author} {\bibfnamefont {P.}~\bibnamefont {Zarka}},\
  }\href {\doibase 10.1051/0004-6361/201424104} {\bibfield  {journal} {\bibinfo
   {journal} {Astron. Astrophys.}\ }\textbf {\bibinfo {volume} {569}},\
  \bibinfo {pages} {A86} (\bibinfo {year} {2014})},\ \Eprint
  {http://arxiv.org/abs/1408.1333} {arXiv:1408.1333 [astro-ph.EP]} \BibitemShut
  {NoStop}%
\bibitem [{\citenamefont {Badenes}\ and\ \citenamefont
  {Maoz}(2012)}]{Badenes:2012ak}%
  \BibitemOpen
  \bibfield  {author} {\bibinfo {author} {\bibfnamefont {C.}~\bibnamefont
  {Badenes}}\ and\ \bibinfo {author} {\bibfnamefont {D.}~\bibnamefont {Maoz}},\
  }\href {\doibase 10.1088/2041-8205/749/1/L11} {\bibfield  {journal} {\bibinfo
   {journal} {Astrophys. J.}\ }\textbf {\bibinfo {volume} {749}},\ \bibinfo
  {pages} {L11} (\bibinfo {year} {2012})},\ \Eprint
  {http://arxiv.org/abs/1202.5472} {arXiv:1202.5472 [astro-ph.SR]} \BibitemShut
  {NoStop}%
\bibitem [{\citenamefont {Balestra}\ \emph {et~al.}(1985)\citenamefont
  {Balestra} \emph {et~al.}}]{Balestra:1985wk}%
  \BibitemOpen
  \bibfield  {author} {\bibinfo {author} {\bibfnamefont {F.}~\bibnamefont
  {Balestra}} \emph {et~al.},\ }\href {\doibase 10.1016/0370-2693(85)91227-4}
  {\bibfield  {journal} {\bibinfo  {journal} {Phys. Lett.}\ }\textbf {\bibinfo
  {volume} {165B}},\ \bibinfo {pages} {265} (\bibinfo {year}
  {1985})}\BibitemShut {NoStop}%
\bibitem [{\citenamefont {von Ballmoos}(2014)}]{vonBallmoos:2014zza}%
  \BibitemOpen
  \bibfield  {author} {\bibinfo {author} {\bibfnamefont {P.}~\bibnamefont {von
  Ballmoos}},\ }\bibfield  {booktitle} {\emph {\bibinfo {booktitle}
  {{Proceedings, 11th International Conference on Low Energy Antiproton Physics
  (LEAP2013): Uppsala, Sweden, June 10-15, 2013}}},\ }\href {\doibase
  10.1007/s10751-014-1024-9} {\bibfield  {journal} {\bibinfo  {journal}
  {Hyperfine Interact.}\ }\textbf {\bibinfo {volume} {228}},\ \bibinfo {pages}
  {91} (\bibinfo {year} {2014})},\ \Eprint {http://arxiv.org/abs/1401.7258}
  {arXiv:1401.7258 [astro-ph.HE]} \BibitemShut {NoStop}%
\bibitem [{\citenamefont {Tseliakhovich}\ and\ \citenamefont
  {Hirata}(2010)}]{Tseliakhovich:2010bj}%
  \BibitemOpen
  \bibfield  {author} {\bibinfo {author} {\bibfnamefont {D.}~\bibnamefont
  {Tseliakhovich}}\ and\ \bibinfo {author} {\bibfnamefont {C.}~\bibnamefont
  {Hirata}},\ }\href {\doibase 10.1103/PhysRevD.82.083520} {\bibfield
  {journal} {\bibinfo  {journal} {Phys. Rev.}\ }\textbf {\bibinfo {volume}
  {D82}},\ \bibinfo {pages} {083520} (\bibinfo {year} {2010})},\ \Eprint
  {http://arxiv.org/abs/1005.2416} {arXiv:1005.2416 [astro-ph.CO]} \BibitemShut
  {NoStop}%
\bibitem [{\citenamefont {Affleck}\ and\ \citenamefont
  {Dine}(1985)}]{Affleck:1984fy}%
  \BibitemOpen
  \bibfield  {author} {\bibinfo {author} {\bibfnamefont {I.}~\bibnamefont
  {Affleck}}\ and\ \bibinfo {author} {\bibfnamefont {M.}~\bibnamefont {Dine}},\
  }\href {\doibase 10.1016/0550-3213(85)90021-5} {\bibfield  {journal}
  {\bibinfo  {journal} {Nucl. Phys.}\ }\textbf {\bibinfo {volume} {B249}},\
  \bibinfo {pages} {361} (\bibinfo {year} {1985})}\BibitemShut {NoStop}%
\bibitem [{\citenamefont {Carr}\ and\ \citenamefont
  {Hawking}(1974)}]{Carr:1974nx}%
  \BibitemOpen
  \bibfield  {author} {\bibinfo {author} {\bibfnamefont {B.~J.}\ \bibnamefont
  {Carr}}\ and\ \bibinfo {author} {\bibfnamefont {S.~W.}\ \bibnamefont
  {Hawking}},\ }\href@noop {} {\bibfield  {journal} {\bibinfo  {journal} {Mon.
  Not. Roy. Astron. Soc.}\ }\textbf {\bibinfo {volume} {168}},\ \bibinfo
  {pages} {399} (\bibinfo {year} {1974})}\BibitemShut {NoStop}%
\bibitem [{\citenamefont {Carr}(1975)}]{Carr:1975qj}%
  \BibitemOpen
  \bibfield  {author} {\bibinfo {author} {\bibfnamefont {B.~J.}\ \bibnamefont
  {Carr}},\ }\href {\doibase 10.1086/153853} {\bibfield  {journal} {\bibinfo
  {journal} {Astrophys. J.}\ }\textbf {\bibinfo {volume} {201}},\ \bibinfo
  {pages} {1} (\bibinfo {year} {1975})}\BibitemShut {NoStop}%
\end{thebibliography}%
\end{document}